\documentclass[article]{aa}
\usepackage{txfonts}
\usepackage{graphicx}
\usepackage{xcolor}
\usepackage{ulem}
\bibpunct{(}{)}{;}{a}{}{,}
\usepackage{amsmath}
\usepackage{subcaption}

\newcommand{\kms}{km~s$^{-1}$}

\begin{document} 

%

  \title{Modeling the CO outflow in DG Tau B: \\ Swept-up shells versus perturbed MHD disk wind.}

   \subtitle{}

   \author{A. de Valon
          \inst{\ref{inst1}}
          \and
          C. Dougados 
          \inst{\ref{inst1}}
          \and
          S. Cabrit
          \inst{\ref{inst2}\and\ref{inst1}}
          \and
          F. Louvet
          \inst{\ref{inst1}\and\ref{inst3}}
          \and 
          L. A. Zapata
          \inst{\ref{inst4}}
          \and 
          D. Mardones
          \inst{\ref{inst3}}
          }

   \institute{Univ. Grenoble Alpes, CNRS, IPAG, 38000 Grenoble, France \label{inst1}
             \and
             Observatoire de Paris, PSL University, Sorbonne Universit\'{e},  CNRS UMR 8112, LERMA,  61 Avenue de l'Observatoire, 75014 Paris, France \label{inst2}
             \and 
             Departamento de Astronomia de Chile, Universidad de Chile, Santiago, Chile
             \label{inst3}
             \and
             Instituto de Radioastronomia y Astrofisica, Universidad Nacional Aut\'onoma de M\'exico, P.O. Box 3-72, 58090, Morelia, Michoac\'an, M\'exico
             \label{inst4}
            }

   \date{}

 
  \abstract
   {
   The origin of outflows and their exact impact on disk evolution and planet formation remain crucial open questions. DG Tau B is a Class I protostar associated with a rotating conical CO outflow and a structured disk. Hence it is an ideal target to study these questions.} 
   {We aim to characterize the morphology and kinematics of the DG Tau B outflow in order to elucidate its origin and potential impact on the disk.}
   {Our analysis is based on Atacama Large Millimeter Array (ALMA) $^{12}$CO(2-1) observations of DG Tau B at 0.15$^{\prime\prime}$ (20 au) angular resolution. We developed a tomographic method to recover 2D (R,Z) maps of vertical velocity $V_{\rm Z}$ and specific angular momentum $j=R\times V_{\phi}$. We created synthetic data cubes for parametric models of wind-driven shells and disk winds, which we fit to the observed channel maps.}
   {Tomographic analysis of the bright inner conical outflow shows that both $V_{\rm Z}$ and $j$ remain roughly constant along conical surfaces, defining a shear-like structure. We characterize three different types of substructures in this outflow (arches, fingers, and cusps) with apparent acceleration. Wind-driven shell models with a Hubble law fail to explain these substructures. In contrast, both the morphology and kinematics of the conical flow can be explained by a steady conical magnetohydrodynamic (MHD) disk wind with foot-point radii $r_0  \simeq 0.7 - 3.4$ au, a small magnetic level arm parameter ($\lambda \le 1.6$), and quasi periodic brightness enhancements. These might be caused by the impact of jet bow shocks, source orbital motion caused by a 25 M$_{\rm J}$ companion at 50 au, or disk density perturbations accreting through the wind launching region. The large CO wind mass flux (four times the accretion rate onto the central star) can also be explained if the MHD disk wind removes most of the angular momentum required for steady disk accretion.
   }
   {
   Our results provide the strongest evidence so far for the presence of massive MHD disk winds in Class I sources with residual infall, and they suggest that the initial stages of planet formation take place in a highly dynamic environment.
   }
   
   
   

   \maketitle

\section{Introduction}

Understanding the origin of protostellar flows is a key element to our full comprehension of the star formation process. Protostellar flows come in two components: high speed collimated jets and slower, often less collimated winds and outflows. We focus here on the slow molecular outflows which are traditionally associated with the earlier stages of star formation. However, they have also recently been detected around more evolved Class II systems \citep[e.g.,][]{pety_hh30,louvet_hh30_2018,lopez_ringed_2020}. Despite their ubiquity, the exact origin of molecular outflows, their link to the high-velocity jets, and their impact on the young forming star and disk are still crucial open questions. 
  
Two main paradigms are currently considered. The first traditional model describes these slow outflows as swept-up material, tracing the interaction between an inner jet or a wide-angle wind with the infalling envelope or parent core. These models have been mainly used for interpreting outflows from Class 0 and I stars  which are still surrounded by massive envelopes \citep{zhang_episodic_2019,Shang_unified_2020,lee_co_2000}. However, on small scales (less than a few 1000 au), recent observations have revealed rotating molecular outflows to originate from well within the disk at all evolutionary stages from Class 0 to Class II \citep[e.g.,  ][]{Launhardt2009, zapata_kinematics_2015, bjerkeli_resolved_2016,tabone_alma_2017,Hirota2017,louvet_hh30_2018,Zhang_Rotation_2018,Lee2018-HH211,de_valon_alma_2020,Lee_First_2021}. 
These observations suggest an alternative paradigm by which these slow molecular outflows, at least at their base, would trace matter directly ejected from the disk, by thermal or magnetic processes.
In support of this interpretation, the flow rotation signatures are consistent with an origin from disk radii $r_0 \simeq 1-50$ au (see references above), 
where \citet{panoglou_molecule_2012} have shown that magnetic disk winds could remain molecular.

These two paradigms imply different evolutions for the disk. Jet- and wind-driven shell models predict that an important mass is swept up from the envelope, impacting the reservoir of matter infalling onto the disk. Disk wind models predict an extraction of mass from the disk and, in the case of 
magnetohydrodynamic (MHD) models, an extraction of angular momentum which could drive disk accretion \citep{bai_magneto-thermal_2016}. 
Evaluating the contributions of each of these mechanisms to the slow molecular outflow emission requires high-angular resolution studies of the molecular outflow base. The flux of angular momentum extracted by the rotating molecular wind is estimated in only two sources so far, HH30 and HH212, and it is found sufficient to drive disk accretion at the current observed rate in both cases \citep{louvet_hh30_2018, Tabone2020}.

Recent high-angular resolution observations also reveal striking signatures of multiple CO shells in a few sources \citep{zhang_episodic_2019,lopez_ringed_2020}. Under the classical paradigm where CO outflows trace swept-up shells, they would require short episodic wind and jet outbursts every few 100 yrs. Characterizing and understanding the origin of these variabilities could bring critical insights into the star formation dynamics.

We present here an analysis of the DG~Tau~B CO outflow, based on recent Atacama Large Millimeter Array (ALMA) observations at 0.15$^{\prime\prime}$ resolution by \citet{de_valon_alma_2020} (hereafter DV20). DG~Tau~B is a Class I 1.1 M$_{\odot}$ protostar located in the Taurus cloud ($\approx$ 140 pc) 
and is associated with a bipolar atomic jet \citep{mundt_jets_1983} and a strongly asymmetric CO outflow first mapped by \citet{mitchell_dg_1997}. The bright redshifted CO outflow lobe displays a striking bright and narrow conical shape at its base. \citet{zapata_kinematics_2015} detect rotation signatures in the same sense as the disk. The ALMA observations by DV20 clearly confirm rotation in the bright inner conical redshifted lobe and show that it is surrounded by a wider and slower outflow. Residual infall signatures are detected at opening angles $\ge 70^{\circ}$, almost tangent to the disk surface.
In addition, DV20 report striking substructures in the CO channel maps at different line-of-sight velocities, reminiscent of the nested layers recently identified in HH46/47 by \citet{zhang_episodic_2019} and suggesting variability or interaction processes.
The exquisite levels of detail provided by these new ALMA observations 
provide a prime opportunity to distinguish between swept-up and disk wind origins.

In Sect. \ref{sec:description}, we recall the main properties of the DG Tau B outflow and characterize the three types of substructures visible in the channel maps. In Sect. \ref{sec:tomo} we present a model-independent analysis of the inner conical outflow component, which allowed us to retrieve 2D maps of the expansion velocity $V_{\rm Z}$ and specific angular momentum $j=R V_{\phi}$. We compare these overall properties with parametric models of wind-driven shells (in Sect. \ref{sec:WDS})  and disk winds (in Sect. \ref{sec:DW}). We discuss our results and their implications for the origin of the CO outflows in DG Tau B  in Sect. \ref{sec:Discussion}. Section \ref{sec:conclusion} summarizes our conclusions.

\section{Summary of outflow structure}
\label{sec:description}

\begin{figure*}
    \resizebox{\hsize}{!}{\includegraphics{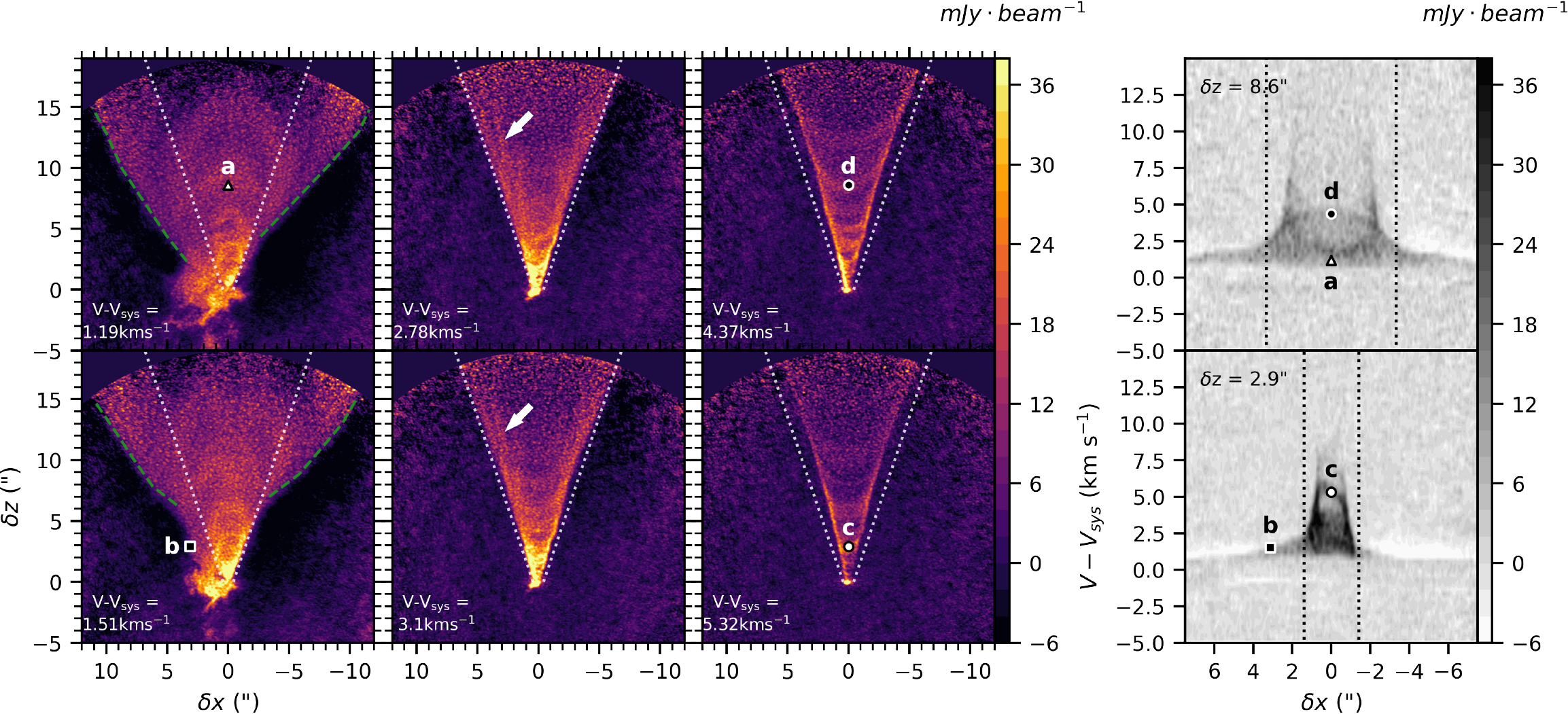}}
    \caption{ Substructures in the DG Tau B redshifted CO outflow.
   {\sl Left panels:} $^{12}$CO  channel maps at selected line-of-sight velocities. The white dashed line traces the $\theta=17^{\circ}$ outer limiting cone of the inner outflow as defined by DV20. {\sl Right panels}: Transverse PV diagrams across the flow averaged over a slice of $\Delta Z= 0.2^{\prime\prime}$. Black dotted lines indicate the outer limits of the conical outflow at the specified height.
    The triangle (a) and the two circle symbols (c and d) highlight the on-axis height of one arch and two different cusps, respectively. The square symbol (b) is located in the extended outer flow. The symbols are represented both on the channel maps and PV diagrams.}
    \label{fig:discrete}
\end{figure*}

Figure \ref{fig:discrete} summarizes the main properties of the DG~Tau~B redshifted CO outflow as identified in DV20. A narrow, limb brightened conical outflow is visible in the channel maps at $(V-V_{\rm sys}) > 2.0$ \kms. Its opening angle decreases from 17$^{\circ}$ at ($V-V_{\rm sys}) = 2.8$ \kms~ until 12$^{\circ}$ at $(V - V_{\rm sys})\ge 5.0$ \kms. 
The sheer-like velocity gradient across the conical layer is best seen in transverse position-velocity (hereafter PV) cuts (Fig.~\ref{fig:discrete}), where the flow width clearly narrows down at higher velocity, up $(V - V_{\rm sys}) > 5.0$ \kms.
The conical outflow is surrounded by a slower and wider outflowing component visible at $(V-V_{\rm sys}) < 2$ \kms. This outer flow is  visible in PV diagrams as an extended pedestal with a shallower velocity gradient (see Fig.~\ref{fig:discrete}).

Another striking property of the DG Tau B redshifted outflow, revealed by the ALMA observations in DV20, are brightness enhancements visible in channel maps. These various substructures, illustrated in Fig.~\ref{fig:discrete}, can be classified in three types:

    Bow-shaped intensity enhancements are visible at low velocities $(V-V_{\rm sys})=0.88-1.51$ \kms~(see Fig. \ref{fig:discrete}, left panels). We refer to these substructures as arches. The radial extent of the biggest arch is larger than the inner conical outflow, implying that this arch is at least partially formed outside of the conical outflow. The arch seems to increase in height with increasing velocity although this phenomenon is difficult to quantify due to the limited spectral sampling. 
    
    At intermediate velocities, from $(V-V_{\rm sys})=2.46-3.42$ \kms, thin quasi vertical lines (see white arrow in Fig.~\ref{fig:discrete}) are visible inside the conical outflow, close to the edge.  They are almost vertical at $(V-V_{\rm sys})=2.46$ \kms~and more open at higher projected velocity until they become almost tangent to the edge of the conical outflow at $(V-V_{\rm sys})~=~3.42$ \kms.  We refer to these substructures as fingers.
   
    At high velocities, from $(V-V_{\rm sys})=3.1$ to almost 7 \kms, multiple U-shaped structures are visible inside the conical outflow. We refer to these substructures as cusps. The contrast of these cusps is maximal at $(V-V_{\rm sys})=4.37$ \kms~and decreases with increasing velocity. The cusps show a signature of {\sl apparent acceleration}: their projected distance from the source increases with increasing projected velocity. 

\begin{table}[h!]
\caption{\label{tab:discrete} Characteristics of the observed arches and cusps}
\begin{tabular}{cccc}
\hline\hline
\multicolumn{4}{c}{Arches}                             \\ \hline
Name & position at  & Radial extension at & Aspect Ratio at  \\
       & $V_{\rm A}$\tablefootmark{a}(")&  $V_{\rm A}$\tablefootmark{a}(") & $V_{\rm A}$\tablefootmark{a}\\ \hline 
   A0     &14.9$\pm$ 0.2&9.7$\pm$ 0.3 &1.4\\
     A1     &9.4$\pm$ 0.2&? &?\\
   A2     &6.2$\pm$ 0.2&1.5$\pm$ 0.3 &1.2\\
   A3     &3.7$\pm$ 0.2&1$\pm$ 0.2 &1.4\\

       \hline\hline
\multicolumn{4}{c}{Cusps}                             \\ \hline
Name & position at & $N_{\rm chan}$   &  derivative  \\
       & $V_{\rm U}$\tablefootmark{b}(") &  &   ("/\kms ) \\ \hline 
U0 &11.9 $\pm$ 0.2         &(5)     &  3.2 $\pm$ 0.5      \\
U1 &8.2  $\pm$ 0.2        &(6)    &  2.2 $\pm$ 0.3     \\
U2&6.5  $\pm$ 0.1        &(4)     &  1.4 $\pm$ 0.5      \\
U3 &5.0 $\pm$ 0.1        &(6)     &  0.7 $\pm$ 0.4     \\
U4 &3.6  $\pm$ 0.1        &(3)     &  0.5 $\pm$ 0.2      \\
U5 &2.2  $\pm$ 0.1        &(4)     &  0.4 $\pm$ 0.2      \\\hline
\end{tabular}
\tablefoot{\tablefoottext{a}{$(V-V_{\rm sys})=1.19$~\kms }
\tablefoottext{b}{$(V-V_{\rm sys})=4.37$~ \kms}}
\end{table}

Table \ref{tab:discrete} lists the  characteristics of the main arches and cusps. We identify four arches (A0 to A3) and six cusps (U0 to U5). On the channel map at $(V-V_{\rm sys})=1.19$ \kms~we  derived the maximal height of each arch on axis (at $\delta x=0$) and the maximal radial extension. We divided these two values to derive the arch aspect ratio. 
The cusps are also characterized from the channel maps at 
$ 3.73 \boldsymbol{\le} (V-V_{\rm sys}) \le 5.32$ \kms~(See Fig. \ref{fig:u_tracking}). 
At higher velocities, the cusps could not be characterized because the outflow signal-to-noise ratio (S/N) decreases drastically. Moreover, the region at $\delta z < 2.2^{\prime\prime}$  was not studied because the cusps locations are complex to identify due to overlapping structures.
We  derived the cusp reference height on-axis on the channel map at ($V-V_{\rm sys}$)~=~4.37 \kms. 
The apparent acceleration of each cusp, in ($^{\prime\prime}$)/\kms~listed in Table~1, was obtained by measuring the average spatial shift of the cusp between two consecutive channel maps (taking as error bar the rms dispersion between measurements in different channels).

Internal discrete structures are also visible in transverse PV-diagrams as pseudo-ellipses (see Fig. \ref{fig:discrete}).  The top and bottom of the ellipses seem to match with respectively the top of some arches and bottom of some cusps  (see Fig. \ref{fig:discrete}). This potentially indicates that cusps and arches are linked to the same phenomenon. We present a model-dependent study of these ellipses in Sect.~\ref{sec:DW}.

\section{Tomography of the inner conical outflow}
\label{sec:tomo}

In this section, we develop a model-independent method that allows us to recover the dynamics and the morphology of the inner conical outflow component. This method assumes that the outflow is axisymmetric. We later discuss possible departures from axisymmetry and their implications on the analysis conducted here.

\subsection{Method}

\begin{figure}[htp]
\centering
\includegraphics[clip,width=0.7\columnwidth]{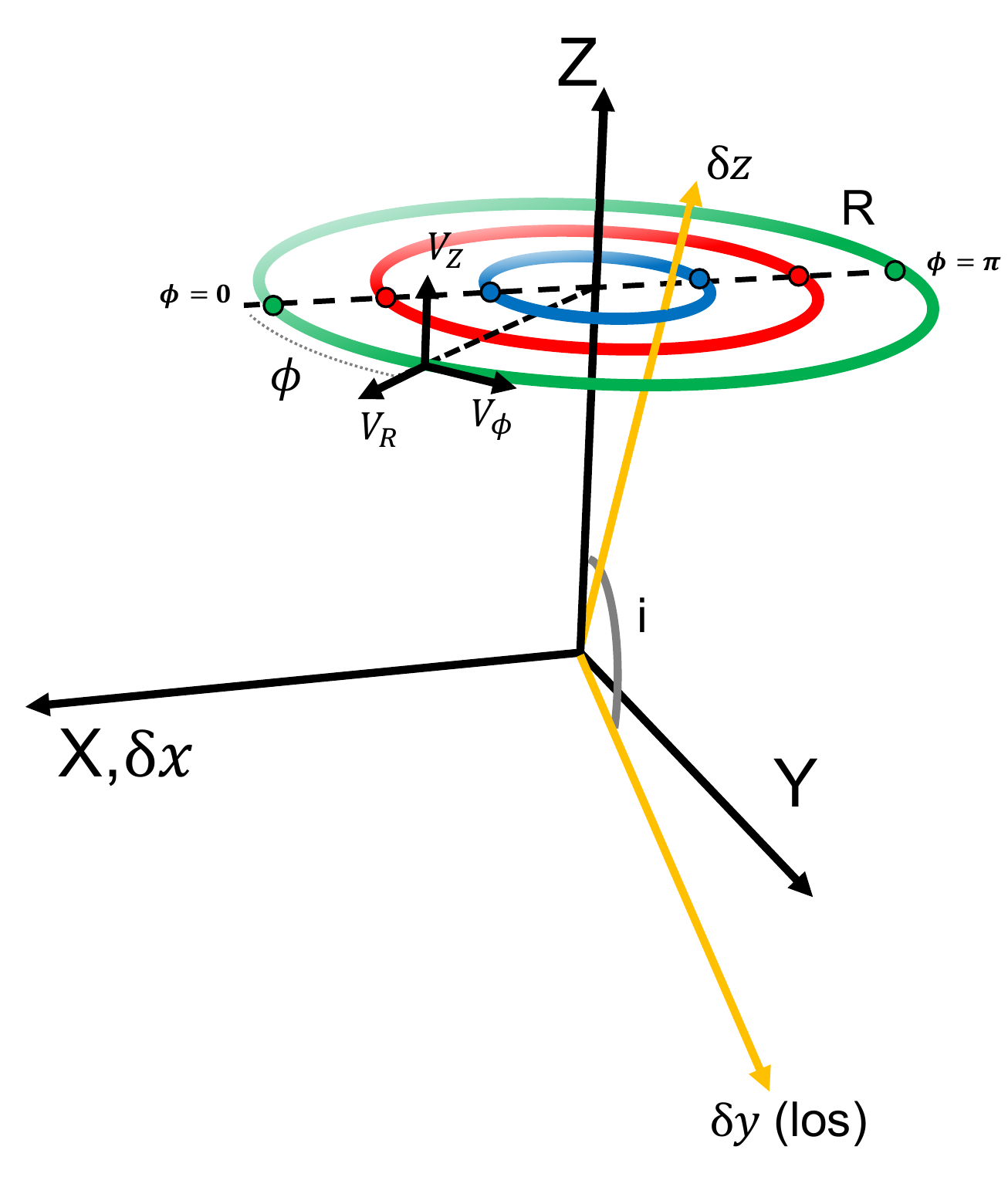}
\includegraphics[clip,width=\columnwidth]{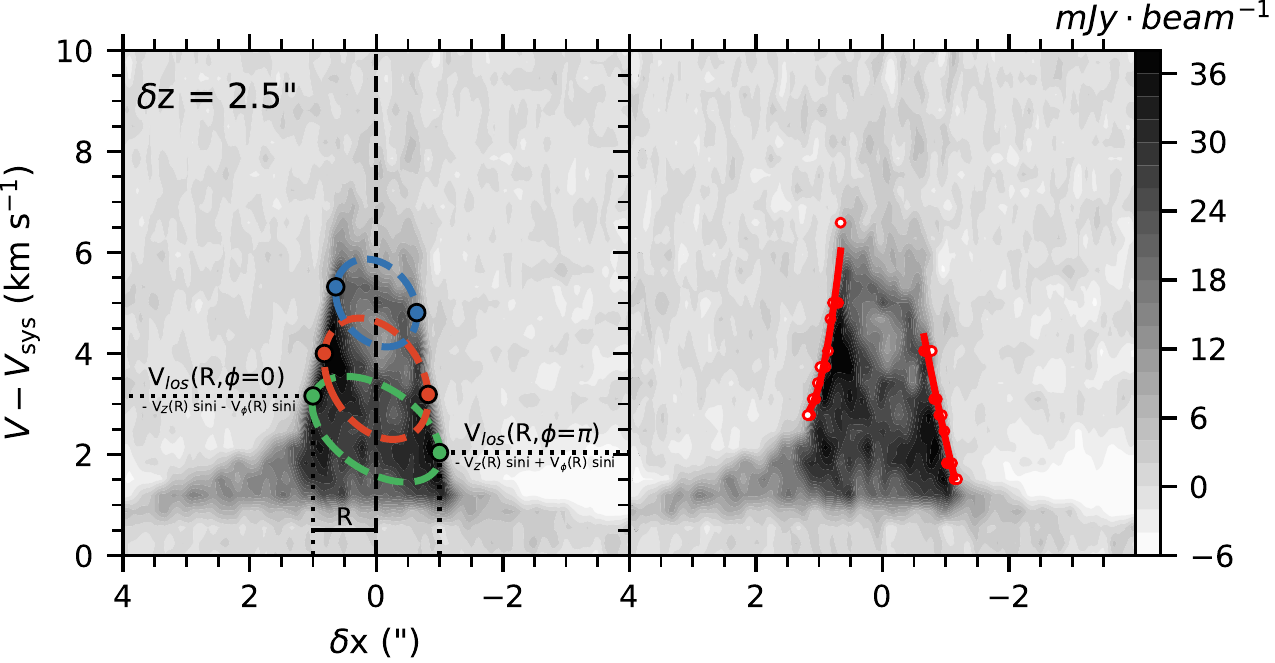}
\caption{ Principle of tomographic reconstruction method. {\sl Top panel}: 3D representation of the two reference systems used: the outflow  $(X,Y,Z)$ in black and the observer  $(\delta x,\delta y,\delta z)$ in yellow. The colored circles illustrate the emitting rings defined by five parameters (see text). The colored dots trace the locations at $\phi=0$ and $\phi=\pi$ along the emitting rings. {\sl Bottom panels}: Transverse PV diagrams at $\delta z = 2.5^{\prime\prime}$ across the flow axis and averaged over a slice of $\Delta z=0.2^{\prime\prime}$. In the left panel is shown the schematic projection of the colored rings and corresponding colored dots in the PV diagram. Because of the flow inclination, the rings are in fact projected at slightly different heights. Their real projection is studied in Sect. \ref{sec:ellipse_incl}. The white dots in the right panel illustrate the outer limits of the PV diagram. The red curve shows the polynomial fitting of the two edges.}
\label{fig:referential}
\end{figure}

We  followed \citet{louvet_hh30_2018} who modeled the outflow of HH30 at a given vertical offset by an emitting ring with radius R and  extended their method to take into account the inclination of the outflow. For this purpose, we defined the outflow and the observer reference systems (see Fig. \ref{fig:referential}). 
On the outflow reference system, $\Vec{Z}$ is defined by the outflow axis and $\Vec{X}$ is tangent to the plane of sky. The observer reference system is defined by $\Vec{\delta z}$ the projection of the outflow axis onto the plane of the sky, $\Vec{\delta y}$, the line-of-sight direction, and $\Vec{\delta x}=\Vec{X}$, in the plane of the sky. 
The inclination of the outflow $i$ is then defined by the angle between $\Vec{\delta y}$ and $\Vec{Z}$. In the case of edge-on disks such as HH30, the two reference systems are identical.  We  modeled one layer of the outflow at a specified height Z by an emitting ring of radius R and azimuthal angle $\phi$ (see Fig. \ref{fig:referential}) with $X=R\cos{\phi}$ and $Y=R\sin{\phi}$. For each ring the velocity components are defined in cylindrical coordinates with: $V_{\rm R}(R,Z)$, $V_{\rm Z}(R,Z)$ and $V_{\phi}(R,Z)$ (see Fig. \ref{fig:referential}). Hence, an emitting ring is defined by 5 parameters: $Z$, $R$, $V_{ \rm R}(R,Z)$, $V_{\rm Z}(R,Z)$ and $V_{\phi}(R,Z)$.

The observational coordinates on a  position-position-velocity (PPV) data cube are defined by the projection of the outflow on the plane of sky ($\Vec{\delta x},\Vec{\delta z}$) and the projected velocities on the line of sight $V_{los}=-\Vec{V}\cdot\Vec{e_{\rm y}}$ with redshifted velocities considered as positive. This depends on $R$,$Z$ and $\phi$ as:
\begin{eqnarray}    
    \label{eqx}
    \delta x   & = &R \cos{\phi}\\ 
    \label{eqz}
    \delta z & =& Z\sin{i}-R\sin{\phi}\cos{i}\\
    \label{eqv}
    V_{\rm los} & = & -V_{\rm z}\cos{i} - V_{\phi}\cos{\phi}\sin{i}
    - V_{\rm R}\sin{\phi}\sin{i}
.\end{eqnarray}

A transverse PV diagram corresponds to a pseudo-slit of the data cube perpendicular to the flow axis. This corresponds to a solution of Eqs. \ref{eqx},\ref{eqz},\ref{eqv} with $\delta z=cst$.
In the case of edge-on flows, a ring traces a perfect ellipse in the PV diagram. A fit of these ellipses give complete information about the morphology and dynamics of the outflow as shown by \citet{louvet_hh30_2018}. 

In the case of an inclined outflow such as DG Tau B, different rings overlap on the transversal PV diagrams (see Fig. 2 left). Hence it  was not possible to fit them individually. However constraints on some of the ring parameters  could be derived from characterizing the outer limits of the PV diagram. The radius of the ring corresponds on the first order to $\delta x_{max} = \delta x(\phi \approx 0,\pi)$. In addition, the  projected velocities at the edge of the ellipses $V_{\rm los}(\delta x_{\rm max})$  allow one to recover both $V_{\rm Z}(R,Z)$ and $V_{\phi}(R,Z)$ from the following equations: 
\begin{eqnarray}    
    \label{eqx2}
    Z   & = &\frac{\delta z}{\sin{i}}\\ 
    \label{eqz2}
    R   & = &\delta x_{\rm max} \\ 
    \label{eqvz}
    V_{\rm Z} & \simeq &\frac{ V_{\rm los}(\delta x_{\rm max})+V_{\rm los}(-\delta x_{\rm max})}{-~2\cos{i}}\\
    \label{eqvphi}
    V_\phi & \simeq &\frac{ V_{\rm los}(\delta x_{\rm max})-V_{\rm los}(-\delta x_{\rm max})}{2\sin{i}}
.\end{eqnarray}

 By consequence, characterizing the outer limits of the PV diagram along multiple heights allowed us to recover a 2D map of the expansion velocity $V_{\rm Z}$ and specific angular momentum $j=R V_{\phi}$. In the following, we  used the inclination derived from DV20 at $i = 117^{\circ}\pm 2$. To characterize the outer shape of the transverse PV diagram at a given $\delta z$, we  derived the maximal projected velocity for each value of $\pm\delta x$. Numerically, we  computed the gradient of the emission profile at a fixed $\delta x$ and  localized its maximum. We also derived an uncertainty
on $V_{\rm los}$ which is found to vary in the range 0.05 to 0.2~\kms. We used in the following a mean value of 0.1~\kms. Figure \ref{fig:referential} illustrates our method to  determine $V_{\rm los}$ on the edges of the PV diagram.This procedure  failed at low radii (or high velocities) because of the low S/N and the almost vertical profile that  generated high uncertainties on the velocity estimate. 
The determination of the velocity  was also limited by our spectral resolution of 0.3~\kms. We  fit the two edges for each PV diagram with a polynomial curve (see Fig.~\ref{fig:referential}). We also  applied a Gaussian filter with $\delta z$ using a standard deviation of 0.16$^{\prime\prime}$. 

\subsection{Results}

\begin{figure*}
    \resizebox{\hsize}{!}{\includegraphics{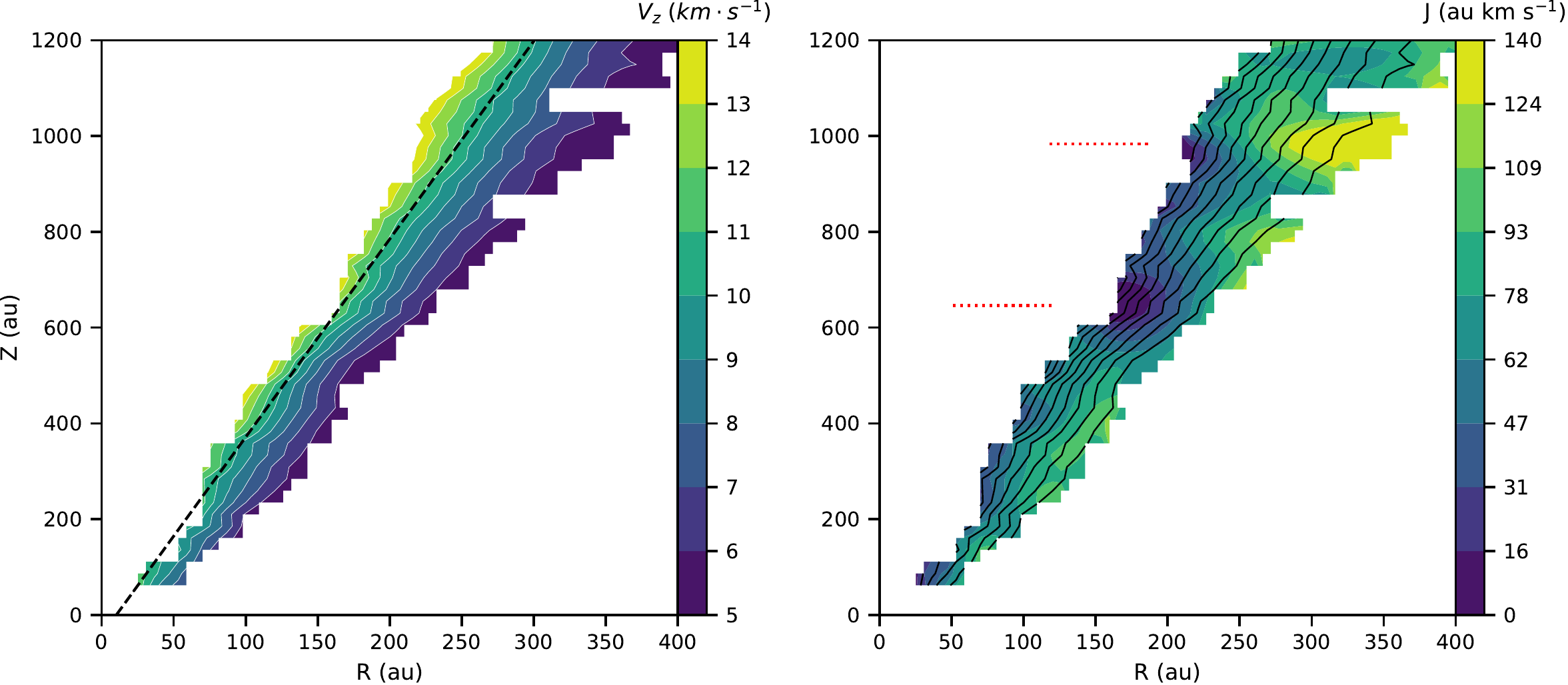}}
    \caption{Tomographic maps of $V_{\rm Z}$ (left panel) and specific angular momentum $j$ (right panel) in the outflow referential. The black dashed line traces the conical fit of the region $V_{\rm Z}=10-11$~\kms. The white (resp. black) contours in the left (resp. right) panel show $V_{\rm Z}$ contours. The red dashed lines indicate the height of the two extrema in specific angular momentum.}
    \label{fig:VZ_J}
\end{figure*}

Using Eqs. \ref{eqz2}, \ref{eqx2}, \ref{eqvz}, and \ref{eqvphi}, a tomographic map of $V_{\rm Z}(R,Z)$ and $V_{\rm \phi}(R,Z)$ could be recovered. We show the specific angular momentum $R V_{\rm \phi}(R,Z)$ instead of the rotation alone as this is more meaningful in the understanding of the dynamics. Figure \ref{fig:VZ_J} shows the resulting tomographic map of $V_{\rm Z}$ and $R V_{\phi}$ in the outflow referential. The tomography efficiently traces the conical shape visible on the channel maps. Curves of constant $V_{\rm Z}$ trace conical surfaces with semi-opening angles varying from 12$^{\circ}$ for the highest velocities to 17$^{\circ}$ for the lowest velocities. $V_{\rm Z}$ radially decreases from $\approx$ 14~\kms~to 5~\kms. This range of velocities is conserved until at least $Z=1200$ au.

\begin{figure}
    \resizebox{\hsize}{!}{\includegraphics{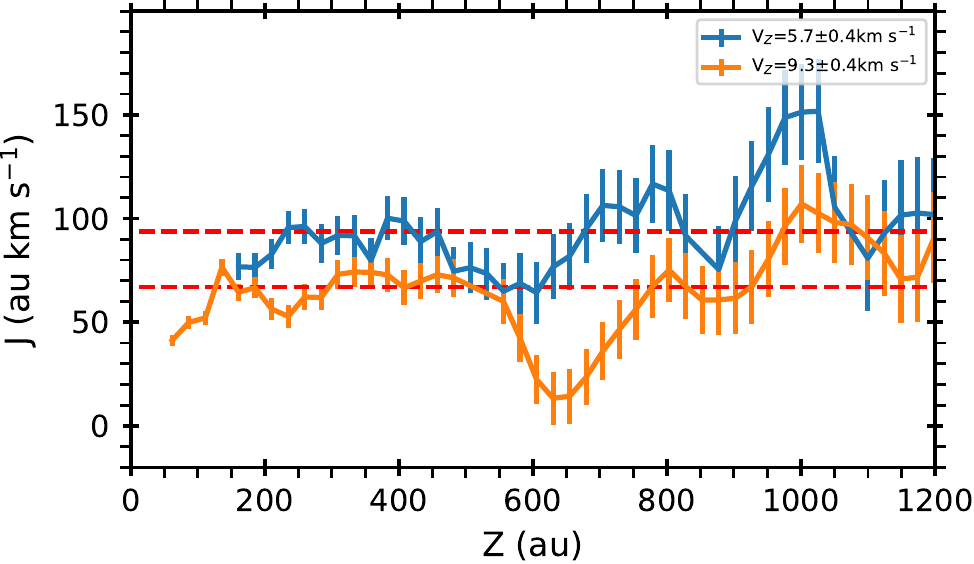}}
    \caption{Specific angular momentum $j$ along curves of constant $V_{\rm Z}$. The corresponding range of $V_{\rm Z}$ is shown in the box.  The uncertainty of specific angular momentum was obtained by propagating the $V_{\rm proj}$ uncertainty. The red dashed line corresponds to the median value of each curve.}
    \label{fig:XV_J}
\end{figure}

The specific angular momentum derived from the tomographic study varies from 0 to 140 au~\kms~ and is consistently in the same sense as the disk rotation. At $Z < 500$~au the specific angular momentum increases with radius from $\approx 30$~au~\kms~in the inner radius to $\approx 100$~au~\kms~on the outer radius.
The specific angular momentum is also roughly constant on conical lines of constant $V_{\rm Z}$ until $Z \approx 500$ au (see Fig.~\ref{fig:XV_J}). Our average value around 70~au~\kms~is consistent with the previous estimate of DV20 of $\approx$~65~au~\kms.

Two extrema in the specific angular momentum map can be observed at $Z=550-800$ au and $900-1100$ (see Fig.~\ref{fig:XV_J}). At the lowest altitude, the specific angular momentum reaches zero while at the highest altitude the specific angular momentum increases up to $j>140$~au~\kms. These irregularities are also visible on channel maps. They correspond to regions where bumps in the cones are observed: toward $\delta x<0$ at $\approx 4.5^{\prime\prime}$, $\delta x>0$ at $\approx 6.5^{\prime\prime}$ (see the channel map $(V-V_{\rm sys})= 4.37$~\kms~on Fig. \ref{fig:discrete}). These bumps may be due to local radial displacements of the outflow axis, due for example to wiggling. We study the impact of small amplitude wiggling in \ref{sec:wigg}.

\subsection{Limitations and biases}

In this section, we discuss the different biases and limitations of this tomographic study. Firstly,
this study could not recover the radial velocity component $V_{\rm R}$ as it impacts mostly the size of the ellipse at $x=0$ where all the ellipses are stacked. A model-dependent study to characterize this radial velocity will be achieved in Sect. \ref{sec:DW}.
Furthermore, in order to apply this method, it is critical that the centers of the rings 
are not significantly displaced from $\delta x=0$. Such displacements can be induced by a poorly estimated outflow position axis (PA) or by outflow axis wiggling.  We derived the PA of the redshifted outflow in Sect. \ref{sec:axis} at
$\rm PA =295^{\circ} \pm 1^{\circ}$. This value is in very good agreement with the disk rotation axis $\rm PA=115.7^{\circ} \pm 0.3^\circ$ determined by DV20.
We also  determined an upper limit of 0.5$^{\circ}$ for the wiggling of the CO outflow axis. 
We discuss in Appendix \ref{sec:wigg} the impact of possible low-amplitude wiggling on our results.

The different biases  were also computed.
We  show in Appendix \ref{sec:ellipse_incl} that assuming that the maximal radial extent  corresponded to $\phi = 0,\pi$  was partially inaccurate, and could introduce a bias in the estimate of $R$, $Z$, $V_{\rm Z}(R,Z)$ and $V_{\phi}(R,Z)$. The effect of ellipse stacking and its effect on the estimate of the velocities  was studied in Appendix \ref{sec:ellipse_stack}. We evaluate at $\lesssim$ 20\% the potential bias in our estimate of the conical outflow dynamics. Our estimates of $V_{\rm Z}$ and specific angular momentum are overestimated and underestimated respectively (see Appendix~\ref{sec:ellipse_stack}). We estimate the bias on $R$ and $Z$ to be respectively $< 1.5 \%$ and $< 3 \%$, resulting in an error $< 3 \%$ on our estimate of the opening angle $\theta$.

The highly asymmetric pedestal emission visible on the two sides of the PV diagram at $(V-V_{\rm sys}) < 2$~\kms~traces the outer region (see Fig.~\ref{fig:discrete}). DV20  show that this region is outflowing and surrounds the conical outflow.
We  did not attempt to apply the tomographic method to this region.
From its morphology in the channel maps, we  derived for this component an opening angle $> 30^{\circ}$. Such a large opening angle produces a bias of $\approx 60\%$ in the estimation of $V_{\rm Z}$ using our reconstruction method (see Appendix~\ref{sec:bias_tomo}). Moreover, this pedestal may potentially be explained by the top of one large ellipse, with the extremal region located at velocities $<~1$~\kms, absorbed by the medium. In that case, our reconstruction method is not applicable. By consequence, we  did not apply our method for line-of-sight
velocities $(V-V_{\rm sys})< 2$~\kms.   In the following, we investigate to which extent  wind-driven shells and disk winds can account for both the conical velocity stratification determined here and the striking substructures (arches, fingers, cusps) identified in Sect. 2.

\section{Wind-driven shell modeling}
\label{sec:WDS}

The traditional interpretation proposed for CO molecular outflow cavities around young stars is that they trace shells of ambient material swept up 
by a wide-angle wind or by jet bow shocks \citep[see][for reviews]{Cabrit_models_1997,lee_co_2000,Arce_molecular_2007}. 
In this section, we investigate the simplest and most widely used model to interpret CO outflow observations, namely the wind-driven shell (hereafter WDS)  solution of \citet{lee_co_2000} where the shell is a parabola that expands radially in all directions with a velocity proportional to the local 
distance from the source  (hereafter referred to as the Hubble law). Such a shell structure is predicted under a set of specific 
conditions in the wind and ambient medium\footnote{It is obtained when a wide-angle wind with velocity varying with angle as $V_w \propto \cos\theta$ and density varying as $\propto 1/(r^2 \sin^2\theta)$ sweeps-up a static, flattened isothermal core with density $\propto \sin^2\theta/r^2$, and they mix instantly in the shell. We note that the radial shell expansion results from instant mixing, while the Hubble law derives from the identical radial fall-off of wind and ambient density (both $\propto 1/r^2$), which yields a shell speed that is constant over time \citep{shu_star_1991}. Finally, the parabolic shell shape derives from the combined $\theta$-dependencies of the densities and wind speed  \citep{Lee_Hydrodyamic_2001}.}(see Sect. 6 for details). This simple WDS model is recently shown by \citet{zhang_episodic_2019} to reproduce several features of the multiple CO shell  structures at the base of the HH46/47 molecular outflow. Therefore it is natural to investigate whether the same WDS model can also reproduce the morphology and kinematics of the DG~Tau~B outflow, on smaller spatial scales.

Following \citet{lee_co_2000}, the parabolic morphology and the radial  Hubble-law kinematics of the shell can be empirically described by two parameters, $C$ and $\tau$, through:
\begin{equation}
\label{eq:wds}
\centering
Z= C \times R^2 \qquad \qquad V_{\rm Z}=\frac{Z}{\tau} \qquad \qquad V_{\rm R}=\frac{R}{\tau},
\end{equation}
where $\tau$ is the age of the shell, $C$ is the inverse size of the parabola at $\theta = 45\degr$ (where $Z = R = 1/C$), and
the product $\tau C$ defines the shell expansion speed at each polar angle $\theta = \arctan(R/Z)$ through:
\begin{eqnarray}
\label{eq:vel-theta-wds}
V_{\rm Z}(\theta) & = \frac{1}{\tau C} \left(\frac{Z}{R}\right)^2 = \left({\tau C \tan^2\theta}\right)^{-1} \\
V_{\rm R}(\theta) & = \frac{1}{\tau C} \left(\frac{Z}{R}\right) = \left(\tau C \tan\theta\right)^{-1}.
\end{eqnarray}

The above equations always produce ellipses in both channel maps and transverse PV diagrams \citep{lee_co_2000}. This is a direct result of the assumed  Hubble law, where the shell velocity vector is proportional to the position vector. A channel map at a given line-of-sight velocity is then equivalent to making a cut through the shell at a given depth along the line of sight and this cut is shaped as an ellipse. Similarly, a transverse PV diagram has the same (elliptical) shape as a cut through the shell at the corresponding projected height.
On channel maps, the ellipse is projected at increasing distances from the source with increasing velocity, due to the Hubble law. Similarly, on transverse PV diagrams, the mean velocity of the ellipse increases with the distance of the PV cut from the source \citep[see Figs. 24 and 26 in][]{lee_co_2000}. 
In Appendix \ref{sec:Aspect_ratio}, we  derived analytical formulae for the center of the ellipse in channel maps as well as for its aspect ratio. Interestingly, we find that the ellipse aspect ratio only depends on the inclination $i$ and is equal to $1/|\cos{i}|$ for the classical model of Eq. \ref{eq:wds} (see Appendix \ref{sec:Aspect_ratio}). In the following, we attempt to fit with this model first the low-velocity outer flow component, and then the bright conical outflow and its discrete structures.

\subsection{Low-velocity outer flow }
\label{sec:outer}

\begin{figure*}
    \resizebox{\hsize}{!}{\includegraphics{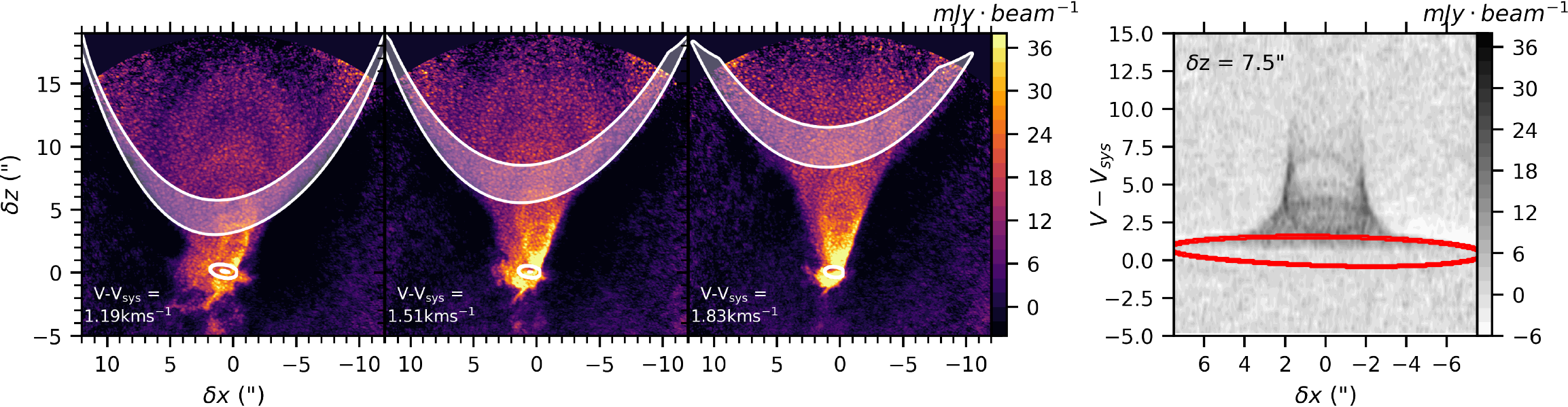}}
    \caption{ Comparison of the low-velocity outer CO outflow with a classical WDS model. {\sl Left panels}: $^{12}$CO individual channel maps at different line-of-sight velocities tracing the low-velocity outer CO outflow. The white contours trace the model of a WDS of parabolic shape defined by $C=10^{-3}$ au$^{-1}$, dynamical age $\tau=6000$ yr and specific angular momentum $j = 250\pm 50$~ au \kms~with an inclination of $i=117^{\circ}$. {\sl Right panel}: Transverse PV cut at $\delta z =7.5^{\prime\prime}$ averaged over a slice of $\Delta Z=0.2^{\prime\prime}$. The red ellipse traces the WDS model. ($V-V_{\rm sys}$) units are  \kms.}
    \label{fig:outer}
\end{figure*}

The wide and low-velocity outflow at $(V-V_{\rm sys}) < 2$ \kms\ shows several properties suggestive of a "classical" parabolic WDS with a Hubble law. Its outer border in channel maps has a parabolic shape, and it exhibits a larger offset from the origin at higher line-of-sight velocities (see Fig.1 left panels).  
Although such WDS models do not usually consider rotation, we  included rotation to properly fit the large left-right asymmetry observed in the channel maps. To reduce the number of free parameters, we  considered that the specific angular momentum $j$ is the same at all positions of the swept-up shell. $C$  was then fixed by the global parabolic shape of the cavity, $\tau$ by its spatial shift between the channel maps, and $j$ by its global left-right asymmetry. 

Figure \ref{fig:outer} (left panels) shows that the outer contour of the low-velocity outflow and its increased spatial offset with velocity are well fit by a WDS obeying Eq.~\ref{eq:wds} with parameters $C = 10^{-3}$ au$^{-1}$, $\tau$ = 6000 years and $j = 250 \pm 50$ au \kms. The rightmost panel in Fig. \ref{fig:outer} shows that this shell model reproduces well the most extended, lowest velocity emission of the broad pedestal in transverse PV cuts; the predicted blue-shifted emission from the front side of the shell falls very close to systemic velocity, consistent with its nondetection in our data. On the other hand, our assumption of a thin parabolic shell does not match the observed outflow thickness at high altitudes.This discrepancy is visible at $\delta z\approx 15^{\prime\prime}$ on Fig.~\ref{fig:outer} where the observed width of the emissive outer layer is $\approx 3^{\prime\prime}$, significantly larger than predicted by our model. The shell should actually have a thickness  $\simeq 3^{\prime\prime} \simeq 500$~au.

\subsection{Conical outflow and discrete structures}
\label{sec:WDS_conical}

\begin{figure}
    \resizebox{\hsize}{!}{\includegraphics{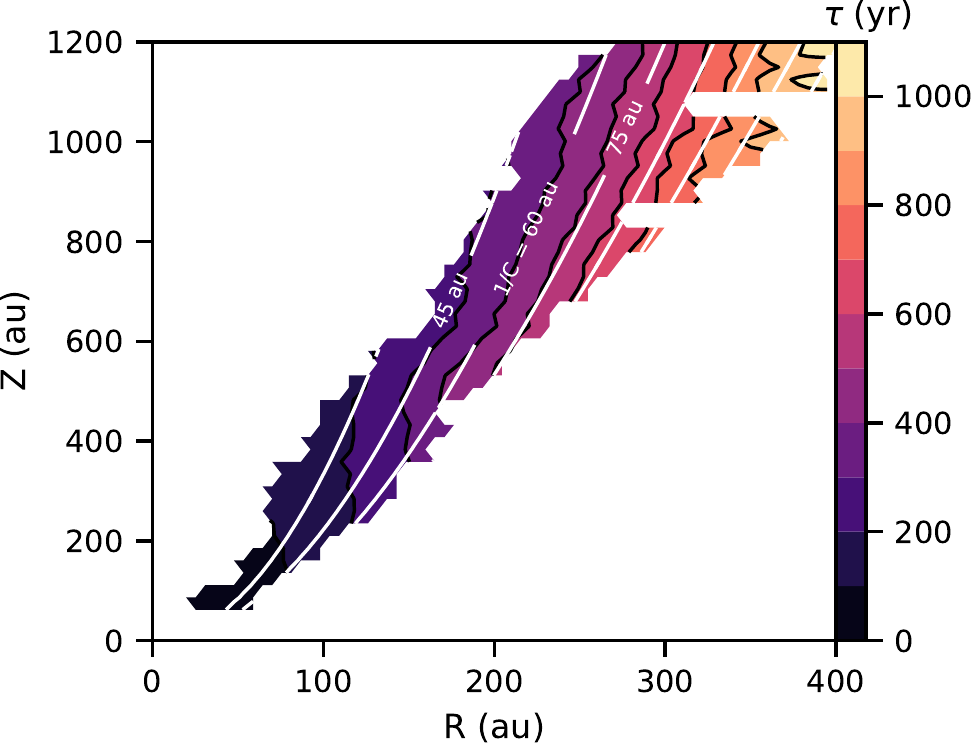}}
    \caption{Dynamical times ($\tau=Z/V_{\rm Z}$) derived from the tomographic map of $V_{\rm Z}$ are shown in color. The white lines correspond to a parabolic curve with $1/C$ varying from 30 to 135~au, in steps of 15~au.}
    \label{fig:tomo_tauC}
\end{figure}

In this section, we attempt to model the conical outflow and its discrete structures (arches, cusps, fingers, described in Sect. \ref{sec:description}) by a stacking of several parabolic wind-driven shells with a Hubble velocity law. Several qualitative features are suggestive of such a model: 
the loop shapes of the arches at low velocity are reminiscent of the ellipses predicted in channel maps (see Appendix \ref{sec:Aspect_ratio}),
the apparent acceleration of the cusps (increased altitude with increasing velocity) is reminiscent of the predicted Hubble law dynamics, and
finally, in the conical flow studied by tomography, contours of constant $\tau = Z/V_{\rm Z}$ 
follow quasi-parabolic curves above $Z \simeq 400$~au (see Fig. \ref{fig:tomo_tauC}).
Hence we investigate below whether the conical outflow could be made of successive nested parabolic wind-driven shells, 
where the apparent continuous aspect of the tomography would be an artifact of our limited spatial and spectral sampling,
and the discrete structures (arches, cusps, fingers) would trace a few individual shells brighter than average.

Contrary to the slow outer flow modeled in Sect. \ref{sec:outer}, the left-right asymmetry in these faster flow regions is small. Therefore, we  neglected rotation when fitting WDS models to the channel maps. We set the WDS axis inclination equal to the large-scale disk inclination derived from ALMA studies ($i=63^\circ \pm 2^\circ$, DV20), leading to $i=180\degr - 63\degr = 117\degr$ in the redshifted lobe.  

Here, we find that the "classical" WDS model of \cite{lee_co_2000}
encounters a major problem, as shown in the top row of Fig. \ref{fig:loops}: the predicted aspect ratio of ellipses in channel maps, $A=1/|\cos{i}|$ (see Appendix \ref{sec:Aspect_ratio}) is too large ($\simeq 2.2$).
In order to match the observed aspect ratio of the arches ($\approx1.4$), the inclination of the shell axis
should be $i \approx
135\degr$\ instead of $i=117\degr$. In the WDS model, however, the direction of shell elongation is not arbitrary but must 
follow the direction of both highest wind density (traced by the axial jet) and lowest ambient density (traced by core flattening). Proper motions of jet knots in DG Tau B imply a jet axis inclination of $i \geq 65^{\circ}$ for the blue-shifted lobe \citep{eisloffel_imaging_1998}, hence $i \leq 115^{\circ}$ for the redshifted lobe. This limit agrees within 2\degr\ with the large-scale disk inclination determined by ALMA ($i = 63\degr \pm 2\degr$, DV20), which should follow the core flattening. Therefore, we can exclude a shell axis at $i \approx 135\degr$\ as a solution to the ellipse aspect-ratio problem of the WDS model of \cite{lee_co_2000}. 

\begin{figure*}
    \resizebox{\hsize}{!}{\includegraphics{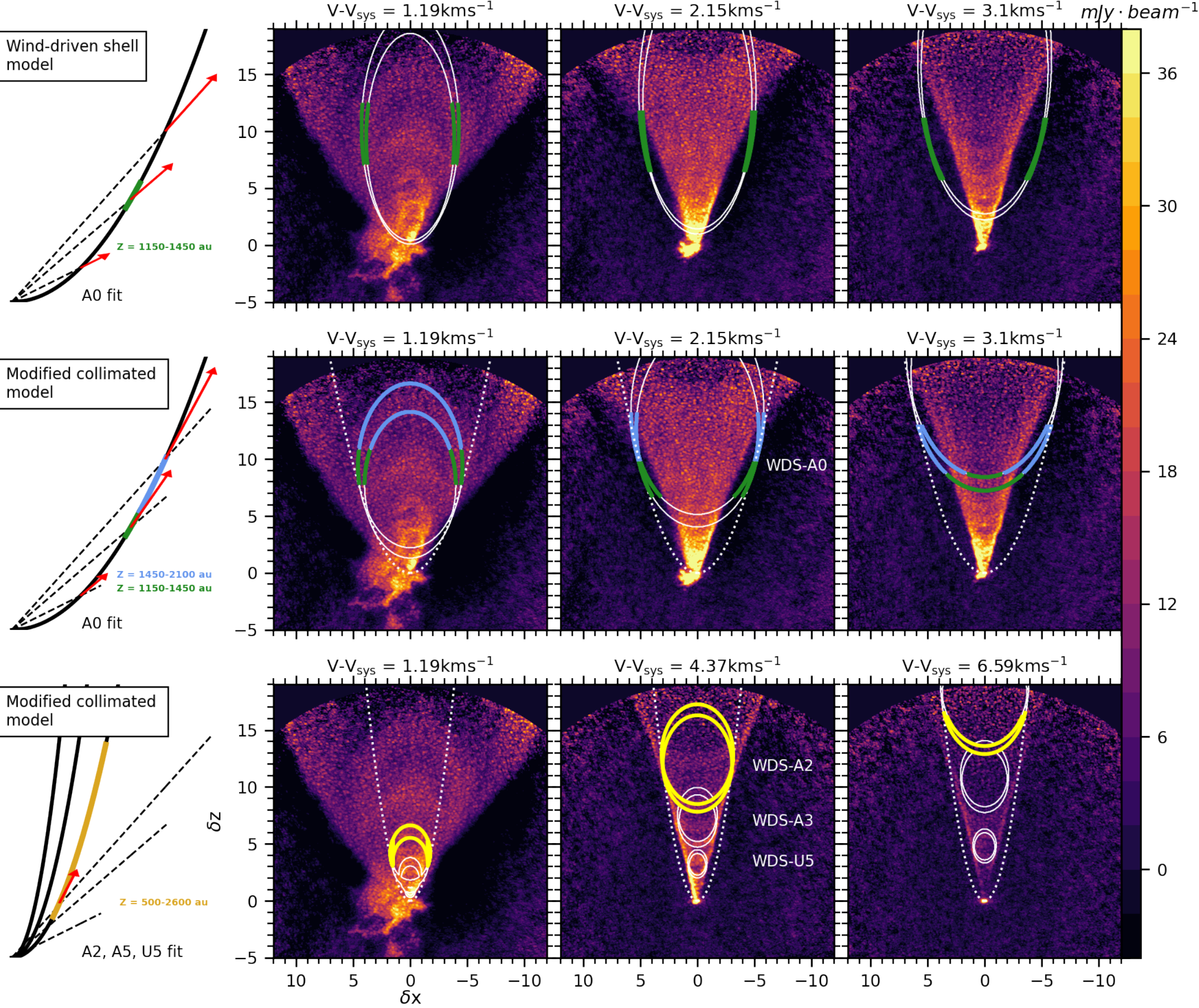}}
    \caption{Comparison of $^{12}$CO channel maps at three different line-of-sight velocities (color maps) with predicted ellipses for parabolic WDS (white contours). The model used on each row is sketched in the left-most panel. {\sl Top row:} Classical model with radial Hubble expansion \citep[see Eq. \ref{eq:wds} and][]{lee_co_2000}. {\sl Middle and bottom rows}: Modified model with "collimated" expansion (see Eq. \ref{eq:wds2}). Green, blue, and yellow contours highlight specific height ranges, indicated in the first column. White dotted lines outline the radial boundary defined by the shell ellipses with increasing velocity. Model parameters are listed in Table \ref{tab:WDS_models}. $\delta x$ and $\delta z$ units are arcseconds.}
    \label{fig:loops}
\end{figure*}

Since the ellipse aspect ratio in channel maps does not depend on $\tau$ nor $C$ (see Appendix \ref{sec:Aspect_ratio}) the only possibility to reduce it without changing the shell parabolic shape is to modify the shell dynamics. The maximum height of the ellipse is reached on-axis ($\delta x=0$) where the projected velocity is greatly affected by the radial velocity component $V_{\rm R}$ (see Eq. \ref{eqv} with $\phi=\pm\frac{\pi}{2}$). To keep a small number of model parameters, we thus chose to add an ad hoc free parameter $\eta$ that  modified the radial velocity as:
\begin{equation}
\label{eq:wds2}
\centering
Z=CR^2 \qquad \qquad V_{\rm Z}=\frac{Z}{\tau} \qquad \qquad V_{\rm R}=\eta\frac{R}{\tau}.
\end{equation}

In Appendix \ref{sec:Aspect_ratio}, we show that the ellipse aspect ratio in this modified model is set at  $A=(\eta~ \tan^{2}{i}+1)| \cos{i} |$. 
Since we  wanted to reduce the aspect ratio, we  needed $\eta < 1$. In other words, we  needed a velocity vector that is more collimated (forward-directed) than the  radial shell expansion in the original WDS model of \citet{lee_co_2000}. We refer to this ad hoc model as "modified collimated WDS."

As shown in the second row of Fig. \ref{fig:loops}, the shape of the largest arch A0 at $(V-V_{\rm sys}) = 1.19$ \kms\ is well fit by a modified collimated WDS with $\eta=0.6$. The smallest two arches A2,A3 and the smallest cusp U5 are also well fit by a collimated WDS with $\eta=0.5$ (Fig. \ref{fig:loops}, bottom row). The parameters of these best-fit solutions are listed in Table \ref{tab:WDS_models}. 
The inferred shell dynamical times have typical intervals of $\Delta \tau \simeq 300-750$ yrs, 
similar to those inferred by \citet{zhang_episodic_2019} in HH46-47.
The fit velocity values $V_{\rm Z}$ at $\theta = 14\degr$ (in the region of the conical flow) are also listed\footnote{We note that $\tan{14\degr} \sim 1/4$ hence $V_{\rm Z}(14\degr) \simeq 16/(\tau C)$, cf. Eq. \ref{eq:vel-theta-wds}}.
Not surprisingly, their range of $\simeq 7.5-16$ \kms\ is similar to our "model-independent" tomographic results for $V_{\rm Z}$ in the conical flow region.

\begin{table}
\caption{\label{t2} Parameters of parabolic wind-driven shells with collimated Hubble law fit to arches and cusps in Fig.~\ref{fig:loops}.}
\centering
\begin{tabular}{ccclc}
\hline\hline
Name & $C$  & $\tau$ & $\eta$ & $V_{\rm Z}(\theta = 14\degr)\tablefootmark{a}$ \\
of model & (au$^{-1}$) &  (yr) &  & (km s$^{-1})$ \\
\hline
WDS-A0 & 0.003 & 1600 & 0.6 & 16 \\
WDS-A2 & 0.01 &850 & 0.5 & 9 \\
WDS-A3 & 0.02 &500 & 0.5 & 7.5 \\
WDS-U5 & 0.04 &220 & 0.5\tablefootmark{b} & 8.5 \\
\hline
\end{tabular}
\label{tab:WDS_models}
\tablefoot{The collimated parabolic WDS model is described by Eq. \ref{eq:wds2}: $C$ is the inverse characteristic parabola size, $\tau$ the shell age along $Z$, $\eta$ the $V_{\rm R}$ reduction factor ($\eta =1$ for a radial flow).\\ 
\tablefootmark{a} $V_{\rm Z}(\theta) = 1/(\tau C \tan^2\theta)$ (see Eq. \ref{eq:vel-theta-wds}).\\
\tablefootmark{b}{In the case of WDS-U5, we do not have any constraint on the aspect ratio, and set $\eta$ at 0.5.}}
\end{table}

\begin{figure}
    \resizebox{\hsize}{!}{\includegraphics{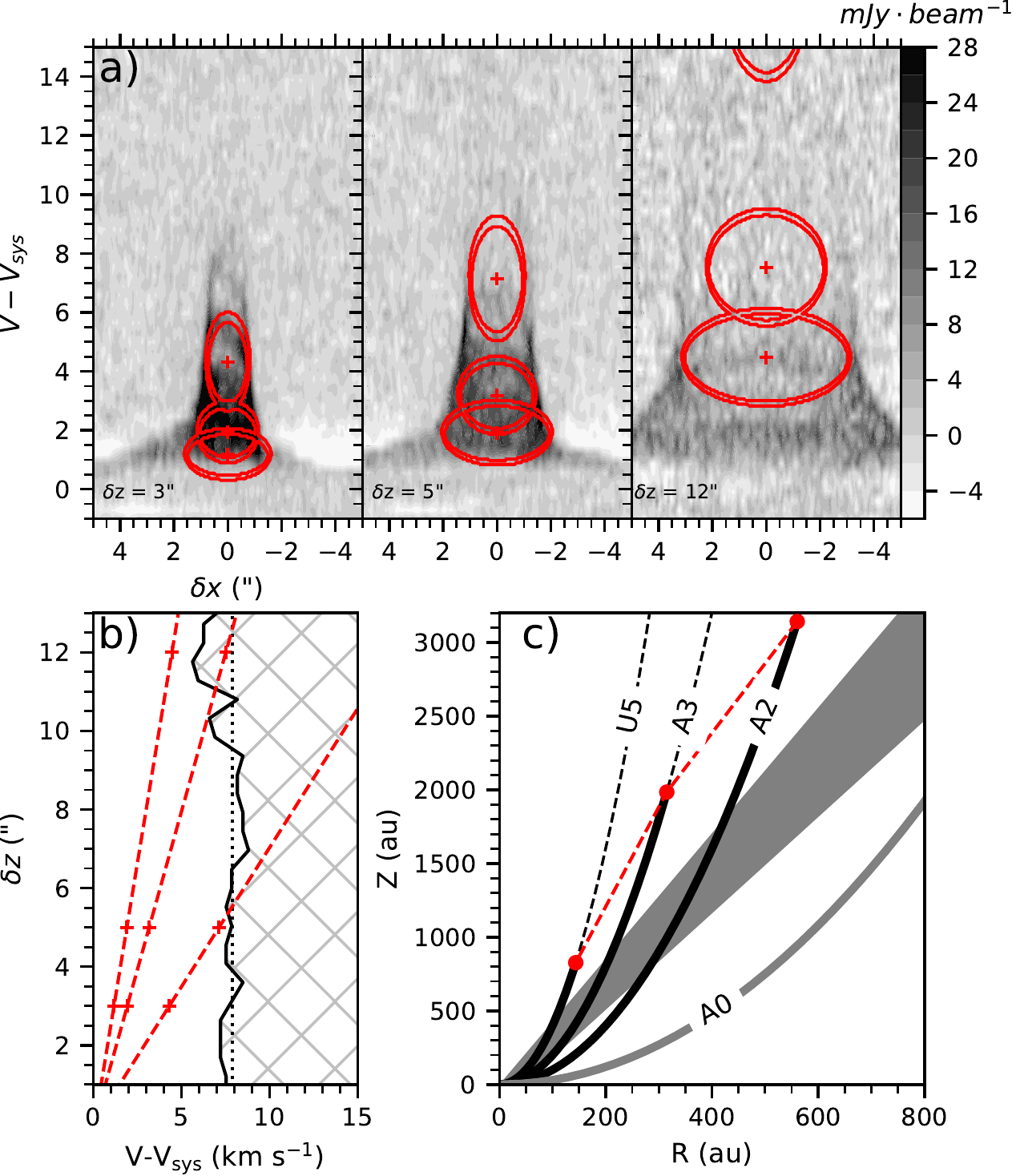}}
    \caption{  Kinematic evolution along the flow axis. {\sl Panels a)} Position-velocity diagrams averaged over a slice of width $\Delta Z=0.2^{\prime\prime}$ at three different $\delta z$ positions along the flow. The red contours trace solutions of collimated WDS used to fit A2, A3, and U5 (third line on Fig. \ref{fig:loops}). {\sl Panel b)} In black is represented the maximal velocity of emission of the PV diagram. In red is shown the center velocity of the ellipses on the PV diagrams. {\sl Panel c)} The gray region highlights the conical outflow domain. The black lines represent the parabolic morphology of the three solutions. The red dots show for each solution the region where the projected velocity reach $V_{\rm max}$.}
    \label{fig:Vmax_HL}
\end{figure}

The Hubble law in the WDS model predicts an ever-increasing shell speed at higher altitudes, 
until it reaches the polar wind speed, which is $\simeq 125$ \kms~according to the redshifted jet speed in DG Tau B \citep{eisloffel_imaging_1998}. 
In contrast, the maximum line-of-sight velocity with detectable emission in our transverse PV cuts (averaged between the left and right sides) is found to stay roughly constant with altitude at $V_{\rm max} \simeq  8\pm1$ \kms~(see Fig. \ref{fig:Vmax_HL}b).  Therefore, all successive wind-driven shells should be truncated, or have their CO emission strongly suppressed, above the point where they reach $V_{\rm Z} = V_{\rm max}/\cos{i} \simeq 18$ \kms. This velocity limit is close to the molecule dissociation limit $\simeq$ 20 \kms\ in dense hydrodynamical shocks \citep[see e.g.,][]{Wilgenbus_ortho_2000}. Therefore, the disappearance of CO emission above a certain speed might be explained by shock-dissociation of ambient CO.
Figure~\ref{fig:Vmax_HL}c shows that the corresponding truncation region for the best-fit WDS models in Table \ref{tab:WDS_models} has a rough conical shape with $\theta \simeq 9\degr$. However, our ad hoc modified collimated WDS model meets two serious issues, detailed below. 

The model predicts a full ellipse in each channel map (white contours in Fig. \ref{fig:loops}), which is not observed. In contrast, discrete structures highlight only a portion of ellipse, depending on the velocity range (see Fig. \ref{fig:loops} and Sect. \ref{sec:description}): the ellipse top at low-velocities (arches), ellipse flanks at mid-velocities (fingers), and ellipse bottom at high-velocities (cusps).  We find that a transition from arches at low velocity to cusps at high velocity can only be obtained if emission is restricted to a range of heights from $z_{\rm min}$ to $z_{\rm max}$, as illustrated by the colored contours in Fig.~\ref{fig:loops}. Serious discrepancies still remain with observations, however: In the broadest shell, WDS-A0, the extents of Arch A0 and Cusp U1 require inconsistent ranges of emitting heights, and the predicted "fingers" at intermediate velocity are much wider than observed (see blue and green contours in middle row of Fig.~\ref{fig:loops}). In the smaller inner WDS, full ellipses are still predicted in intermediate velocity channels, which are not observed (see yellow contours in bottom row of Fig.~\ref{fig:loops}). The same problems remain even if we adopt conical shapes for the shells instead of parabolae, hence the above discrepancies appear intrinsically linked to the assumed  Hubble-law dynamics.

Another serious issue is that the best-fitting value of $\eta$ in our modified collimated WDS models is always close to 0.5 (see Table \ref{tab:WDS_models}). A ratio $V_{\rm R}/V_{\rm Z} = 0.5 R/Z$ corresponds to a velocity vector locally tangent to the parabola. Hence the shell is not expanding but stationary. The physical justification for the Hubble law in the WDS model, namely a shell expanding at constant speed over time \citep{shu_star_1991}, is then no longer applicable. If the emitting material is moving parallel to the shell, a velocity increasing in proportion to distance would require, instead, a constant accelerating force of unknown nature operating out to z=3000~au, which is totally unphysical.

In summary, we find that only the outer faint, low-velocity flow in DG Tau B can be reproduced with the parabolic WDS model with radial Hubble law proposed by \citet{lee_co_2000}. In contrast, the bright conical outflow at mid to high velocity,
although reminiscent of WDS models because of the apparent acceleration of its discrete structures, cannot be explained by such models, even when ad hoc modifications to the kinematics, emissivity range, and shape
are introduced. The model faces important issues which seem intrinsically linked to the  Hubble-law dynamics. We discuss in Sect. \ref{sec:Discu_WDS} the implications of these results in the context of more general wind or jet-driven shell scenarios.

\section{Disk-wind modeling}
\label{sec:DW}

In this section, and alternatively to the WDS models considered in Sect. \ref{sec:WDS}, we investigate a simple kinematical disk wind model for the DG Tau B redshifted outflow
where the conical morphology visible in Fig. \ref{fig:VZ_J} would trace the trajectory of CO molecules ejected from the disk. Although we cannot derive $V_{\rm R}$ in a model independent way, from the external contours of transverse PV diagrams (see Sect. \ref{sec:tomo}), we showed in Sect. \ref{sec:description} the existence of brighter elliptical structures visible on transverse PV diagrams.
 Assuming that they trace each a specific layer of the flow along which $V_{\rm Z}$ and $V_{\rm R}$ stay roughly constant with height, we  fit these ellipses to  derive both the $V_{\rm R}$ and $V_{\rm Z}$ components of the velocity (see Appendix C). Figure \ref{fig:ellipse_fit} shows that the derived velocity directions are parallel to the conical contours  of constant $V_{Z}$.
This comforts our hypothesis that the trajectory of the outflow  follow lines of constant $V_{\rm Z}$. From this hypothesis, we  derived the collimation and kinematics of the streamlines using the tomography and  created a synthetic data cube of the conical disk wind outflow.

\subsection{Steady disk-wind model}
\label{sec:Steady_DW}

We  made the assumption that the flow is axisymmetric and that the matter has reached its terminal velocity and has a constant poloidal velocity along its trajectory. We  fit this trajectory by a conical surface defined by an angle $\theta$ from the Z-axis and an anchoring radius $r_{0,\rm geo}$. 
We  extracted from the tomography the specific angular momentum, $j= R \times V_{\phi}$, along curves of constant $V_{\rm Z}$ (see Fig. \ref{fig:XV_J}) and  derived a median value for each streamline. We  defined the uncertainty of this value as the standard deviation of specific angular momentum. We also  computed the poloidal velocity $V_{\rm P}=V_{\rm Z}/\cos{\theta}$.

\begin{figure}
    \resizebox{\hsize}{!}{\includegraphics{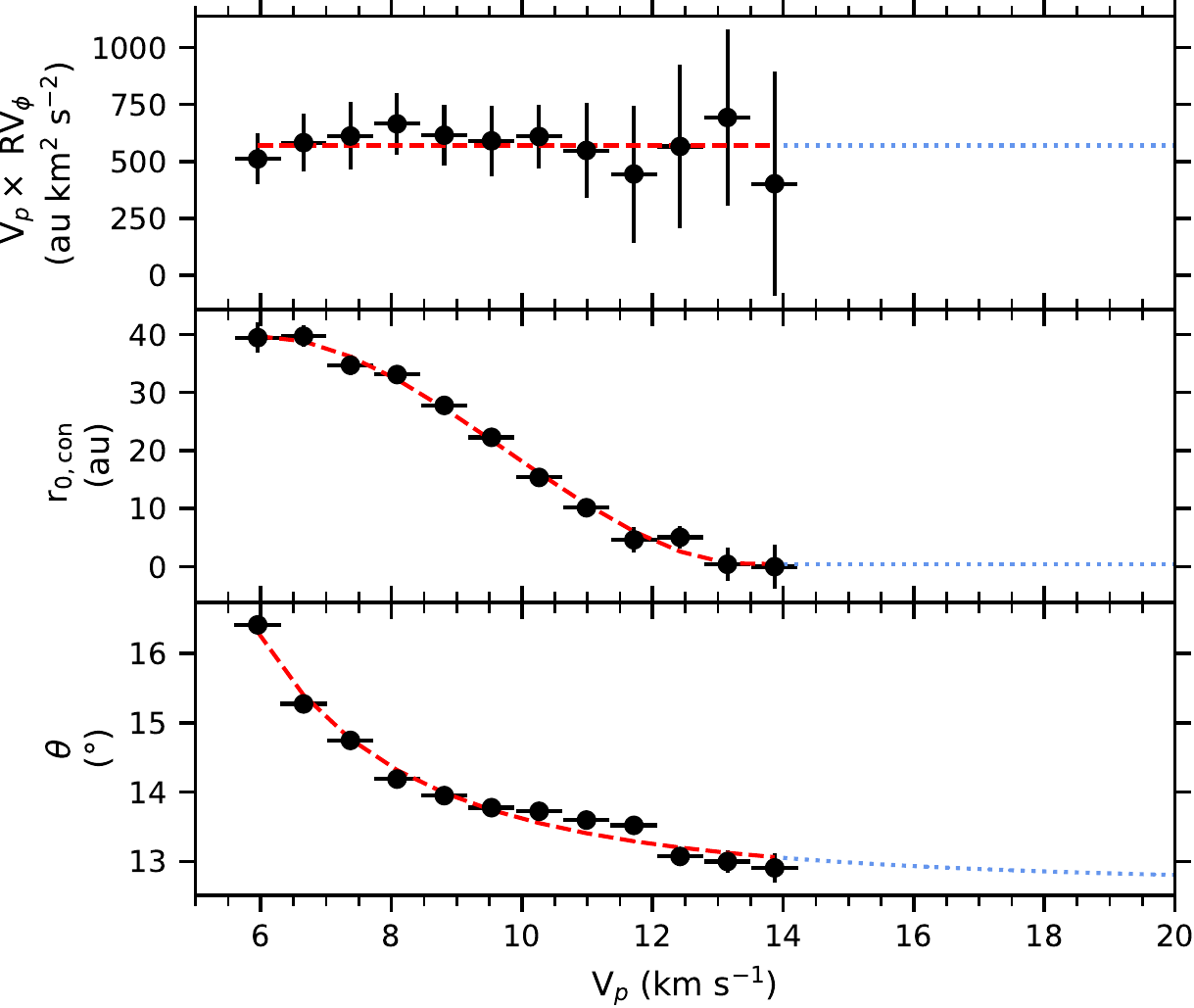}}
    \caption{ Disk-wind properties derived from the tomography along lines of constant $V_{\rm Z}$.  Red dashed curves show fits as a function of the poloidal velocity of the streamline $V_{\rm p}$: the anchoring radius of the streamline $r_{0,\rm con}$ was fit by a third-order polynomial, the angle of the streamline with the flow axis $\theta$  by a power law and the product $V_{\rm P} \times j$ by a constant value of 570 au km$^{2} s^{-2}$. Blue dotted lines represent the extrapolation used to model the high-velocity component that could not be mapped by tomography due to its low S/N.}
    \label{fig:Etude_XV_data}
\end{figure}

Figure \ref{fig:Etude_XV_data} represents the derived values of  $r_{0,\rm geo}$, $\theta$ and $V_{\rm P}\times j$ for each streamline of constant $V_{\rm P}$.  We fit the variation of anchoring radius $r_{0,\rm con}$ with the poloidal velocity by a polynomial law. The variation of ejection angle $\theta$ was fit by a power law. $V_{\rm P}\times j$ was taken constant at 570$\pm$ 50 au km$^{2}$ s$^{-2}$ for all the streamlines. The fits were achieved using nonlinear least squares, The equations are as follow:
\begin{eqnarray}
    r_0 &\simeq &0.18-5.33V_{\rm p} +44.29V_{\rm p}^2-73.23V_{\rm p}^3\\
    \theta  & \simeq& 12.63+348.57V_{\rm p}^{-2.55}\\
    j\times V_{\rm p} &\simeq& 570
.\end{eqnarray}

\begin{figure*}
    \resizebox{\hsize}{!}{\includegraphics{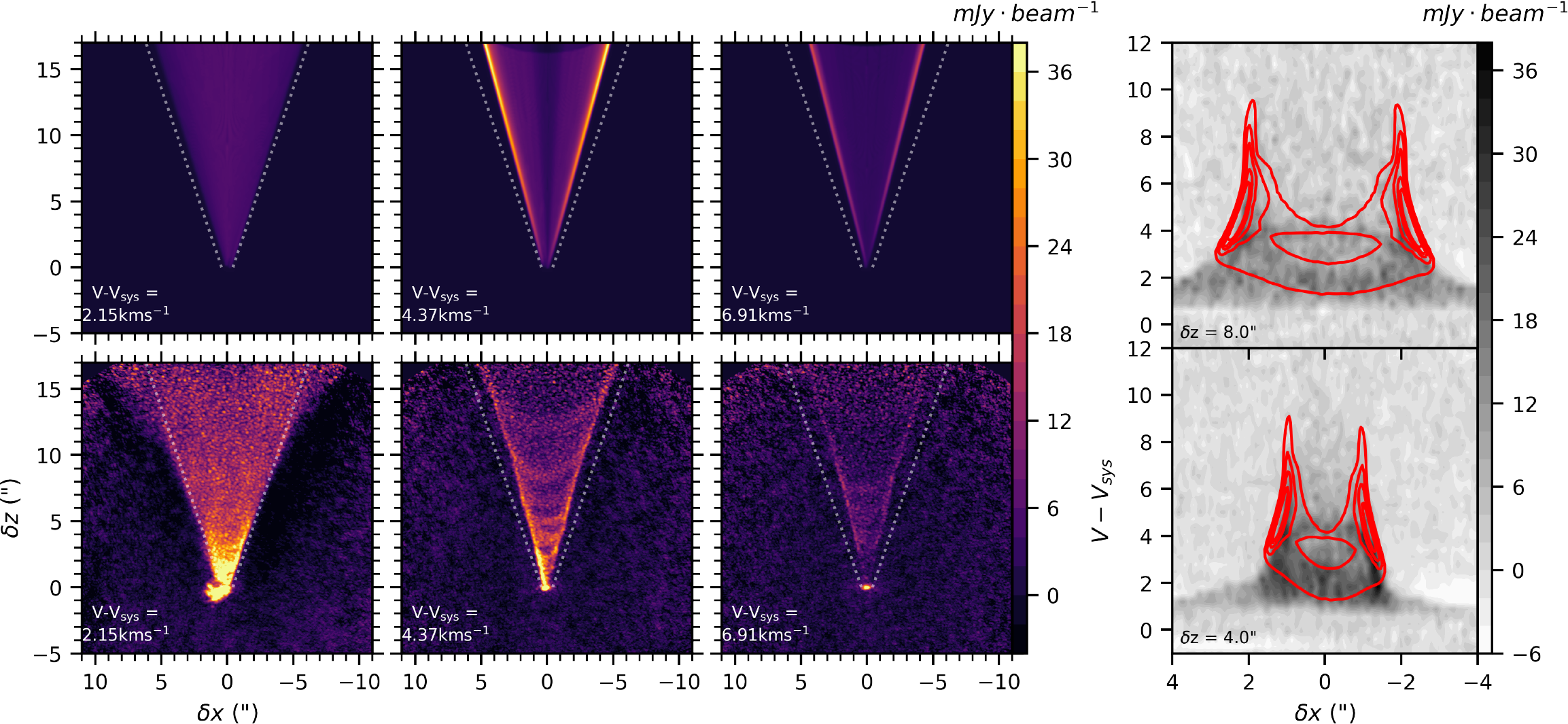}}
    \caption{ Comparison of observations with steady disk-wind model.
    {\sl Left panels}: Individual $^{12}$CO channel maps computed from the global disk-wind model at selected line-of-sight velocities (top row) are compared to observations (bottom row). The color scale is the same for all the channel maps. {\sl Right panels}: Transverse PV diagrams at two positions $\delta z$ along the flow and averaged over a slice of width $\Delta Z = 0.2^{\prime\prime}$. The background grayscale image shows the observations and the red contours trace the predictions from the disk-wind model. $(V-V_{\rm sys})$ units are \kms.}
    \label{fig:synthetic}
\end{figure*}

Here $r_0$ is in au, $V_{\rm P}$ in \kms, $\theta$ in degrees, and $j$ in au~\kms.
We modeled the disk wind with axisymmetric conical streamlines with the dynamics and morphology laws derived in Fig. \ref{fig:Etude_XV_data} and created a synthetic data cube of the conical outflow. We set the external and slower layer at $V_{\rm p} = 6$~\kms, corresponding to the smallest value that could be mapped with our tomography (see Sect. \ref{sec:tomo}. We set the internal, faster velocity at 20~\kms. The parameters in the velocity range $V_{\rm p}=14-20$~\kms, not covered by the tomography, were determined from an extrapolation of our fits (blue dotted line in Fig.~\ref{fig:Etude_XV_data}). This extrapolation was done in order to describe the almost-vertical high-velocity component not described by the tomography due to insufficient S/N. For each layer, we set the initial value of $V_{\phi} (R)$ assuming $V_{\rm p} \times j = 570$~au~km$^{2}$ s$^{-2}$ (see Fig. \ref{fig:Etude_XV_data}).  We assumed optically thin emission throughout the outflow, which is justified by the observed ratios of $^{13}$CO/$^{12}$CO (see DV20). We did not consider a variation of emissivity with radius of ejection nor with height (see Sect.~\ref{sec:Global_model}). Proper modeling would require CO chemistry and temperature profiles, which is well beyond the scope of this paper. Projection and beam convolution effects were also taken into account.

Figure \ref{fig:synthetic} shows synthetic channel maps and PV diagrams for our model compared with  observations. The global morphology of the outflow at $(V-V_{\rm sys}) >2$~\kms~ as well as its variation with line-of-sight velocity are well recovered as expected since we use the tomographic results to constrain the wind collimation and kinematics. This model does not attempt to describe the extended outflow surrounding the cone at low velocities $(V-V_{\rm sys}) = 1.15$~\kms\ (see Sect. \ref{sec:description}, Fig. \ref{fig:synthetic}). To describe completely this extended low-velocity component, we would need to extrapolate the  disk-wind model at larger ejection radii. However, as this component falls at absorbed cloud velocities, we are not able to derive model-independent constraints on the dynamics of this component. Proper modeling would require time-consuming and uncertain parameter space exploration. We choose therefore to focus in the following on the conical outflow; nonetheless, the disk wind could be more extended radially than we describe with our current modeling.

Although effective to describe the global morphology of the outflow, our simple axisymmetric and steady  disk-wind model does not reproduce  the different substructures identified in our observations: cusps and arches and the local deviations of specific angular momentum at $Z \approx 600$ and 1000 au. In the following subsections, we discuss two small perturbations of our  disk-wind models which could  explain the various substructures observed. 

\subsection{Wiggling of the flow axis}
\label{sec:wigg}

Although we do not detect a clear signature of wiggling in our data, we cannot exclude a small amplitude wiggling of the CO axis $< 0.5^{\circ}$ (see Appendix A). A wiggling of the outflow axis could explain the variations observed on the specific angular momentum tomographic map. Indeed, in order to determine the specific angular momentum using Eq. \ref{eqvphi}, we assumed that the center of the layer is located at $\delta x=0$ at all heights. 
If the center is shifted toward $\delta x>0$ or $\delta x<0$, the  specific angular momentum computed with our method will be respectively higher or lower than the true value. This effect is more critical if the PV diagram shows a strong velocity gradient, which is the case for DG Tau B. In this section, we investigate this effect, and show that small  amplitude wiggling may also create the substructures observed (cusps, fingers and arches).

\begin{figure*}
    \resizebox{\hsize}{!}{\includegraphics{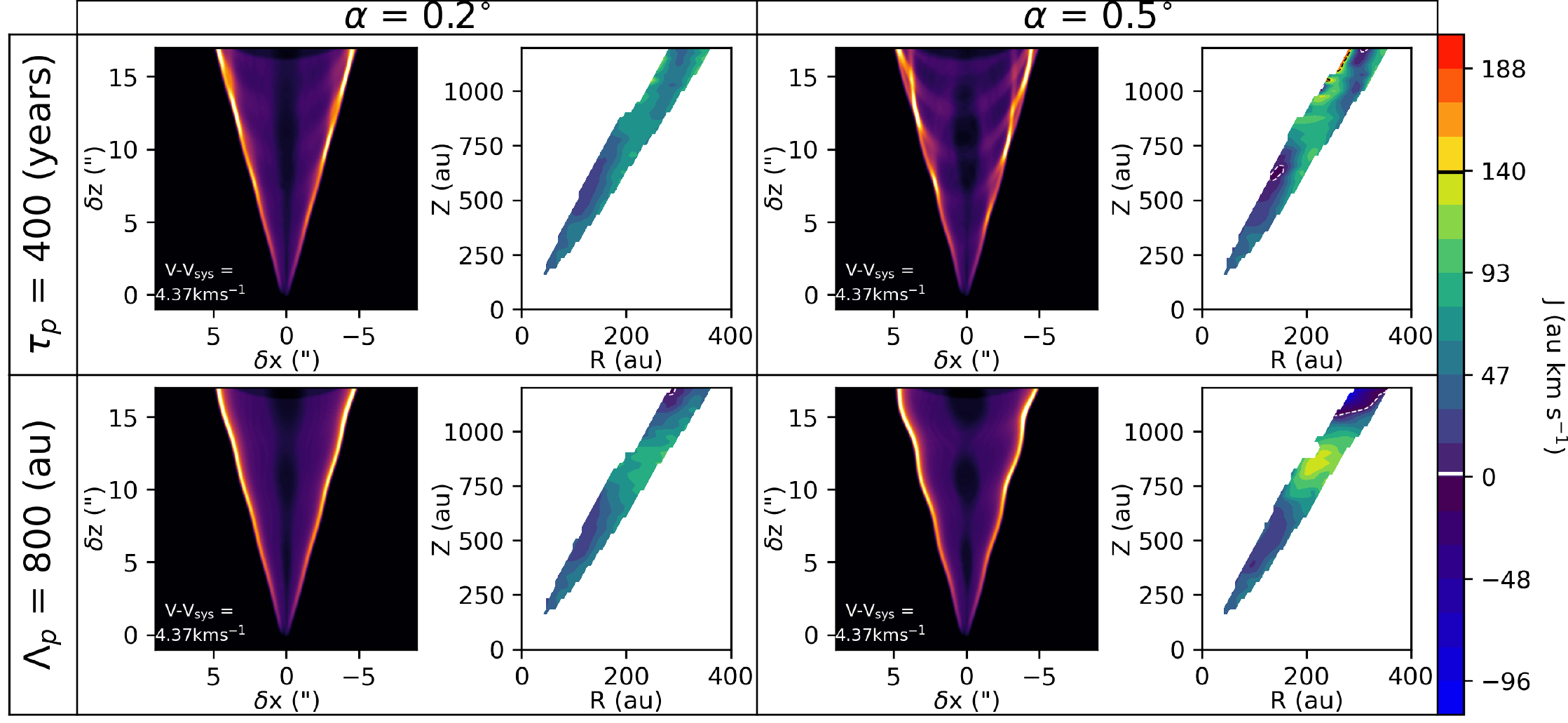}}
    \caption{Synthetic data cubes computed from the generic  disk-wind model with flow axis precession: with a precession angle of 0.2$^{\circ}$ (left panels) and 0.5$^{\circ}$ (right panels) and assuming a constant precession period $\tau_{\rm p}=400$~yrs (top row) or a constant precession spatial wavelength $\Lambda_{\rm p}=800$~au (bottom row). For each model are shown both a channel map at $(V-V_{\rm sys}) = 4.37$~\kms~ and the corresponding tomography of $j$ derived with the method described in Sect. \ref{sec:tomo}. The color scale of $j$ ranging from 0 to 140 au \kms~ is identical to the color scale of Fig. \ref{fig:VZ_J}.}
    \label{fig:wigg}
\end{figure*}

We modifed the  disk-wind model presented in Sect. \ref{sec:Steady_DW} to add a precession of the outflow axis.  Each conical layer of the outflow precesses  with an angle $\alpha$ and a precession period $\tau_{\rm p}$. This is an extension of the model developed by \citet{masciadri_herbigharo_2002} for jets, modified to take into account the conical morphology of the outflow and inclination to the plane of the sky.
We modeled both a prograde and retrograde precession. However, due to the small value of $\alpha$ in our models, the two models give very similar results. We present here results for the prograde model only. We first assumed that all  disk-wind layers precess with the same $\alpha$ and the same precession period $\tau_{\rm p}$. Due to the velocity shear across the outflow, the spatial period $\Lambda_{\rm p}=\tau_{\rm p}\times V_{\rm p}$  then varies between layers according to the poloidal velocity.  We also investigated a precession model where the spatial period $\Lambda_{\rm p}$ is constant across all streamlines. A constant $\Lambda_{\rm p}$ corresponds to a variation of precession period as $\tau_{\rm p} \propto V_{\rm p}^{-1} \propto r_0^{0.5}$. We visually fit $\tau_{\rm p}$ and $\Lambda_{\rm p}$ to best reproduce the location of the two extrema variations in the specific angular momentum map separated by $\simeq$ 400 ~au. For each model, we computed synthetic data cubes and derived the  specific angular momentum map using the same method used in Sect.~\ref{sec:tomo} for the observations.

Fig.~\ref{fig:wigg} shows the resulting channel maps and the specific angular momentum maps for the two precession models (constant $\tau_{\rm p}$ and constant $\Lambda_{\rm p}$) with two different precession angles (0.2 and 0.5$^{\circ}$) compatible with the upper limit derived for the CO outflow axis wiggling in Annex~\ref{sec:axis}. Precession models with constant $\tau_{\rm p}=400$~yrs successfully reproduce the channel maps morphology, in particular the cusps at high-velocity and arches at low velocities (see Fig. \ref{fig:CM_DW}). A best match to the intensity contrast is obtained for $\alpha=0.5^{\circ}$. However, the resulting map of specific angular momentum $j$ is not fully consistent with our observations. Indeed, as $\Lambda_{\rm p}$ is different for each layer, the perturbations of specific angular momentum are not localized at one specific height, as in the observations. 

The modified model with constant $\Lambda_{\rm p}=800$~au for all layers better reproduce the positions of the two extrema at $Z = 600$ and $Z = 1000$~au in the specific angular momentum map. However, the cusps have a lower intensity contrast than observed, even with the maximum allowed wiggling angle of 0.5$^{\circ}$. In addition, this model predicts clear detectable wiggling on the edges of the cone in channel maps, which is not seen in the observations. A model in between these two extremes, that is to say with $\alpha \simeq 0.5\degr$ and a precession period $\tau_{\rm p}$ increasing more slowly than $r_0^{0.5}$ may better account for all observational properties.

A remaining discrepancy with observations is that none of the  wiggling models reproduce the short spacing of $\simeq 1.5^{\prime\prime}=200$~au between the inner cusps, as well as the apparent increase of cusp separation with distance from the source (see Table~\ref{tab:discrete}, Fig.~\ref{fig:CM_DW}), although this latter effect is mostly seen in the farthest cusp A0 and may result from a lack of sensitivity. The constant $\Lambda_{\rm p}$ model predicts a projected separation between the cusps of $\Lambda_{\rm p} \sin(i)$, corresponding to $\simeq 5^{\prime\prime}$, while the constant $\tau_{\rm p}$ model a twice smaller separation typically.

However, we stress that our wiggling models are probably too simplistic as they do not take into account the (magneto)-hydrodynamical interactions between the layers. \citet{masciadri_herbigharo_2002} have shown that simulations depart rapidly from analytical solutions in the case of jet wiggling due to precession.  This difference is potentially even greater with a shearing outflow. Dedicated numerical simulations are required to fully test this scenario.
Nonetheless, this model is a promising candidate to explain the variation of specific angular momentum along the DG~Tau~B outflow.
We discuss in Sect.~6 possible wiggling mechanisms and their implications.

\subsection{Emissivity enhancements in the disk wind}
\label{sec:Density_DW}

In this section, we investigate an alternative model where  cusps and arches are created by localized emissivity enhancements in the conical disk wind. We first derived their location from the tomography, assuming that they are axisymmetric, and then created a synthetic data cube to compare with our observations.

\begin{figure}
    \resizebox{\hsize}{!}{\includegraphics{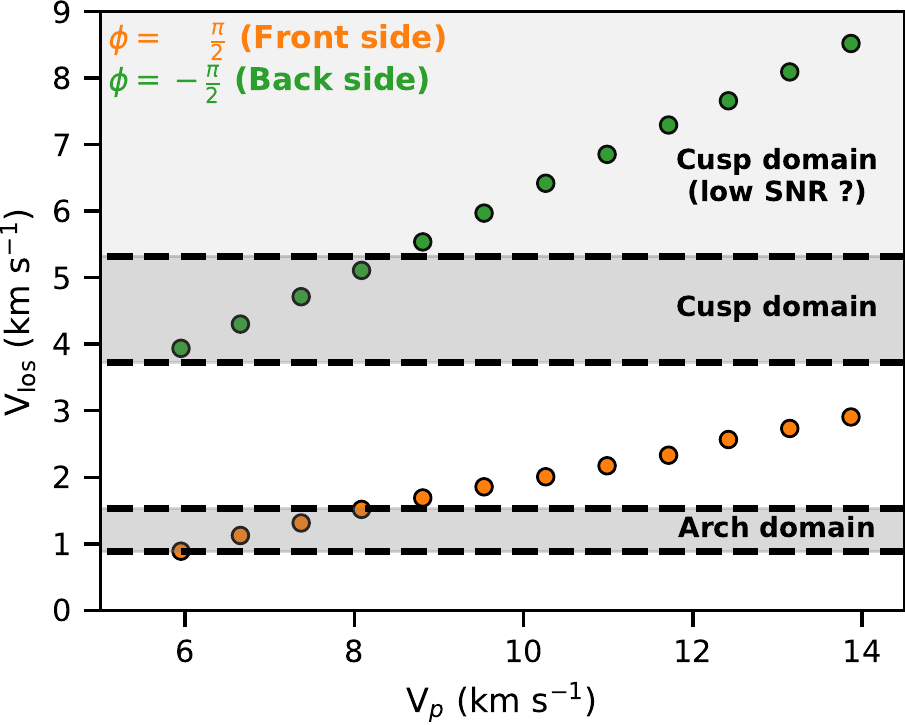}}
    \caption{Line-of-sight velocity on-axis (at $\delta x = 0$ ) of the model  disk-wind flow surfaces as a function of their poloidal velocity. Orange circles show the front side ($\phi = \frac{\pi}{2}$), and green circles show the back side ($\phi = - \frac{\pi}{2}$) of the disk-wind flow surfaces. In gray, we show the line-of-sight velocity range where the arches and cusps are observed (see Annex~\ref{sec:Caract_discrete}). In light gray, is represented the low S/N domain where the cusps could not be characterized. }
    \label{fig:Projection_domain}
\end{figure}

Figure \ref{fig:Projection_domain} shows the projected velocity on-axis at  $\delta x = 0$  for the front side and the back side ($\phi = \pm \frac{\pi}{2}$) of each conical wind layer in our model, predicted from Eq. \ref{eqv}. We also represent the domain of line-of-sight velocities where arches and cusps positions could be measured in channel maps as described in Sect.~\ref{sec:description}. The cusps observed at $(V-V_{\rm sys})>5$ \kms\ could not be characterized in Sect.~\ref{sec:description} because of low S/N.

Figure~\ref{fig:Projection_domain} shows that if the observed substructures are due to axisymmetric emissivity enhancements in the conical outflow, the cusps come from the back side ($\phi \approx - \frac{\pi}{2}$)  and the arches from the front side ($\phi \approx + \frac{\pi}{2}$) of the enhanced {\sl ring}. Figure~\ref{fig:Projection_domain} also shows that wind layers with poloidal velocities $V_{\rm p}> 9$ \kms\ are not located on the arch domain, and are located on the low S/N cusp domain, which would require weaker emissivity enhancements in the fastest internal layers.

Each cusp was characterized by its projected height at $\delta x =0$ at a specific projected line-of-sight velocity $V_{\rm los}$. The projected height of the cusp corresponds to an equation $Z(R)$ derived from Eq. \ref{eqz}, with $\phi =-\frac{\pi}{2}$. Similarly, the projected velocity of the cusps corresponds to a poloidal velocity as shown in Fig.~\ref{fig:Projection_domain} and assuming $V_{\rm R}$, a conical line of constant $V_{\rm Z}$ in the tomography. As a result, we can associate the cusp observed in a given channel map with a $(R,Z)$ location on the $V_{\rm Z}$ tomography map. This location would be the intersection between the $Z(R)$ equation derived from the projected height, and the conical line derived from the projected velocity. 

\begin{figure}
    \resizebox{\hsize}{!}{\includegraphics{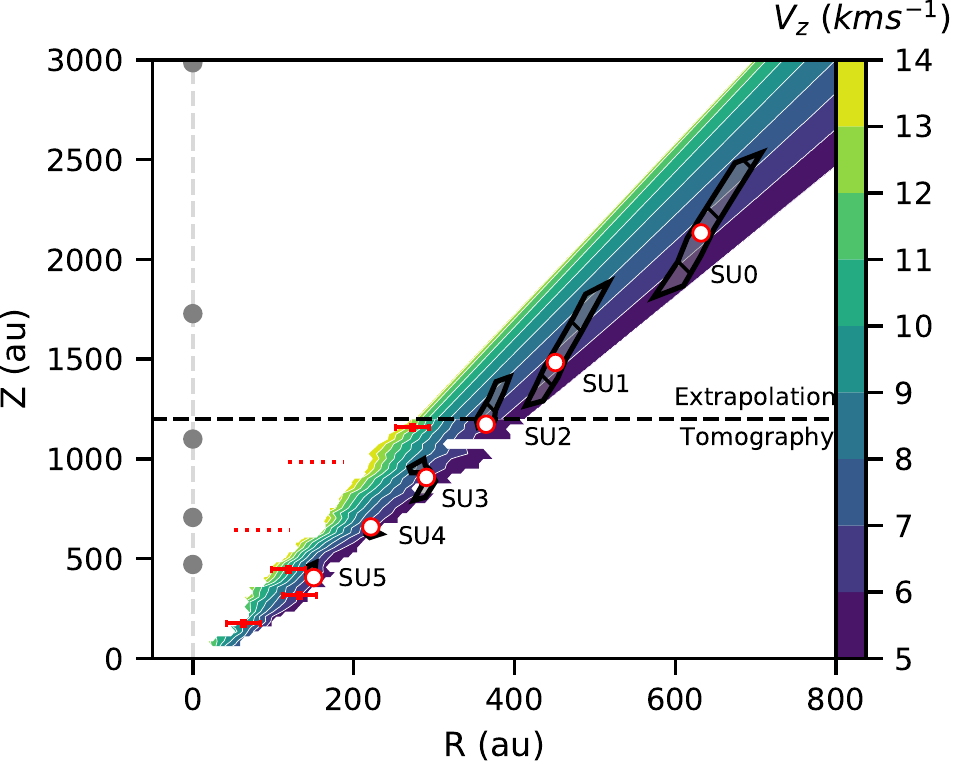}}
    \caption{Tomographic map of $V_{\rm Z}$ from Fig. \ref{fig:VZ_J} extrapolated from $Z=1200$~au to $Z=3000$~au using the conical wind model from Fig. 9 . Gray hatched regions named SU0-5 represent the locus of emissivity enhancements required to explain the cusps on-axis positions at  $V-V_{\rm sys}= 3.73-5.32$~\kms~(see Fig.~\ref{fig:u_tracking}). White dots represent the solutions for the cusps at $V-V_{\rm sys}= 4.37$~\kms. Red dotted lines indicate the height of the two extrema in specific angular momentum (from Fig.~\ref{fig:VZ_J}). The gray points located on the $R=0$ axis represent the positions of jet knots observed by \citet{podio_tracing_2011}. The red dots indicate the positions with associated errors of the ellipses identified in transverse PV diagrams (see Sect. \ref{sec:ellipse_fit}) and indicate that emissivity enhancements extend to the inner streamlines.}
    \label{fig:tomo_discrete}
\end{figure}

Figure \ref{fig:tomo_discrete} shows the tomography of the conical outflow using equations from Fig.~\ref{fig:Etude_XV_data} for extrapolation at $Z > 1200$~au. The white dots correspond to the solutions (named SU0 to SU5) for the cusps heights on axis identified in the channel map at $(V-V_{\rm sys})=4.37$~ \kms. We were able to follow six different cusps in up to six different channel maps (see Fig. \ref{fig:u_tracking}), which allowed us to reconstruct the shape for some of these enhancements, shown as hatched areas in Fig.~\ref{fig:tomo_discrete}. As mentioned in Appendix \ref{sec:Caract_discrete}, the cusps are also visible at higher line-of-sight velocities, but could not be characterized reliably due to their lower S/N. Consequently, the hatched area shown in Fig. \ref{fig:tomo_discrete} should be extended toward the inner streamlines. Moreover, density enhancements could also be present at $\delta z < 2.2 ^{\prime \prime}$ but were not identified by our procedure.The apparent acceleration of the cusps seen in channel maps can be readily reproduced, in this disk-wind model, by axisymmetric emissivity enhancements that cross obliquely the flow streamlines. The upward shift of each cusp at higher velocity (apparent acceleration) is simply a result of the velocity shear across flow streamlines. It does not require a Hubble-law dynamics in the underlying flow.

Interestingly enough, the density enhancements SU3-4 are located close to the two extrema of specific angular momentum, at $Z=600$ and 1000 au, suggesting a possible link.  We could not derive the shape of the emissivity enhancements for $V_{\rm p}$ > 9 \kms, due to the low S/N of the cusps at the corresponding projected velocities $(V-V_{\rm sys}) > 5.32$ \kms~(see Fig. \ref{fig:Projection_domain}). 

\begin{figure}
    \resizebox{\hsize}{!}{\includegraphics{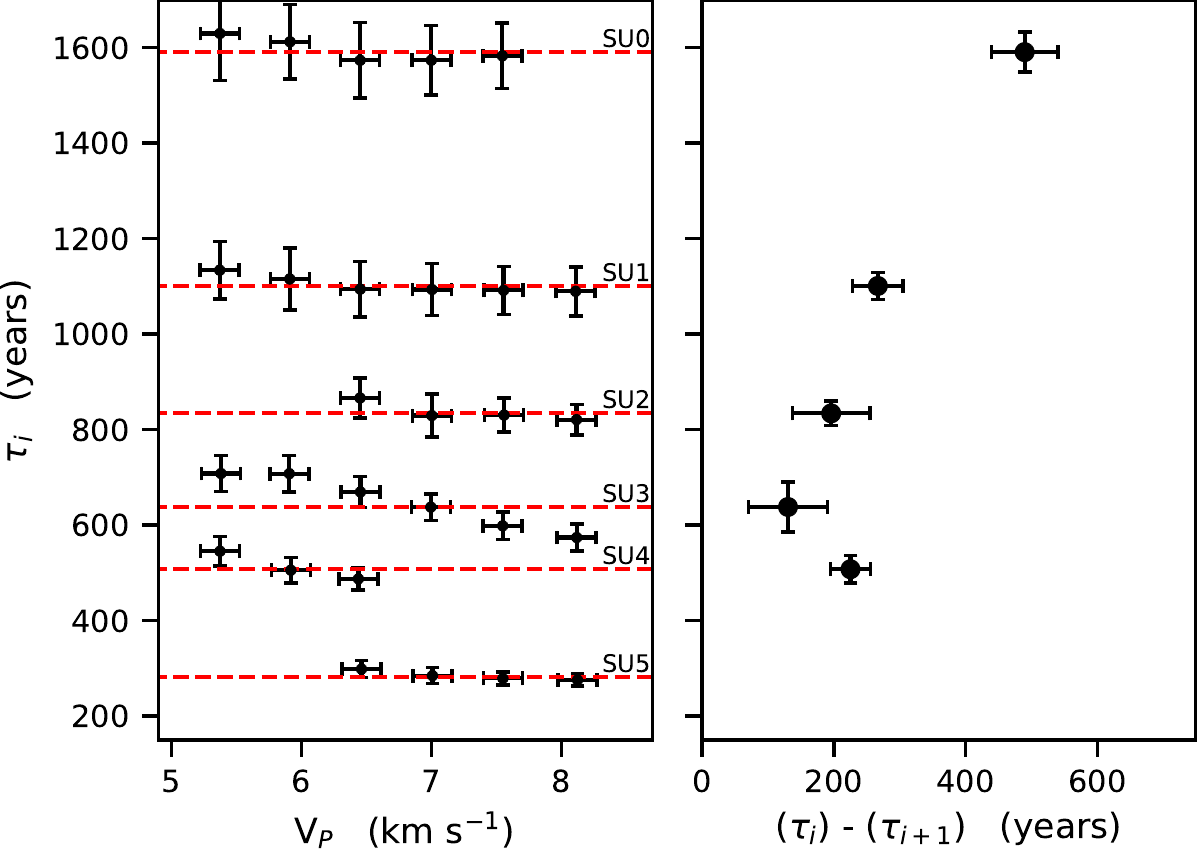}}
    \caption{Dynamical evolution of the enhancements.{\sl Left panel}: Dynamical age $\tau=\frac{Z}{V_{\rm Z}}$ of the enhancements SU0-5 as a function of $V_{\rm P}$. The red dotted line represents the mean value. {\sl Right panel}: Dynamical age versus  timescale between adjacent cusps ($\tau_{i}-\tau_{i+1}$). The uncertainties are determined by propagating the errors on the cusp positions.}
    \label{fig:discrete_tau_vz2}
\end{figure}

We computed the dynamical age of these emissivity enhancements with $\tau_{\rm w}=\frac{Z}{V_{\rm Z}}$. This value would give the true dynamical age of the enhancement if it is created by variability of ejection from the disk surface. The derived values of $\tau_{\rm w}$ for cusps visible on multiple channel maps are shown in Fig.~\ref{fig:discrete_tau_vz2}. The dynamical age is almost constant within each cusp  with only a slight decrease with increasing poloidal velocities, especially for U3 \& U4. The difference in dynamical ages between two successive cusps varies from 190 to 490 years and roughly increases linearly with the age.

\begin{figure*}
    \resizebox{\hsize}{!}{\includegraphics{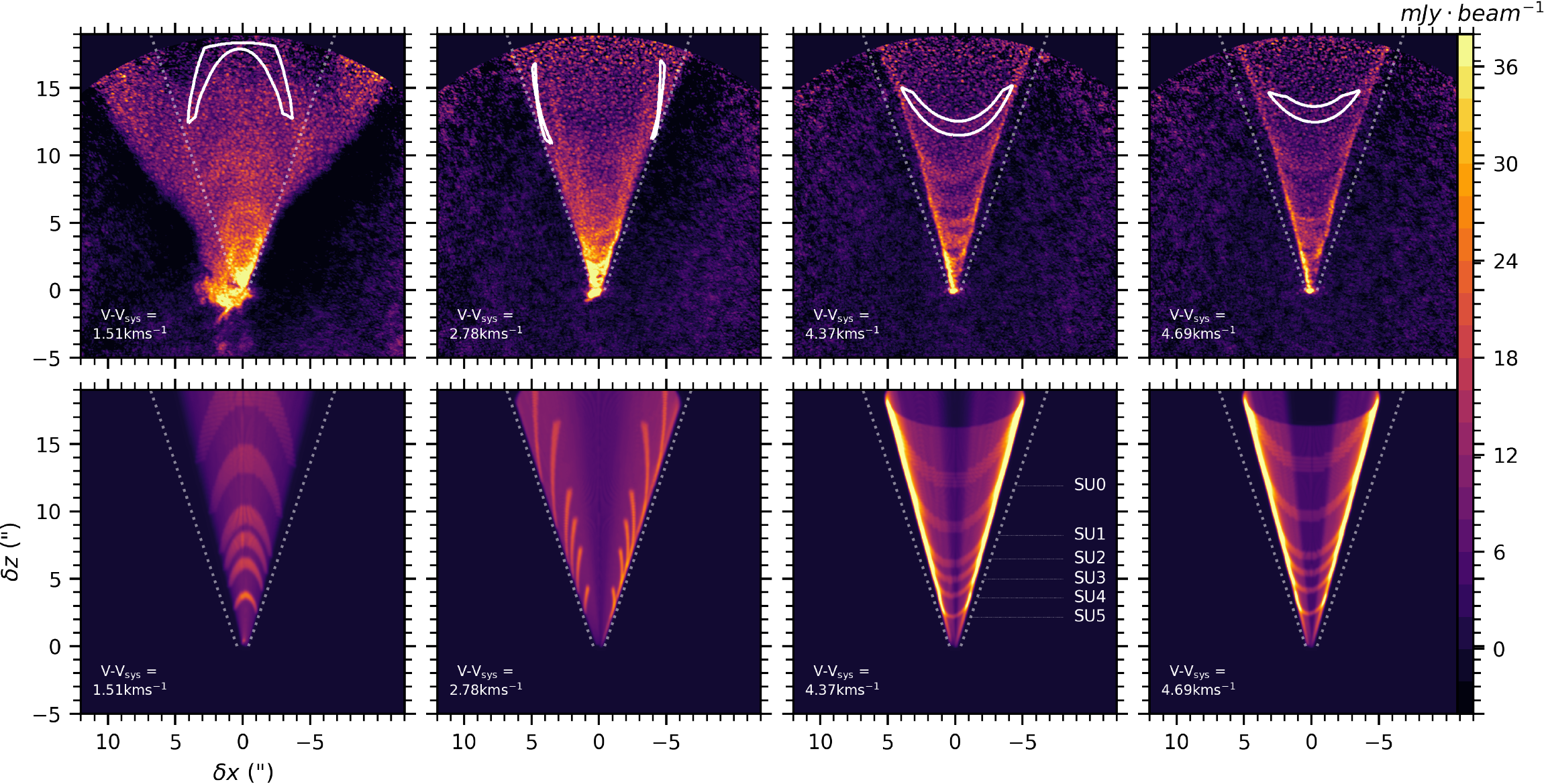}}
    \caption{$^{12}$CO channel maps at selected line-of-sight velocities for the steady conical disk-wind model (bottom row) and observations (top row). White contours highlight the predicted emission from the modeled density enhancement SU0. The intensity color scale is the same for all channel maps.}
    \label{fig:Burst_model}
\end{figure*}

These solutions were determined using only the properties of the cusps on-axis (at $\delta x=0$), corresponding to $\phi =-\frac{\pi}{2}$. In order to check if the reconstructed enhancements were consistent with the full cusp morphology in all channel maps, we developed a 3D model where
we modified the emissivity profile of the synthetic disk-wind model presented in Sect. \ref{sec:Steady_DW}.
We multiplied the underlying DW emissivity profile by Gaussian components representing the emissivity enhancement.
We used the mean $\tau_i$ values derived from Fig.~\ref{fig:discrete_tau_vz2} to determine the location of the  enhancements along each wind streamline. In order to reproduce the observed width and intensities in the channel maps, we set the full width at half maximum of the Gaussian component at 23 years and the maximal emissivity enhancement at $G_{\rm i}= 3$. We introduced these enhancements only in the external layers of the outflow ($V_{\rm P} < 9$~\kms), 
since we did not have constraints on their location at higher velocities.
Figure \ref{fig:Burst_model} shows computed and observed channel maps at different line-of-sight velocities. The contour of the modeled  enhancement SU0 is also represented on top of the observations. This model successfully reproduces the morphology of the cusps  as well as their apparent offset from source with increasing velocity.  Interestingly enough, the model used to reproduce the cusps also matches the arches at low velocity, with the locations of the apexes consistent with our observations (see Fig. \ref{fig:CM_DW}). The intensity contrast is also roughly recovered with our $G_{\rm i}$ value, meaning that the outflow brightness is locally multiplied by  three.

The channel map at ($V-V_{\rm sys})=1.51$~ \kms~ shows that only the top of the arches is reproduced. Indeed, the flanks of the largest arches are wider than the conical flow region modeled by our disk wind in Fig. 9.
In order to fit completely the arches, we would need to extend the disk-wind model to larger radii and opening angles and lower poloidal velocities as suggested in Sect. \ref{sec:Steady_DW}, and extend the density enhancements to these regions. However, due to cloud absorption at low velocities $< 2$ \kms, we do not have model-independent tomographic constraints on the streamline shape and kinematics in this slow external flow. Therefore we cannot determine a reliable solution for the emissivity enhancements producing these arch flanks.

\section{Discussion}
\label{sec:Discussion}

In this section, we use the results of our 
parametric modeling of the flow (Sect.~\ref{sec:WDS},\ref{sec:DW}) 
and tomographic study to critically discuss two possible origins for the small-scale redshifted CO outflow in DG Tau B: 
1) a stacking of multiple shells swept-up  by an inner wind (or jet), without any contribution from an extended disk wind (Sect.~\ref{sec:Discu_WDS}),
2) an extended disk wind (Sect.~\ref{sec:Discu_Steady}), with internal perturbations causing the observed substructures (Sect.~\ref{sec:Discu_Pertu}). 
We find that our measurements of rotation put stringent constraints on each of these scenarios, and we discuss physical implications for the ejection process and its relation to the disk accretion process.

\subsection{Stacking of wind-driven shells}
\label{sec:Discu_WDS}

\begin{figure}
    \resizebox{\hsize}{!}{\includegraphics{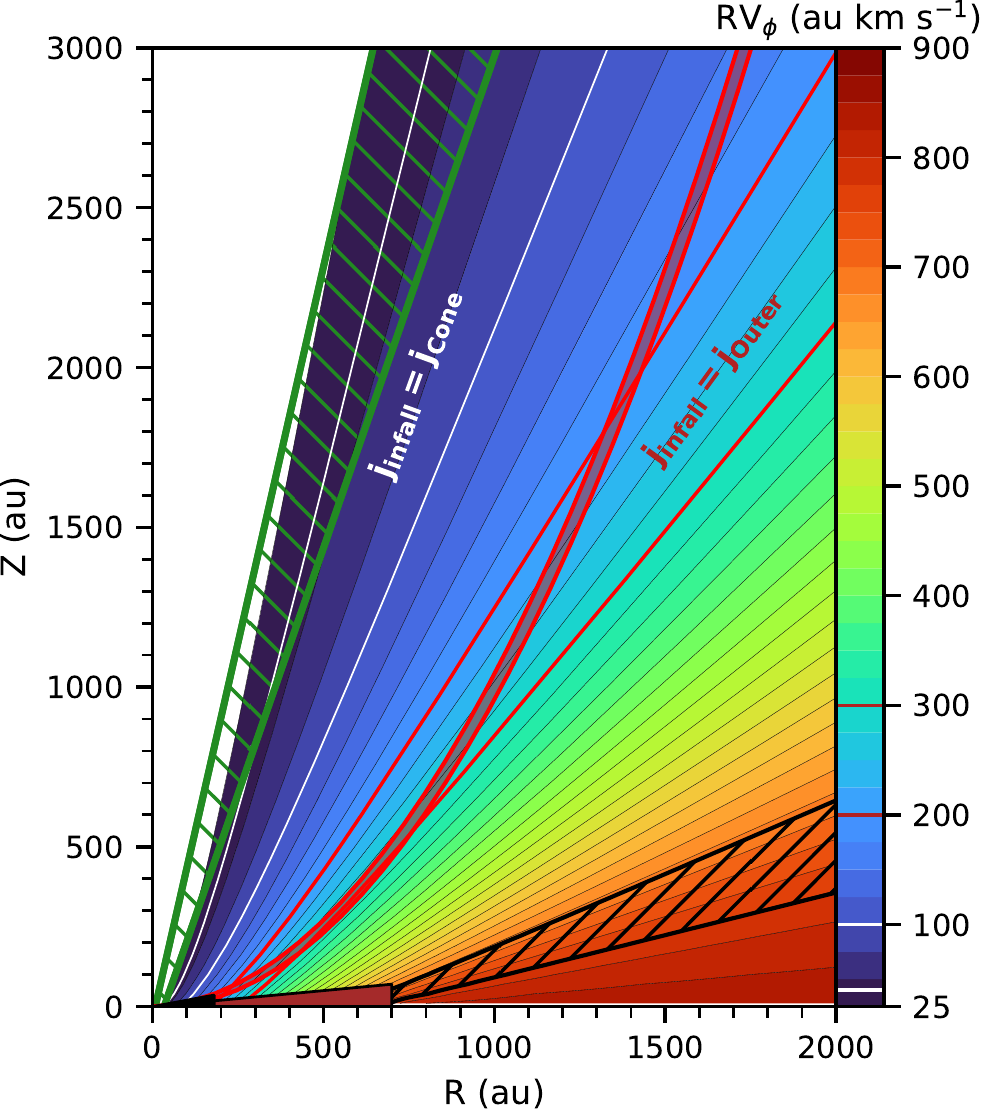}}
    \caption{Specific angular momentum map (in color) for the model of ballistic rotating infall \citep{ulrich_infall_1976} with $R_{\rm d}=700$ au and $M_\star = 1.1 M_\odot$. The angular momentum varies from 25 au \kms~to 900 au \kms.
    The green hatched and red filled regions represent respectively the limits of the conical outflow derived from the tomography in Sect. \ref{sec:tomo} and the parabolic shape of the classical Hubble-law WDS model fit to the external CO outflow.The white and red contours outline the infalling streamlines with a specific angular momentum similar to the conical outflow $j=40-100$~au~\kms~(white), and to the WDS model $j=200-300$~au~\kms~(red). The black hatched region outline the streamlines with initial $\theta_0 = 70 \pm 5^{\circ}$ reproducing the infall signatures seen in DG Tau B (DV20).}
\label{fig:mix}
\end{figure}

We discuss here a scenario where the redshifted CO outflow in DG~Tau~B can be accounted for by a stacking of multiple swept-up shells resulting from the interaction of an episodic wide-angle wind (or jet) with the ambient medium. 
Such a scenario is recently proposed by \citet{zhang_episodic_2019} 
to reproduce the multiple shell structures at the base of the HH46/47 outflow. 
They find good agreement with the WDS model of
\citet{lee_co_2000,Lee_Hydrodyamic_2001}, namely a parabolic layer undergoing radial expansion
following a Hubble law $\vec{V} \propto \vec{r}$. Below we summarize the key successes and failures encountered in Sect. \ref{sec:WDS} by the same WDS model when applied to the various flow components in DG Tau B, and we show that our  rotation measurements raise additional issues for this model in terms of cavity refilling.

\subsubsection{Outer flow component}
\label{disc:outer-wds}

In Sect. \ref{sec:outer}, we find that the morphology and kinematics of the low-velocity outflow can be well reproduced by the simple WDS model of \citet{lee_co_2000}. We now examine whether a WDS origin is physically consistent with the large specific angular momentum inferred from our modeling, $j_{\rm outer} \simeq 250\pm 50$ au \kms,  in the same sense as the disk (see Sect. \ref{sec:outer}).

Studies of rotation signatures in pre-stellar and proto-stellar cores show that specific angular momentum stops decreasing with radius below scales $\leq 3000$~au and becomes roughly constant  \citep{Ohashi_rotation_1997,Gaudel_angular_2020}.
This "plateau" is interpreted as the region where infall motions start to dominate, and specific angular momentum is roughly conserved along streamlines. 
Depending on the object, the specific angular momentum in the "plateau" is $\approx$ 40--400 au \kms\ \citep[$0.2-2 \times 10^{-3}$~\rm km ~\rm s$^{-1}$~pc,][]{Gaudel_angular_2020}.
Our estimate for the outer flow in DG Tau B, $j_{\rm outer} \simeq 250\pm 50$ au \kms\ falls well within this range. In addition,
infall signatures are identified around DG Tau B at large polar angles $\theta \simeq 70\degr$ (DV20). 
No such signatures are seen at smaller polar angles, but rotational flattening predicts lower envelope densities there \citep{ulrich_infall_1976}, hence they might be too faint for detection. It thus appears promising to consider that infalling material might dominate the rotation in the outer CO layer.

Strictly speaking, the  Hubble law assumed in the WDS model of \citet{lee_co_2000} is only valid for a static ambient medium with a $1/r^2$ radial density decrease\footnote{A static medium ensures that, after full mixing, the shell expands in the same radial direction as the wind, while the $1/r^2$ ambient density decrease ensures that the expansion speed is constant over time (the density ratio between the wind and ambient medium being independent of radius); both properties then together yield the "Hubble law" $\vec{V} = \vec{r} / \tau$ \citep{shu_star_1991,Lee_Hydrodyamic_2001}}. A rotating infalling envelope, in contrast, has a 
non-radial motion and a flatter radial density law $\propto 1/r^{1.5}$ \citep{ulrich_infall_1976}.
However, the calculations of \citet{lopez-vazquez_angular_2019} for such an ambient medium show that the WDS expansion remains quasi radial, except close to the mid-plane, and with almost constant speed after 200 yrs. Our simple model in Sect. \ref{sec:outer} thus remains roughly valid if the wind expands into an infalling envelope.

Using the shell rotation speeds computed by \citet{lopez-vazquez_angular_2019}, we expect a shell specific angular momentum close to that of the infalling material immediately ahead of it. Therefore, we consider in Fig.~\ref{fig:mix} the spatial distribution of specific angular momentum in a rotating, free-falling envelope from \citet{ulrich_infall_1976}, with a centrifugal radius $R_d=700$~au and central mass $M_{\star}=1.1 M_{\odot}$ appropriate to DG Tau B (DV20). The predicted $j$ values along the fit parabolic outer flow boundary are very similar to the observed one, $j_{\rm outer}$. This detailed comparison confirms that an infalling envelope in DG Tau B, if present up to polar angles $\theta \simeq 30\degr$, could provide enough specific angular momentum to explain the rotation in the outer CO layer. 

However, this analysis is highly simplistic. First, we consider here that the specific angular momentum of the envelope is fully transferred to the entrained layer. This assumption gives an upper limit for the shell rotation velocity, as turbulent mixing would decrease its specific angular momentum. Secondly, the spherical and ballistic infall model used here \citep{ulrich_infall_1976} does not take into account the effects of pressure gradients or the magnetic field. Pressure gradients could potentially increase the specific angular momentum of the infalling envelope at large polar angles, through "pushing" outer infalling streamlines toward the axis, while the magnetic field would decrease it due to magnetic breaking. Dedicated numerical simulations of the interaction of an infalling material with an inner wind component  and taking into account all these effects are needed to fully test the entrainment scenario for the outer CO layer.

A serious issue with this interpretation, however, is the young inferred shell age, $\tau=V_{\rm Z} / Z = 6000$~yrs (Sect. \ref{sec:outer}), much younger than the true age of DG~Tau~B. A first way out of this "short age problem" would be that the interface between wind and envelope in DG~Tau~B is not expanding, but static. A static shell may form when mixing between the wind and the ambient material is not instantaneous, as assumed in most WDS models, but very gradual. Shocked ambient material is then slowly entrained along the shell surface by the shocked wind in a thin turbulent mixing-layer. The static shell shape and mixing-layer properties were recently computed in the case of a free-falling rotating envelope by \citet{Liang_steady_2020}. Using again $R_d = 700$ au and $M_\star = 1.1M_\odot$ for DG Tau B (DV20), the specific angular momentum in the mixing-layer is predicted to be $j \approx 0.15 \sqrt{G M_\star R_d}$ $\simeq 120$~au~\kms, twice lower than estimated in the outer flow. Therefore, a static wind/envelope interface does not seem able to explain the rotating outer flow.

A second way out of the short age problem would be that the outer flow component does not trace the first wind encounter with the infalling envelope, but a more recent wind outburst from 6000 yrs ago. To provide its high angular momentum material to the young shell, however, the infalling envelope should somehow manage to penetrate and "refill" all the older shells created  by  previous (unobserved) wind outbursts. Whether such an efficient cavity refilling by the envelope is physically possible on $\leq 6000$ yr timescales is a difficult open question, well outside the scope of the present paper. As shown below, the issue of cavity "refilling" becomes even more acute when the WDS scenario is applied to the inner conical flow.

\subsubsection{Inner conical outflow}
\label{disc:inner-wds}

In contrast to the outer flow component, we find
in Sect. \ref{sec:WDS_conical} that the morphology and kinematics of the inner conical outflow and its bright substructures (arches, fingers and cusps) 
cannot be reproduced by the model of parabolic WDS with radial Hubble law used by \citet{lee_co_2000} and \citet{zhang_episodic_2019}, even after several ad hoc modifications. We identify two serious issues:
(1) The observed aspect ratio of arches in channel maps at low velocity is significantly shorter than predicted by the original model (ellipse aspect ratio = $1/\cos{i}$); it can be reproduced by a more collimated WDS model where the flow is parallel to the parabola; but that is no longer physically consistent with a Hubble-law velocity field, which requires an expanding shell \citep{shu_star_1991}.
(2) The shell models fitting the arches at low-velocity and cusps at high velocity do not agree with the observed emission morphology in channel maps at mid-velocities, predicting fingers that are broader than the cone, or full ellipses that are not seen.

We find that assuming a conical shell instead of a parabolic one, but keeping a Hubble law, still creates the same problems. They appear intrinsically caused by the Hubble-law velocity field, regardless of the shell detailed morphology. 
Therefore, any model where the shell is expanding quasi-radially and at nearly constant speed over time will fail to reproduce our observations. This, in particular, discards all models where the wide-angle wind and the ambient medium share a similar power-law in $r$, and where they instantly mix in the shell \citep[e.g., the models of][]{lopez-vazquez_angular_2019,shang_unified_2006}.

Simulations including magnetic field \citep{Wang_temperature_2015,Shang_unified_2020} and stationary solutions \citep{Liang_steady_2020} show the formation of a shear layer along the shell more in line with DG Tau B. However the first model predicts a shell anchoring radius increasing with time, while the second has a shell anchored near the centrifugal radius $R_d \simeq 700$ au (DV20). This is inconsistent with the small observed anchoring radius of the DG Tau B conical outflow ($\le 50$ au, DV20). Alternative models of swept-up shells exist involving an infalling sheet \citep{Cunningham_wide_2005} or a jet instead of a wide angle wind \citep[e.g., in][]{Downes_jet_2007}, but they have no analytical solutions. Therefore, dedicated numerical simulations would be required to test them in DG Tau B.

In the following, we show that the specific angular momentum measured in the conical outflow by tomography, $j_{\rm cone} \simeq 40-100$ au \kms\ (see Fig. \ref{fig:VZ_J}), raises additional issues for the swept-up shell scenario. We first note that a wide-angle "X-wind" cannot explain the observed rotation in the conical outflow; with a launching radius $\simeq 0.05-0.1$ au from the central protostar and a wind magnetic lever arm parameter $\lambda = j_X / j_{\rm Kep} \simeq 3$ \citep{Shang_synthetic_1998}, its specific angular momentum is predicted at $j_X \approx 20-30$~au~\kms, a factor two to three times less than observed. 
In addition, the wind cannot dominate the swept-up shell mass unless it is slower than twice the shell speed\footnote{ram pressure equilibrium between the reverse shock in the wind and the forward shock in the static ambient medium imposes $\rho_{\rm w} (V_{\rm w} - V_{\rm s})^2 = \rho_a V_{\rm s}^2$, where $V_{\rm w}$ and $V_{\rm s}$ are the wind and shell speeds, and $\rho_{\rm w}$ and $\rho_a$ are the wind and ambient density. The mass-flux entering the shell from the wind side will then dominate over the swept-up mass if $(V_{\rm w} - V_{\rm s}) < V_{\rm s}$.}, which in the present case would require $V_{\rm w} \leq 15$ \kms\ (see Table~\ref{tab:WDS_models}). This is inconsistent with a wind originating from close to the protostar. The "X-wind" model, for example, has $V_{\rm w} \simeq 150$ \kms~ \citep{Shang_synthetic_1998}.

The low observed expansion speeds $\simeq 6-14$ \kms\ in the conical flow (see Fig.\ref{fig:VZ_J}) is more consistent with jet bow shocks dominating the shell mass. A jet magnetic lever arm parameter $\lambda \simeq 9$ would then provide enough angular momentum. Such a scenario, however, cannot explain why regions of lower speed in the conical flow have inversely higher specific angular momentum (see Fig. \ref{fig:VZ_J}). In a jet bow shock, lower speeds arise where more ambient mass has been swept-up. Assuming the ambient medium provides no angular momentum, the jet angular momentum would get more diluted, and the shell specific angular momentum would drop, instead of increasing. We conclude that the observed rotation in the conical flow cannot come from an inner wind or jet.

To reproduce the observed conical flow rotation in the swept-up shell scenario, we  thus need an external medium with an important angular momentum. The infalling rotating envelope is an obvious candidate. However, the observed specific angular momentum in the conical flow is twice larger than predicted, in the same region, by the Ulrich infalling envelope model  (see Fig. \ref{fig:mix}). In addition, it is unclear how infalling matter could penetrate and "refill" the space between the closely spaced shells producing the cone substructures, especially when the region immediately outside the conical flow is instead in outflow motion (see Fig.~\ref{fig:outer}).

An alternative would be that the swept-up material originates from the rotating disk atmosphere, at radii $R_0 \simeq j^2 / (GM_\star)\simeq$~2-9~au. The problem is then to refill the cleared cavities between shells with a "new" static disk atmosphere. 
We note that our three shell models fitting the substructures in the conical flow have remarkably identical expansion speeds within 1 \kms\ (see last column of Table \ref{tab:WDS_models} for WDS-A2, WDS-A3 and WDS-U5). Assuming that the corresponding wind and jet outbursts were of similar strength, it implies that they met an identical ambient density ahead of them, hence the atmosphere refilling process should be extremely efficient. This appears difficult to achieve unless a large-scale disk wind is present. 

Realistic simulations of the interaction between an episodic inner wind or jet with an infalling rotating envelope and the disk atmosphere will be required to definitely exclude that swept-up shells with enough angular momentum and appropriate kinematics could be generated. However, at this stage and taking into account all the above-mentioned difficulties, we do not favor this scenario as the origin of the small-scale rotating CO outflow in DG Tau B. In the next section, we discuss an alternative scenario where  this rotating outflow traces a (perturbed) extended disk wind.

\subsection{The disk-wind scenario: constraints on the driving mechanism}
\label{sec:Discu_Steady}

We therefore favor the scenario in which the inner CO conical outflow traces matter directly ejected from the disk. We show in Sect.~\ref{sec:DW} that the stratified kinematical structure derived for the  conical outflow is suggestive of a quasi-steady disk-wind. In this section, we discuss constraints on the driving mechanism. Disk winds come in different flavors, depending on the main physical mechanism responsible for driving the flow: pure thermal effects in photo-evaporated disk winds (PDW) \citep{Alexander_dispersal_2014}, cold magneto-centrifugal ejection \citep{Pudritz_DW_2007} or a combination of the two processes in the so-called warm or magneto-thermal disk winds \citep{casse_magnetized_2000,bai_magneto-thermal_2016}. In the following we refer to the two last classes of magnetized disk winds as MHD disk winds (hereafter MHD DW). 

\begin{figure}[th!]
    \resizebox{\hsize}{!}{\includegraphics{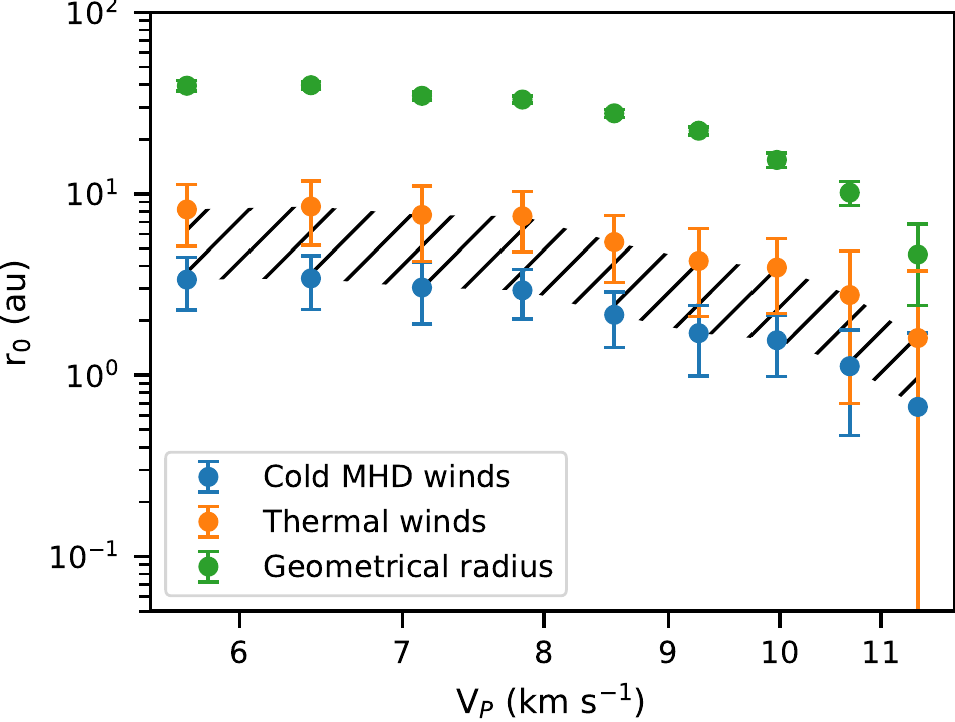}}
    \caption{Launching radius of the streamline as a function of its poloidal velocity assuming: conical extrapolation for the geometrical radius (green circles), steady thermal ejection conserving angular momentum (orange circles) and steady cold magneto-centrifugal ejection (blue circles). See text for more details. We do not derive the launching radius for the largest poloidal velocities due to the large uncertainties.
    }
    \label{fig:r0_value}
\end{figure}

\begin{figure}[th!]
    \resizebox{\hsize}{!}{\includegraphics{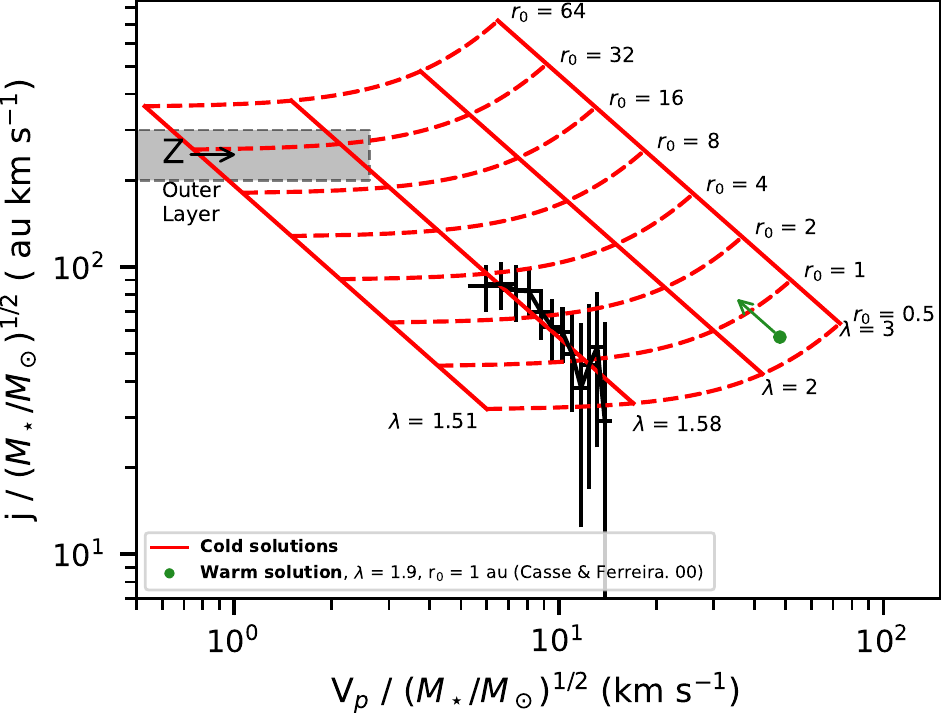}}
    \caption{Specific angular momentum ($j= R \times V_{\phi}$) versus poloidal velocity $V_{\rm P}$ for steady and axisymmetric MHD disk winds.
    The black symbols represent values derived from the tomography and averaged along lines of constant $V_{\rm Z}$, with their uncertainty. Red curves show the expected relation from self-similar cold magneto-centrifugal disk winds with $r_0$ varying from 0.5 to 100~au and $\lambda$ between 1.52 and 2.3 from \citet{ferreira_which_2006}. The green dot corresponds to a warm MHD disk wind solution from \cite{casse_magnetized_2000} with $r_0 = 1$ au and $\lambda = 1.9$. The green arrow shows the path followed by the solution with an increase of $r_0$. The gray box represents the estimated specific angular momentum for the outer low velocity layer on Sect.~\ref{sec:outer}, taking a maximal poloidal velocity corresponding to a height of $Z=3300$~au in the outflow referential.}
    \label{fig:MHDVZJ}
\end{figure}

\subsubsection{Photo-evaporated disk winds} 

In photo-evaporated disk winds, the high energy radiation (UV and X-rays) of the central accreting protostar heats the surface layers of the disk to high temperatures (10${^3}$-10${^4}$ K) up to significant radial distances. Beyond the gravitational radius $r_{\rm g}= (G M_\star)/c_{\rm s}^2$, where $c_{\rm s}$ is the sound speed in the upper disk surface layers, thermal energy exceeds the gravitational binding energy.  Numerical simulations show that significant mass loss starts before $r_{\rm g}$, at the critical radius $r_{\rm c} \simeq 0.1-0.2 r_{\rm g}$.  Consequently, matter is ejected and reaches terminal velocities of typically two to three times $c_{\rm s}$ \citep[][and references therein]{Alexander_dispersal_2014}. The exact properties of the wind depends on the dominant source of high-energy irradiation. 
Extreme-UV (EUV) heating  creates an isothermal ionized layer on the disk surface with temperature $T \simeq 10^4$~K ($c_{\rm s}=10$ \kms) which drives a fast wind ($\sim 35$~\kms) with mass-loss rates 10$^{-9}$-10$^{-10}$ M$_{\odot}$ yr$^{-1}$ 
\citep{font_photoevaporation_2004,Wang_hydrodynamic_2017}.
X-ray irradiation results in cooler and slower flows ($c_{\rm s} \simeq 3-5$ \kms, $v\simeq 15-20$ \kms) but penetrates at higher densities and therefore can drive mass-loss rates up to a few 10$^{-8}$ M$_{\odot}$ yr$^{-1}$ \citep{Picogna_dispersal_2019}. 
Non ionizing far-UV (FUV) heating mostly drives slow mass-loss ($v\simeq 18$ \kms) from the outer disk regions.

The conical morphology of the DG~Tau~B CO outflow  matches expectations from both self-similar PDW models by \cite{clarke_self-similar_2016} and hydrodynamical simulations by \cite{owen_protoplanetary_2011} and \citet{Wang_hydrodynamic_2017}. 
In such disk-wind solutions, the specific angular momentum is conserved along the streamlines and is equal to the Keplerian value at the foot-point radius. The observed values of $j$ in the CO conical outflow imply footpoint radii in the range $r_0=1.6$ to 8.2~au.
Figure \ref{fig:r0_value} shows that these foot-point radii are lower (by a factor $\simeq$ three) than the ones derived from a straight extrapolation of the conical morphology on large scales. This requires a larger opening angle of the streamlines at the base, consistent with the self-similar models of \citet{clarke_self-similar_2016}. 

The values of $r_0$ are in line with expectations from EUV dominated PDW models.  Indeed, for a 1 M$_{\odot}$ star and $c_{\rm s} \simeq 10$ \kms, expected in EUV heated PDW, the gravitational radius $r_{\rm g} \simeq 10$~au and significant mass loss starts at $r_{\rm c} \simeq 1-2$~au. 
On the other hand, observed terminal velocities of $V_{\rm p}= 4-16$ \kms~ require sound speeds $\le$ 2-8 \kms~ at radii $r_0 \le 10$ au, excluding EUV-driven models.

The large derived mass flux of $1.7-2.9 \times 10^{-7}$ M$_{\odot}$~yr$^{-1}$ for the CO conical flow (DV20) within $r_0\le 10$ au  however excludes FUV driven winds, which fail by at least one order of magnitude  \citep{Wang_hydrodynamic_2017}. We show below that it is also inconsistent with the latest X-ray driven models.
Indeed, recent X-ray driven photo-evaporation models of \cite{Picogna_dispersal_2019} predict mass loss rates up to 10$^{-7}$ M$_{\odot}$ yr$^{-1}$ for stellar X-ray luminosities $L_{\rm X} \ge 10^{31}$~erg~s$^{-1}$. However, these high $L_{\rm X}$ models also predict a stronger contribution of the outer disk regions to the total mass flux due to the increased penetration of X-ray photons. Figure~9 in \cite{Picogna_dispersal_2019} shows that only 10 \% of the total mass flux originates from disk foot-point radii below 10~au. In summary, current PDW models fail to account for the combination of large mass flux and small foot-point radii of $r_0= 1.6-8.2$~au, derived for the DG~Tau~B CO conical flow. 

Last but not least, the survival of CO molecules in such a wind is problematic. The full thermo-chemical computation of \citet{Wang_hydrodynamic_2017} shows that in their fiducial models, CO survives only at the very base of the wind in an intermediate layer on scales $z/r \le 0.6$. However, the models of \citet{Wang_hydrodynamic_2017} are EUV dominated and hence result in warm and fully ionized winds. Similar problems are expected in thermally driven winds launched from the inner disk, which require base temperature greater than 2000 K. Therefore, we conclude that pure thermal processes appear highly unlikely as the main driving mechanism of the DG~Tau~B CO conical wind.

\subsubsection{Magnetic disk winds} 

 Disk winds driven by magnetic forces require a large scale poloidal magnetic field anchored in the disk. This large-scale field exerts a torque on the rotating disk that both ejects matter and removes angular momentum from the disk \citep{blandford_hydromagnetic_1982,Pudritz_DW_2007}. The strength of this torque is characterized by the magnetic lever arm parameter $\lambda \simeq (r_A/r_0)^2$, where $r_A$ is the poloidal Alfven radius and $r_0$ the disk foot-point radius of the streamline. In principle, such disk winds can produce at the same time fast and collimated jets originating from the inner streamlines  and much slower and less collimated winds originating from outer disk radii.  The full kinematics and morphology of these solutions also depend on whether thermal effects are important in the launching regions. 
 Numerical simulations show that the mass loss can be significantly increased when thermal effects are taken into account \citep{casse_magnetized_2000,bai_magneto-thermal_2016}. Such magneto-thermal winds have low to moderate $\lambda$ values but can extract significant mass and angular momentum from the disk.

Terminal velocities depend on both the foot-point radii $r_0$, $\lambda$ and thermal effects.
Under the assumption of steady magnetically driven ejection, the asymptotic values of $V_{\rm p}$ and $V_{\phi}$ are given by \citep[Eqs.~4\&5 in][]{ferreira_which_2006}:
\begin{eqnarray}
    RV_{\phi} & =  & \lambda R_{\rm 0} V_{\phi}(R_{\rm 0})
    \label{eq:j_MHD}\\
    V_{\rm p} & =  & V_{\phi}(R_{\rm 0}) \sqrt{2\lambda-3+\beta}
    \label{eq:vp_MHD}
,\end{eqnarray}
where $\beta$ encompasses all pressure effects, including thermal and turbulent Alfv\'en waves \citep[see][]{ferreira_which_2006}. The streamline foot-point radius can be estimated from these equations assuming cold MHD ejection, ie. negligible thermal effects ($\beta \simeq 0$) following \citep{Anderson_Locating_2003}. Figure \ref{fig:MHDVZJ} traces the relationship between the mean values of $V_{\rm P}$ and $j=R V_{\phi}$ for the various conical layers of constant $V_{\rm Z}$ derived from the tomography. In this figure is also represented the parameter space ($r_0$,$\lambda$) predicted by cold magneto-centrifugal disk-wind models \citep{ferreira_which_2006}. 
The mean poloidal velocities and specific angular momentum coincide with a line of constant $\lambda_{\rm cold} \simeq 1.58$ with foot-point radii $r_{0,\rm cold}=0.7-3.4$~au.  If thermal effects play a dynamical role at the base of the wind, Eqs.~\ref{eq:j_MHD},\ref{eq:vp_MHD} show that the values of $\lambda$ and $r_0$ derived under the cold assumption are respectively upper and lower limits. This effect is illustrated in Fig.~\ref{fig:MHDVZJ} where we plot $V_{\rm P}$ and $j$ from the warmest solution of \cite{casse_magnetized_2000} with $\lambda =1.9$, at $r_0=1$~au. We see that using the cold MHD curves would lead to overestimate $\lambda \simeq 2.5$ and underestimate $r_0 \simeq 0.7$ au. The cold assumption is only valid if $\beta \ll 2\lambda -3 \simeq 0.2$, which would require a cold disk atmosphere and no substantial wind heating. On the other hand,
the low derived upper limit on $\lambda \le 1.6$ is consistent with warm MHD DW models \citep{casse_magnetized_2000,bai_magneto-thermal_2016,Wang_global_2019} or cold MHD DW from weakly magnetized disks  \citep{jacquemin_ide_magneticallydriven_2019}. 

The derived minimum foot-point radius of $r_{\rm in} \ge 0.7$~au for the CO wind streamlines is in  good agreement with the thermo-chemical predictions of \citet{panoglou_molecule_2012}, who demonstrate that CO molecules magnetically launched from foot-point radii $\ge 1$~au survive in the case of accretion rates in the disk $\ge 10^{-7}$~M$_{\odot}$~yr$^{-1}$. Similar results are obtained for warm magneto-thermal wind solutions \citep{Wang_global_2019}. 

The streamline foot-point radii derived from the kinematics are significantly smaller than the radii obtained from direct conical extrapolation (Fig.~\ref{fig:r0_value}), suggesting wider opening angle of the streamlines at their base. This is indeed expected in MHD DW solutions where streamlines originally follow a conical trajectory and recollimate on larger scales due to the hoop stress provided by the azimuthal B-field. The constant opening angle of the streamlines observed out to $Z = 3000$~au suggests that the magnetic hoop stress drops rapidly above $Z \simeq 50$ au.

Contrary to pure thermal disk winds, MHD DW also account for the observed large mass flux in the DG~Tau~B conical CO flow. Wind mass loss rates in the range 10$^{-8}$-10$^{-7}$~M$_{\odot}$~yr$^{-1}$ are predicted by the magneto-thermal wind solutions of \cite{Wang_global_2019}, on the same order as the accretion rate in the underlying disk and increasing with disk magnetization. 

We estimated the local ejection efficiency $\xi$ defined as $\Dot{M}_{\rm acc}(r) \propto r^{\xi}$ \citep{ferreira_which_2006}. We estimated $\xi$ from estimates of the disk accretion rate at the inner launching radius of the CO outflow. Indeed, from the mass conservation across the disk region launching the conical CO outflow ($\Dot{M}_{\rm acc}(r_{\rm in})=\Dot{M}_{\rm acc}(r_{\rm out})-\Dot{M}_{\rm DW}$) and the definition of $\xi$, we derived the following expression:
\begin{equation}
    \frac{\Dot{M}_{\rm DW}}{\Dot{M}_{\rm acc}(r_{\rm in})}=\bigg(\frac{r_{\rm out}}{r_{\rm in}}\bigg)^{\xi}-1
.\end{equation}
We estimated the accretion rate onto the central star by taking 10 \% of the jet mass flux \citep{Ellerbroek_outflow_2013}. \cite{podio_tracing_2011} estimate the red jet mass flux at $6.4 \times 10^{-9}$~M$_{\odot}$~yr$^{-1}$, giving a mass accretion rate onto the star of $\simeq 6 \times 10^{-8}$~M$_{\odot}$~yr$^{-1}$. 
We took this value as a lower limit to $\Dot{M}_{\rm acc} (r_{\rm in})$. From the measured mass flux in the conical CO flow $\Dot{M}_{\rm DW} = 2.3 \times 10^{-7}$~M$_{\odot}$~yr$^{-1}$ and the  disk-wind launching zone $r_{\rm out}/r_{\rm in} \simeq 5 $, we then derived an upper limit on $\xi \le 1$.
The mass flux in the conical wind is $\sim 4$ times the estimated accretion rate onto the star, implying that 80\% of the mass accreting at $r_{\rm out}$ is being ejected before reaching the star.

If the transport of angular momentum in the disk is entirely provided by the torque exerted by the MHD DW, one expects in steady state the following relationship: $\xi=1/(2(\lambda-1))$. The upper limit derived above on $\xi \le 1$  translates into $\lambda \ge 1.5$. This condition is compatible with our upper limit on $\lambda \le 1.58$ derived from the kinematics. Thus the CO mass flux combined with the constraints on launching radii and magnetic lever arm appear compatible with an MHD DW extracting all angular momentum required for the disk to accrete from $r_{\rm out}$ to $r_{\rm in}$. 

If the inner conical outflow is tracing a disk wind, then the outer parabolic outflow cannot be explained by the interaction between the envelope and an inner jet or X-wind. However, the outer outflow could be tracing the interaction between the envelope and outer disk-wind streamlines located outside of the conical outflow ($r_0 > 4$ au). The arches located at least partially outside of the conical outflow as well as the continuous aspect in the PV diagrams suggests an "intermediate" outflow located between the conical outflow and the outer parabolic surface.
Alternatively, the outer flow could also be tracing directly the outer disk-wind streamlines. The derived $j$ and maximal velocity in the outer layer would indicate launching radii $\ge 30$~au and a similar low $\lambda$ value as for the inner cone (see Fig. \ref{fig:MHDVZJ}). Moreover, \citet{bai_magneto-thermal_2016} show that some MHD DW solutions can accelerate until R $\sim$ 100 $r_0$, possibly explaining the apparent acceleration of this component seen in our channel maps. The global outflow would then be a continuous MHD DW originating from 0.7 au to $\ge 30$~au. Unfortunately, the morphology and dynamics of the potential disk wind originating from $r_0 > 4$ au could not be studied in detail due to our limited spectral sampling and the absorption by the surrounding cloud or envelope. However, in that scenario, the origin for the difference of emissivity between the bright conical outflow and the faint outer flow is not clear but could reflect the radial distribution of magnetic field strength or surface density in the underlying disk.

\subsection{Disk-wind scenario: origin of perturbations}
\label{sec:Discu_Pertu}

We discuss here the merits of different models for the origin of the substructures (arches, fingers, and cusps) in the conical flow, observed in the channel maps.

\subsubsection{Perturbation by jet bow shocks}
\label{sec:bowshock}
A first potential explanation for the bright substructures seen in the conical outflow is that the steady disk wind is perturbed by nested bow shock wings created by the propagation of the variable axial jet (see Fig. \ref{fig:Glob_MHD}, scenario B). 
Perturbation of the inner streamlines of a rotating disk wind by a large jet bow shock is recently reported in the much younger system of HH~212 by \citet{Lee_First_2021}. In DG Tau B, this interpretation is supported by the similar spatial spacings between the inferred locations of the perturbations producing the substructures in the CO conical wind, and the axial jet knots identified in optical images, as shown in Fig. 13 (for z=500-3000 au along the flow axis spacings range between 200-1300 au for the jet knots, 300-700 au for the over-densities). Indeed, in the jet-wind interaction scenario, the over-densities trace the point of contact between each bow shock and the outer disk wind so they propagate along the interface at the bow shock propagation speed, which is similar to the inner jet knot propagation speed.

Although the optical knot observations are not synchronous with our ALMA CO observations, and jet knots move away from the source at the jet speed $\simeq 150$ \kms, the general pattern of knot spacing as a function of distance is set by the underlying jet variability properties, 
and thus will tend to remain similar at different epochs in a given jet.
If, in addition, the jet undergoes low-amplitude wiggling (as frequently seen in young stars), perturbations to the disk wind caused by jet bow shocks would be slightly nonaxisymmetric, possibly explaining the apparent distortions in specific angular momentum along disk-wind streamlines using tomography (see Fig.~\ref{fig:XV_J}). Finally, this scenario might also explain the lack of recollimation of the conical disk-wind streamlines, due to the additional internal pressure created by the jet driven bow shocks wings. 

Hydrodynamical simulations of the interaction of a variable inner jet with a 
slower outer disk wind have been recently performed by \citet{tabone_interaction_2018}.
These simulations show the formation of a dense stationary conical layer closing down at the source, created by the stacking of jet bow shock wings in the disk wind. This shell exhibits local over-densities at the positions of individual jet bow shocks, illustrated by the red regions in  Fig. 10 (right panels)  in \citet{tabone_interaction_2018} . These over-densities globally reproduce the observed shapes of the perturbations in the DG Tau B conical layer: for the perturbations closer-in, the density map is dominated by the regions close to the bow shock apex which bend inward, while for the perturbations farther out the density map is dominated by the bow shock wings which bend outward.
This simulation also shows that the conical dense shell mostly retains the velocity of the surrounding disk wind, because the shock is weak and oblique. Therefore, the simulations in \citet{tabone_interaction_2018} does not show the characteristic stratification in
$V_{\rm Z}$ observed in the DG~Tau~B inner conical flow. 
The numerical simulations of  \citet{tabone_interaction_2018} also reproduce the observed trend of similar spacings between the axial jet knots and the over-densities in the contact region (shown in red in their Fig. 10, right panels).
However, the simulations are made in a simplistic configuration where
the outer disk wind is assumed to have a constant vertical velocity of 40\% of the jet speed, uniform density, and no rotation motion. More realistic simulations taking into account the velocity and density gradients across the outer disk wind, and including rotation and magnetic fields, are strongly needed to reliably test this scenario.

\subsubsection{Possible origin of CO outflow axis wiggling}

As discussed in Sect. \ref{sec:wigg}, wiggling of the wind ejection axis
is the only model explored here that can reproduce simply the variations of angular momentum observed along the DG Tau B outflow. From this analysis, we derived estimates of the wiggling period $\tau \simeq 400$~yrs and semi-amplitude wiggling angle $\alpha \simeq 0.5^{\circ}$.

Wind axis wiggling can originate from the precession of the underlying disk angular momentum vector. Disk axis precession can be induced by a mis-aligned companion. In that scenario, orbital periods are expected to be significantly shorter than the disk axis precession period \citep{Terquem_tidal_1998}, hence the companion would be located well within the disk. A companion in a mis-aligned orbit can open a gap in the disk, separating the dynamical evolution of the inner and outer disks \citep{Zhu_inclined_2019}. The inner disk then starts to precess with a period $\tau_{\rm p}$ related to the orbital period of the companion by  $\tau_o/\tau_{\rm p} \simeq 0.37 \mu/\sqrt{1-\mu}$, where $\mu=m_2/(m_1+m_2)$ is the ratio of the companion mass to the total mass of the binary system, and assuming a small mis-alignment $i_{\rm p}$  as suggested by the maximal wind precession angle $\alpha = 0.5\degr$ \citep[see Eq. 27 in][]{Zhu_inclined_2019}. An additional constraint can be obtained by requiring that the semi-amplitude wiggling angle $\alpha$ is dominated by the precession motion, which translates into the condition $V_0 \le V_{\rm Z} \tan(\alpha)$, that is $V_0 \le 0.1$ \kms where $V_0$ is the orbital velocity of the flow source. Combining these two constraints, we derive a companion mass ratio $\mu \le 2.5 \times 10^{-3}$ and binary separation: $a \le 0.5$~au. This separation is smaller than the launching radius of the CO disk wind ($r_0 \ge 0.7$~au, see Fig. \ref{fig:r0_value}).  So the precessing disk launching the CO flow would be outside the orbit of the planetary mass companion, which is inconsistent with the scenario investigated here. Indeed, in such a scenario the outer disk is precessing on much longer timescales than the ones given by the formula above.

\citet{Nixon_tearing_2013} and \citet{Facchini_signature_2018}  investigate circumbinary disk precession around an inner mis-aligned binary system. In such circumstances the inner rim of the circumbinary disk can break from the outer disk and precess. However, large mis-alignments or massive enough companions are necessary for this situation to occur \citep{Facchini_signature_2018}. We show that this is not likely in the DG~Tau~B case. The inner binary system would truncate the disk at  1.5-1.7 times the separation $a$  of the binary \citep{Facchini_signature_2018}. If we take an upper limit of $r_{\rm in}=0.7$~au for the inner circumbinary disk rim, to allow the launching of the CO flow, we get an upper limit of $a \le 0.5$~au for the binary separation, corresponding to an orbital period $\tau_0 \le 0.35$~yrs using Third Kepler's law and a total mass of 1 M$_{\odot}$ for the binary system (DV20). With a small misalignment (less than a few degrees) between the outer disk and the binary orbital planes, suggested by the small wiggling angle of the CO outflow, a mass ratio of the companion $\mu \le 10^{-2}$ would be required to get a precession period of 400~yrs \citep[see Eq. 4 in ][]{Facchini_signature_2018}. Equation~2 in \cite{Facchini_signature_2018} then shows that such a low mass companion combined with a small misalignment will not break the inner circumbinary disk. Therefore this second precession scenario can be ruled out to explain the CO outflow wiggling.

So far we have investigated only precession as the origin of the wiggling of the CO flow axis. However, wiggling of the disk-wind ejection axis can be also induced by the orbital motion of the CO outflow source in a binary system. The equations of motion of the ejected gas will be equivalent to the precession solution investigated so far.  We follow the formulation developed by \citep{masciadri_herbigharo_2002,Anglada_proprer_2007} assuming an orbital plane perpendicular to the outflow axis. From the third Kepler's law of motion, 
$a^3=M_{\rm tot} \tau_o^2$, with $\tau_o=400$~yrs and $M_{\rm tot} =1$ M$_{\odot}$ we can derive the mean separation of the companion at $a\simeq50$~au. On the other hand, from the semi-amplitude of the wiggling ($\alpha \simeq 0.5^{\circ}$) the  orbital velocity of the CO outflow source is constrained at: $V_0 = V_{\rm CO}  \times \tan(\alpha) = 0.1$~\kms, using an average velocity of $\simeq 10$~\kms~for the CO outflow. This in turn gives the orbital radius of the outflow source around the center of mass of the binary $r_0 = 1.3$~au and the ratio $\mu$ between the mass of the companion and the total mass of the system: $\mu = r_0/a = 0.025$. 
Thus a brown dwarf or massive planetary mass companion located at $\simeq 50$~au ($=0.35^{\prime\prime}$) separation would be required to account for the observed wiggling of the CO outflow in the orbital scenario.  
Such a low mass companion could have escaped direct detection so far \citep{Rodriguez_Radio_2012}. 
Strikingly the predicted companion separation is very close to an emission bump at $r=62$~au detected in the continuum emission profile of the disk at millimetric wavelengths \citep{de_valon_alma_2020,garufi_alma_2020}. However no clear gap is detected in the disk emission at this position which would be expected for such a massive companion.

Therefore, the precession scenario is excluded to account for the observed wiggling of the CO flow while the orbital scenario requires a brown dwarf or massive planetary mass companion at 50~au separation, which signature we do not clearly see in the disk yet. We also recall that our wiggling models cannot  account for the observed variable separations between cusps in channel maps. So we conclude that although attractive to explain some of  the substructures observed in the DG~Tau~B CO outflow, the interpretation of the wiggling scenario faces some difficulties.  We discuss below the alternative model where substructures arise from axisymmetric brightness enhancements in the disk wind.

\subsubsection{A variable disk wind}
\label{sec:Discu_Burst}

We show in Sect. \ref{sec:Density_DW} that the cusps, fingers, and a section of the arches can be explained by brightness or density enhancements in the conical outflow. 
The timescales between the density enhancements are typically a few hundred years (see Fig.~\ref{fig:discrete_tau_vz2}). 
Unfortunately, these timescales cannot be directly compared to the ones observed in the DG~Tau~B jet due to the larger jet velocity and its fading brightness at large distances. However, such timescales are observed on younger molecular outflows. The cluster W43-MM1 in \citet{nony_episodic_2020}, CARMA-7 in \citet{plunkett_episodic_2015} as well as HH46/47 in \citet{zhang_episodic_2019} show signatures of variability in molecular outflows with timescales between episodic events typically of a few hundred years. We may be witnessing similar variability in the DG~Tau~B CO outflow.

In the following, we discuss the possibility that these density enhancements are created by variability at the source in the  disk-wind launching regions. For the model presented in Fig.~\ref{fig:Burst_model}, we considered for the sake of simplicity that the density bursts take place simultaneously at all radii in the disk (that is, $\tau_i = cst$ for all layers). A more physical assumption would be to assume that the density burst propagates  radially across the disk with a velocity $V_{\rm prop}$.  If we consider that the burst takes place close to the mid-plane ($Z\approx 0$), 
the expression of the travel time for the density launched from a fixed radius $r_0$ is :
\begin{eqnarray}
\tau (\rm r_0) &=& t-t_{\rm eject}(r_0)\\
t_{\rm eject}(r_0) &=& t_{\rm eject}( 0) + \frac{r_0}{V_{\rm prop}}\\
\label{eq:tau_r0}
\tau (\rm r_0) &=& \tau ( 0) - \frac{\rm r_0}{\rm V_{\rm prop}}
,\end{eqnarray}
where $t$ is the current time and $t_{\rm eject}(r_0)$ is the epoch of density ejection.
Here, $V_{\rm prop}$ is positive when the density burst moves from the inner to the outer disk regions.

\begin{figure}
    \resizebox{\hsize}{!}{\includegraphics{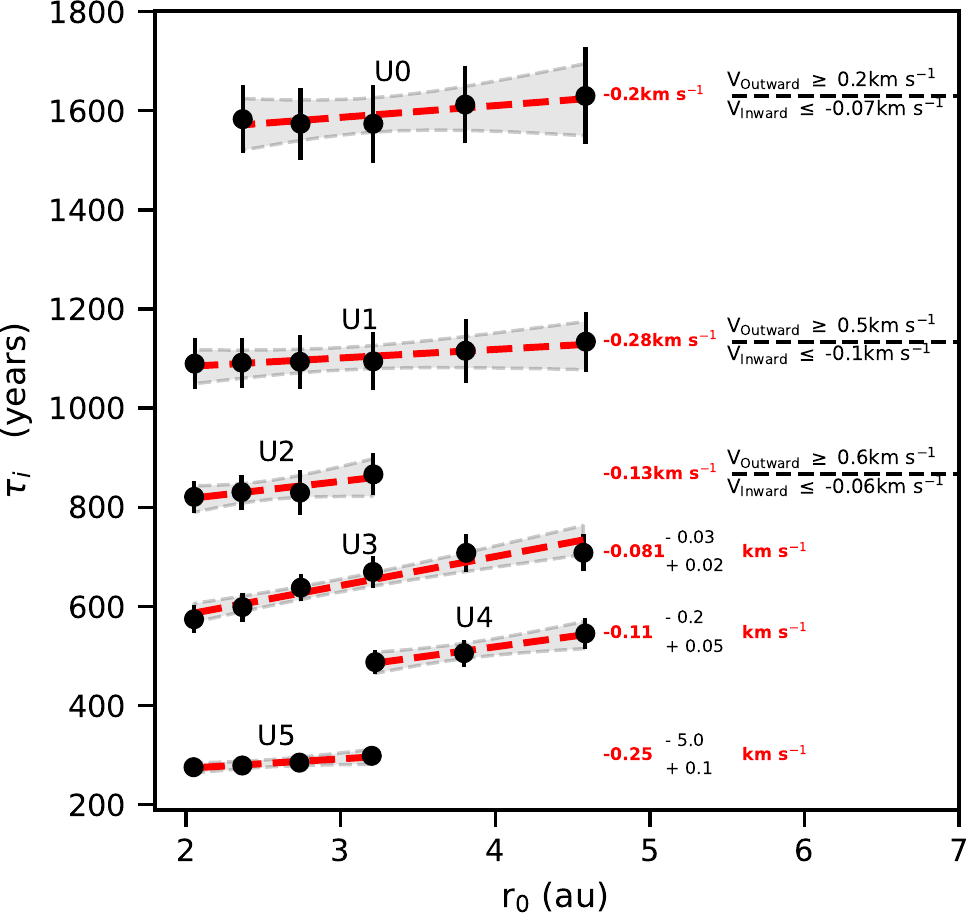}}
    \caption{Dynamical timescales of the density enhancements responsible for the high-velocity cusps in the channel maps, as a function of the launching radius $r_0$  of the streamline under the cold MHD  disk-wind hypothesis (assuming $\lambda = 1.58$). For each cusp, the red dashed line represents the best linear fit of $\tau(r_0)$ and its corresponding velocity.  The one $\sigma$ domain is shown in gray and the uncertainties of the fits are noted on the right.}
    \label{fig:Burst_r0}
\end{figure}

Figure \ref{fig:discrete_tau_vz2} indicates that $\tau$ decreases with increasing poloidal velocity, which is consistent with a density enhancement moving from the outer to the inner regions of the disk.
Figure~\ref{fig:Burst_r0} is a modification of Fig.~\ref{fig:discrete_tau_vz2} in which we transformed the poloidal velocity into the corresponding radius of ejection under the MHD disk-wind hypothesis, using Eq.~8 from \citet{ferreira_which_2006} with $\lambda_{\phi}=1.58$ and $\beta=0$. We did not take into account the uncertainty on the estimation of $r_0$ ($\simeq 20 \%$, see Fig. \ref{fig:r0_value}) as these uncertainties would make impossible the estimation of $V_{\rm prop}$. Nonetheless, this study gave an estimate of the propagation velocity. For each cusp, we traced the profile $\tau (\rm r_0) $ derived from the observed positions of the cusp apex in the different channel maps, as described in Sect.~\ref{sec:Density_DW}.
Linear fits to these profiles with associated $\boldsymbol{\tau}$ uncertainties are also shown. The slopes of these profiles is directly linked to $1/V_{\rm prop}$, and hence give the radial propagation velocities of the density enhancement at the origin.
The average velocity over all cusps is $V_{\rm prop} \approx -0.2$~\kms. The same study could be achieved with PDW models. In that case, the radius of ejection were multiplied by $\lambda^2$~($=2.5$) and therefore $V_{\rm prop}$ values were also multiplied by 2.5, giving an average radial velocity $V_{\rm prop}\approx -0.5$~\kms. From uncertainties on the slopes, we derived uncertainties on $V_{\rm prop}$ values in the case of U3, U4, and U5. In the case of U0,1,2, a horizontal solution ($V_{\rm prop}=\infty$) could not be excluded.  For these last three fits, we derived the minimal radial velocities for the two opposite directions of propagation.

\begin{figure*}
   \centering
    \resizebox{\hsize}{!}{\includegraphics[angle=90,scale=0.4]{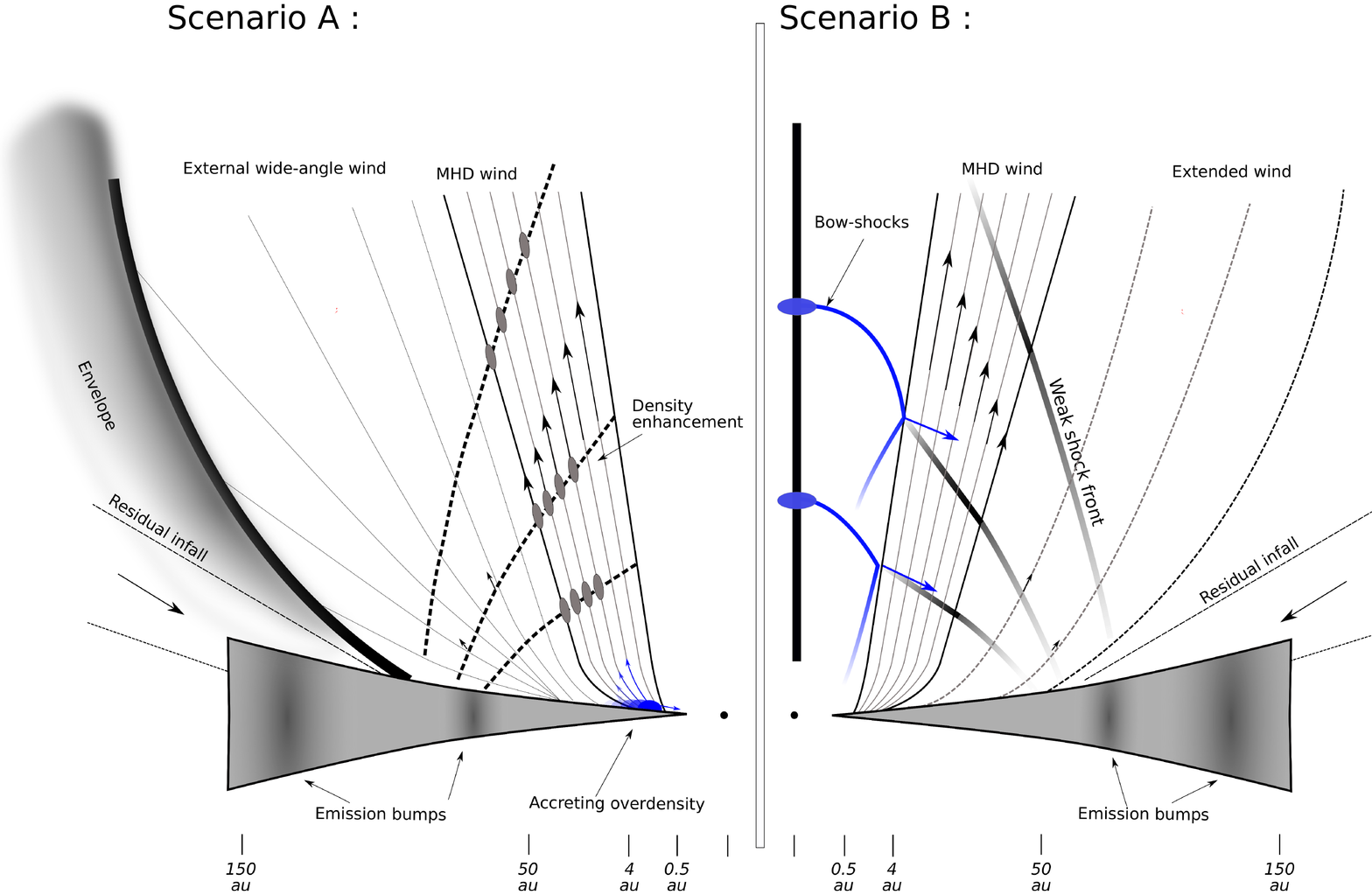}}
    \caption{Schematic scenarios describing the different components of the DG~Tau~B redshifted outflow in the disk-wind paradigm (see text). }
    \label{fig:Glob_MHD}
\end{figure*}

Fig.~\ref{fig:Burst_r0} seems to indicate that the density bursts propagate from the outer to the inner regions of the disk at $\approx |V_{\rm prop}|= 0.2-0.5$~\kms. The accreting density must also propagate toward the inner regions of the conical flow ($r_0 < 2$ au) where the density enhancements were visible but not characterized.
This propagation velocity corresponds to $\approx 0.01-0.04~V_{\rm Kep}(r=2-5~\rm au)$.
Such accretion velocities match expectations for radial surface velocities in MHD wind-driven accretion \citep{riols_dust_2020}.
From this velocity and our estimate of the burst duration of $\simeq$ 23 years, we can constraint the radial extent of the burst propagating along the disk at $\Delta R \approx 1$ au. Figure~\ref{fig:Glob_MHD} illustrates the proposed scenario.

Episodic density bursts propagating inward are obviously reminiscent of Fu Ori and Ex Ori type variable accretion events. Some Fu Ori have burst 
durations $<30$ years \citep[e.g., V1515 Cyg, or V1714 Cyg in][]{hartmann_fu_1996}. But periods are usually assumed to be $\approx$ 10$^{4}- 10^{5}$ years. On the other extreme, Ex Ori have typically bursts of a few months and periods of few years. DG Tau B variability seems to be located between these two extrema. However, the mass accretion rate increase in Fu Ori type events is typically three to four orders of magnitude higher than suggested in the DG Tau B outflow by the moderate factor three emissivity enhancement during the bursts.
Moreover, models of Fu Ori events predict a global accretion affecting the whole vertical structure of the disk during the high state. In magnetically accreting disk models, most of the mass is concentrated in the mid-plane, where the radial accretion velocity is 
$ \simeq V_{\rm Kep}/10^3$ \citep{riols_dust_2020}.
The high propagation velocity combined with the moderate emissivity enhancements derived in DG Tau B suggest that the accretion burst takes place locally on the disk surface and is less extreme than in typical Fu Ori phenomena. 
Such moderate accretion bursts could be due for example to residual infalling envelope material creating a shock wave when infalling into the disk \citep{Hennebelle_spiral_driven_2017}. 
It is important to note, however, that this axisymmetric model does not reproduce the local deviations observed in the specific angular momentum map (Fig. \ref{fig:VZ_J}). Nonaxisymmetric perturbations would be required.

\section{Conclusions}
\label{sec:conclusion}
We present a detailed analysis and modeling of the  ALMA $^{12}$CO(2-1) observations of the DG~Tau~B redshifted outflow published in \citet{de_valon_alma_2020}, with the aim to constrain its origin. Our main conclusions are as follows:

\begin{itemize}
    
    \item We identify three classes of discrete structures visible on the $^{12}$CO channel maps: arches at low velocities, fingers at medium velocities, and cusps at high velocities. Both cusps and arches show apparent acceleration in channel maps.
    
    \item We derived the 2D kinematics of the inner conical outflow using a tomographic method, assuming only axisymmetry of the outflow. 
    We reconstructed 2D maps for both the expansion velocity $V_{\rm Z}$ and specific angular momentum $j=R \times V_{\phi}$. The inner outflow shows a striking $V_{\rm Z}$ shear with faster material closer to the flow axis. Lines of constant $V_{\rm Z}$ are conical from $Z\simeq 50$~au out to $\approx 1200$~au. Specific angular momentum is roughly constant along those lines (except in two localized regions), and it increases outward inversely with $V_{\rm Z}$ from $\simeq 40$ to 100 au~\kms.
   
   \item The lower velocity external CO outflow shows a parabolic morphology, apparent acceleration, and a large specific angular momentum
    of $j=250\pm 50$~au~\kms. This suggests that it is tracing either a swept-up infalling and rotating envelope, or an extended disk wind launched from $\sim 30$ au.
  
    \item The conical outflow and the discrete structures could not be described by wind-driven shells with radial Hubble velocity laws. Such models fail to reproduce at the same time the morphologies of the observed structures (arches, fingers, and cusps) in the channel maps.
    Numerical simulations of the interaction between an episodic jet- and wide-angle wind with an infalling envelope are, however, required to confirm these conclusions.
      
    \item Instead, the conical outflow global morphology and kinematics appear consistent with matter directly ejected from the disk. Constraints on the disk-wind foot-point radii were derived at $r_0=1.6 - 8.2$~au (resp. 0.7 - 3.4~au) in the limiting cases where thermal (resp. magneto-centrifugal) processes dominate. However, none of the current photo-evaporated wind models can reproduce the large observed mass flux ( $2.3 \times 10^{-7}$~M$_{\odot}$~yr$^{-1}$ ) ejected from $r_0 < 10$ au.
    In contrast, an MHD disk-wind model with a constant magnetic level arm parameter $\lambda \leq 1.58$ can reproduce -- at the same time -- the flow velocity and angular momentum, as well as the large mass flux if it extracts most of the angular momentum for disk accretion across the wind launching region. The low lambda value is consistent with recent models of warm or weakly magnetized MHD DW \citep{bai_magneto-thermal_2016, jacquemin_ide_magneticallydriven_2019}.
    
    \item The wiggling of the flow axis may explain both the localized deviations of specific angular momentum and the morphology of the substructures in the conical flow. Orbital motion of the flow source in a binary of separation $\simeq 50$~au with companion mass $2.5 \times 10^{-3}$ M$_{\odot}$ can explain the inferred wiggling period and amplitude. Such a low mass companion could have escaped direct detection so far, but it should produce a gap signature in the continuum dust disk emission, which is not currently detected. In addition, the wiggling scenario fails to account for the variable separation between cusps in channel maps.
    
    \item Alternatively, the substructures observed in the CO channel maps can be explained by a series of mild (a factor three) density perturbations in the wind launching region, propagating inward at a radial velocity of $\simeq 0.2-0.5$~\kms, consistent with the
    surface accretion flow predicted in MHD wind-driven accretion. We derived a typical perturbation width of $\sim 1$ au and intervals of 200-500 years between perturbations. Alternatively the conical morphology and local density enhancements might be explained by the interaction of  inner jet bow shocks with the disk wind \citep{tabone_interaction_2018}; although, further numerical simulations are required to fully test this hypothesis.

\end{itemize}

The discrete structures that are increasingly observed in Class 0 and Class I outflows on larger scales have been usually interpreted in terms of nested shells swept up by an episodic inner wind. In contrast, we have shown in this paper that the substructures in the DG~Tau~B outflow appear best explained by density enhancements at the disk surface and that they propagate in a shear-like MHD disk wind. If confirmed, these results would directly demonstrate the link between accretion and ejection processes in embedded sources.

DG Tau B is a Class I protostar with a structured disk, infalling flows, and possible variable disk wind.
Structures in Class I disks are  assumed to trace early stages of planetary formation processes. Therefore, our results suggest that planetary formation is taking place in a very dynamic environment. The impact of such outflowing and infalling flows on the disk and its evolution remains an open question. Additional models and simulations of variable disk wind are needed in order to fully comprehend its impact on disk evolution and planet formation. 

\begin{acknowledgements}
The authors would like to thank the referee, whose comments helped improve the quality of the paper. This paper makes use of the following ALMA data: ADS/JAO.ALMA\#2015.1.01108.S, ADS/JAO.ALMA\#2017.1.01605.S. ALMA is a partnership of ESO (representing its member states), NSF (USA) and NINS (Japan), together with NRC (Canada), MOST and ASIAA (Taiwan), and KASI (Republic of Korea), in cooperation with the Republic of Chile.The Joint ALMA Observatory is operated by ESO, AUI/NRAO and NAOJ. This work was supported by the Programme National de Physique Stellaire (PNPS) 
and the Programme National de Physique et Chimie du Milieu Interstellaire (PCMI) of CNRS/INSU co-funded by CEA and CNES. FL acknowledges the support of the Marie Curie Action of the European Union (project \textsl{MagiKStar}, Grant agreement number 841276).
\end{acknowledgements}

\bibliographystyle{aa} 
\bibliography{Lib_letter} 

\begin{thebibliography}{68}
\expandafter\ifx\csname natexlab\endcsname\relax\def\natexlab#1{#1}\fi

\bibitem[{{Alexander} {et~al.}(2014){Alexander}, {Pascucci}, {Andrews},
  {Armitage}, \& {Cieza}}]{Alexander_dispersal_2014}
{Alexander}, R., {Pascucci}, I., {Andrews}, S., {Armitage}, P., \& {Cieza}, L.
  2014, in Protostars and Planets VI, ed. H.~{Beuther}, R.~S. {Klessen}, C.~P.
  {Dullemond}, \& T.~{Henning}, 475

\bibitem[{{Anderson} {et~al.}(2003){Anderson}, {Li}, {Krasnopolsky}, \&
  {Blandford}}]{Anderson_Locating_2003}
{Anderson}, J.~M., {Li}, Z.-Y., {Krasnopolsky}, R., \& {Blandford}, R.~D. 2003,
  \apjl, 590, L107

\bibitem[{{Anglada} {et~al.}(2007){Anglada}, {L{\'o}pez}, {Estalella},
  {Masegosa}, {Riera}, \& {Raga}}]{Anglada_proprer_2007}
{Anglada}, G., {L{\'o}pez}, R., {Estalella}, R., {et~al.} 2007, \aj, 133, 2799

\bibitem[{{Arce} {et~al.}(2007){Arce}, {Shepherd}, {Gueth}, {Lee}, {Bachiller},
  {Rosen}, \& {Beuther}}]{Arce_molecular_2007}
{Arce}, H.~G., {Shepherd}, D., {Gueth}, F., {et~al.} 2007, in Protostars and
  Planets V, ed. B.~{Reipurth}, D.~{Jewitt}, \& K.~{Keil}, 245

\bibitem[{{Bai} {et~al.}(2016){Bai}, {Ye}, {Goodman}, \&
  {Yuan}}]{bai_magneto-thermal_2016}
{Bai}, X.-N., {Ye}, J., {Goodman}, J., \& {Yuan}, F. 2016, \apj, 818, 152

\bibitem[{{Bjerkeli} {et~al.}(2016){Bjerkeli}, {van der Wiel}, {Harsono},
  {Ramsey}, \& {J{\o}rgensen}}]{bjerkeli_resolved_2016}
{Bjerkeli}, P., {van der Wiel}, M. H.~D., {Harsono}, D., {Ramsey}, J.~P., \&
  {J{\o}rgensen}, J.~K. 2016, \nat, 540, 406

\bibitem[{{Blandford} \& {Payne}(1982)}]{blandford_hydromagnetic_1982}
{Blandford}, R.~D. \& {Payne}, D.~G. 1982, \mnras, 199, 883

\bibitem[{{Cabrit} {et~al.}(1997){Cabrit}, {Raga}, \&
  {Gueth}}]{Cabrit_models_1997}
{Cabrit}, S., {Raga}, A., \& {Gueth}, F. 1997, in Herbig-Haro Flows and the
  Birth of Stars, ed. B.~{Reipurth} \& C.~{Bertout}, Vol. 182, 163--180

\bibitem[{{Casse} \& {Ferreira}(2000)}]{casse_magnetized_2000}
{Casse}, F. \& {Ferreira}, J. 2000, \aap, 353, 1115

\bibitem[{{Clarke} \& {Alexander}(2016)}]{clarke_self-similar_2016}
{Clarke}, C.~J. \& {Alexander}, R.~D. 2016, \mnras, 460, 3044

\bibitem[{{Cunningham} {et~al.}(2005){Cunningham}, {Frank}, \&
  {Hartmann}}]{Cunningham_wide_2005}
{Cunningham}, A., {Frank}, A., \& {Hartmann}, L. 2005, \apj, 631, 1010

\bibitem[{{de Valon} {et~al.}(2020){de Valon}, {Dougados}, {Cabrit}, {Louvet},
  {Zapata}, \& {Mardones}}]{de_valon_alma_2020}
{de Valon}, A., {Dougados}, C., {Cabrit}, S., {et~al.} 2020, \aap, 634, L12
  (DV20)

\bibitem[{{Downes} \& {Cabrit}(2007)}]{Downes_jet_2007}
{Downes}, T.~P. \& {Cabrit}, S. 2007, \aap, 471, 873

\bibitem[{{Eisl{\"o}ffel} \& {Mundt}(1998)}]{eisloffel_imaging_1998}
{Eisl{\"o}ffel}, J. \& {Mundt}, R. 1998, \aj, 115, 1554

\bibitem[{{Ellerbroek} {et~al.}(2013){Ellerbroek}, {Podio}, {Kaper}, {Sana},
  {Huppenkothen}, {de Koter}, \& {Monaco}}]{Ellerbroek_outflow_2013}
{Ellerbroek}, L.~E., {Podio}, L., {Kaper}, L., {et~al.} 2013, \aap, 551, A5

\bibitem[{{Facchini} {et~al.}(2018){Facchini}, {Juh{\'a}sz}, \&
  {Lodato}}]{Facchini_signature_2018}
{Facchini}, S., {Juh{\'a}sz}, A., \& {Lodato}, G. 2018, \mnras, 473, 4459

\bibitem[{{Fern{\'a}ndez-L{\'o}pez} {et~al.}(2020){Fern{\'a}ndez-L{\'o}pez},
  {Zapata}, {Rodr{\'\i}guez}, {Vazzano}, {Guzm{\'a}n}, \&
  {L{\'o}pez}}]{lopez_ringed_2020}
{Fern{\'a}ndez-L{\'o}pez}, M., {Zapata}, L.~A., {Rodr{\'\i}guez}, L.~F.,
  {et~al.} 2020, \aj, 159, 171

\bibitem[{{Ferreira} {et~al.}(2006){Ferreira}, {Dougados}, \&
  {Cabrit}}]{ferreira_which_2006}
{Ferreira}, J., {Dougados}, C., \& {Cabrit}, S. 2006, \aap, 453, 785

\bibitem[{{Font} {et~al.}(2004){Font}, {McCarthy}, {Johnstone}, \&
  {Ballantyne}}]{font_photoevaporation_2004}
{Font}, A.~S., {McCarthy}, I.~G., {Johnstone}, D., \& {Ballantyne}, D.~R. 2004,
  \apj, 607, 890

\bibitem[{{Garufi} {et~al.}(2020){Garufi}, {Podio}, {Codella}, {Rygl},
  {Bacciotti}, {Facchini}, {Fedele}, {Miotello}, {Teague}, \&
  {Testi}}]{garufi_alma_2020}
{Garufi}, A., {Podio}, L., {Codella}, C., {et~al.} 2020, \aap, 636, A65

\bibitem[{{Gaudel} {et~al.}(2020){Gaudel}, {Maury}, {Belloche}, {Maret},
  {Andr{\'e}}, {Hennebelle}, {Galametz}, {Testi}, {Cabrit}, {Palmeirim},
  {Ladjelate}, {Codella}, \& {Podio}}]{Gaudel_angular_2020}
{Gaudel}, M., {Maury}, A.~J., {Belloche}, A., {et~al.} 2020, \aap, 637, A92

\bibitem[{{Guilloteau} {et~al.}(2011){Guilloteau}, {Dutrey}, {Pi{\'e}tu}, \&
  {Boehler}}]{guilloteau_dual-frequency_2011}
{Guilloteau}, S., {Dutrey}, A., {Pi{\'e}tu}, V., \& {Boehler}, Y. 2011, \aap,
  529, A105

\bibitem[{{Hartmann} \& {Kenyon}(1996)}]{hartmann_fu_1996}
{Hartmann}, L. \& {Kenyon}, S.~J. 1996, \araa, 34, 207

\bibitem[{{Hennebelle} {et~al.}(2017){Hennebelle}, {Lesur}, \&
  {Fromang}}]{Hennebelle_spiral_driven_2017}
{Hennebelle}, P., {Lesur}, G., \& {Fromang}, S. 2017, \aap, 599, A86

\bibitem[{{Hirota} {et~al.}(2017){Hirota}, {Machida}, {Matsushita}, {Motogi},
  {Matsumoto}, {Kim}, {Burns}, \& {Honma}}]{Hirota2017}
{Hirota}, T., {Machida}, M.~N., {Matsushita}, Y., {et~al.} 2017, Nature
  Astronomy, 1, 0146

\bibitem[{{Jacquemin-Ide} {et~al.}(2019){Jacquemin-Ide}, {Ferreira}, \&
  {Lesur}}]{jacquemin_ide_magneticallydriven_2019}
{Jacquemin-Ide}, J., {Ferreira}, J., \& {Lesur}, G. 2019, \mnras, 490, 3112

\bibitem[{{Launhardt} {et~al.}(2009){Launhardt}, {Pavlyuchenkov}, {Gueth},
  {Chen}, {Dutrey}, {Guilloteau}, {Henning}, {Pi{\'e}tu}, {Schreyer}, \&
  {Semenov}}]{Launhardt2009}
{Launhardt}, R., {Pavlyuchenkov}, Y., {Gueth}, F., {et~al.} 2009, \aap, 494,
  147

\bibitem[{{Lee} {et~al.}(2018){Lee}, {Li}, {Hirano}, {Shang}, {Ho}, \&
  {Zhang}}]{Lee2018-HH211}
{Lee}, C.-F., {Li}, Z.-Y., {Hirano}, N., {et~al.} 2018, \apj, 863, 94

\bibitem[{{Lee} {et~al.}(2000){Lee}, {Mundy}, {Reipurth}, {Ostriker}, \&
  {Stone}}]{lee_co_2000}
{Lee}, C.-F., {Mundy}, L.~G., {Reipurth}, B., {Ostriker}, E.~C., \& {Stone},
  J.~M. 2000, \apj, 542, 925

\bibitem[{{Lee} {et~al.}(2001){Lee}, {Stone}, {Ostriker}, \&
  {Mundy}}]{Lee_Hydrodyamic_2001}
{Lee}, C.-F., {Stone}, J.~M., {Ostriker}, E.~C., \& {Mundy}, L.~G. 2001, \apj,
  557, 429

\bibitem[{{Lee} {et~al.}(2021){Lee}, {Tabone}, {Cabrit}, {Codella}, {Podio},
  {Ferreira}, \& {Jacquemin-Ide}}]{Lee_First_2021}
{Lee}, C.-F., {Tabone}, B., {Cabrit}, S., {et~al.} 2021, \apjl, 907, L41

\bibitem[{{Liang} {et~al.}(2020){Liang}, {Johnstone}, {Cabrit}, \&
  {Kristensen}}]{Liang_steady_2020}
{Liang}, L., {Johnstone}, D., {Cabrit}, S., \& {Kristensen}, L.~E. 2020, \apj,
  900, 15

\bibitem[{{L{\'o}pez-V{\'a}zquez} {et~al.}(2019){L{\'o}pez-V{\'a}zquez},
  {Cant{\'o}}, \& {Lizano}}]{lopez-vazquez_angular_2019}
{L{\'o}pez-V{\'a}zquez}, J.~A., {Cant{\'o}}, J., \& {Lizano}, S. 2019, \apj,
  879, 42

\bibitem[{{Louvet} {et~al.}(2018){Louvet}, {Dougados}, {Cabrit}, {Mardones},
  {M{\'e}nard}, {Tabone}, {Pinte}, \& {Dent}}]{louvet_hh30_2018}
{Louvet}, F., {Dougados}, C., {Cabrit}, S., {et~al.} 2018, \aap, 618, A120

\bibitem[{{Masciadri} \& {Raga}(2002)}]{masciadri_herbigharo_2002}
{Masciadri}, E. \& {Raga}, A.~C. 2002, \apj, 568, 733

\bibitem[{{Mitchell} {et~al.}(1997){Mitchell}, {Sargent}, \&
  {Mannings}}]{mitchell_dg_1997}
{Mitchell}, G.~F., {Sargent}, A.~I., \& {Mannings}, V. 1997, \apjl, 483, L127

\bibitem[{{Mundt} {et~al.}(1987){Mundt}, {Brugel}, \&
  {Buehrke}}]{mundt_collimation_1991}
{Mundt}, R., {Brugel}, E.~W., \& {Buehrke}, T. 1987, \apj, 319, 275

\bibitem[{{Mundt} \& {Fried}(1983)}]{mundt_jets_1983}
{Mundt}, R. \& {Fried}, J.~W. 1983, \apjl, 274, L83

\bibitem[{{Mundt} {et~al.}(1991){Mundt}, {Ray}, \& {Raga}}]{mundt_jets_1987}
{Mundt}, R., {Ray}, T.~P., \& {Raga}, A.~C. 1991, \aap, 252, 740

\bibitem[{{Nixon} {et~al.}(2013){Nixon}, {King}, \&
  {Price}}]{Nixon_tearing_2013}
{Nixon}, C., {King}, A., \& {Price}, D. 2013, \mnras, 434, 1946

\bibitem[{{Nony} {et~al.}(2020){Nony}, {Motte}, {Louvet}, {Plunkett},
  {Gusdorf}, {Fechtenbaum}, {Pouteau}, {Lefloch}, {Bontemps}, {Molet}, \&
  {Robitaille}}]{nony_episodic_2020}
{Nony}, T., {Motte}, F., {Louvet}, F., {et~al.} 2020, \aap, 636, A38

\bibitem[{{Ohashi} {et~al.}(1997){Ohashi}, {Hayashi}, {Ho}, {Momose}, {Tamura},
  {Hirano}, \& {Sargent}}]{Ohashi_rotation_1997}
{Ohashi}, N., {Hayashi}, M., {Ho}, P. T.~P., {et~al.} 1997, \apj, 488, 317

\bibitem[{{Owen} {et~al.}(2011){Owen}, {Ercolano}, \&
  {Clarke}}]{owen_protoplanetary_2011}
{Owen}, J.~E., {Ercolano}, B., \& {Clarke}, C.~J. 2011, \mnras, 412, 13

\bibitem[{{Panoglou} {et~al.}(2012){Panoglou}, {Cabrit}, {Pineau Des
  For{\^e}ts}, {Garcia}, {Ferreira}, \& {Casse}}]{panoglou_molecule_2012}
{Panoglou}, D., {Cabrit}, S., {Pineau Des For{\^e}ts}, G., {et~al.} 2012, \aap,
  538, A2

\bibitem[{{Pety} {et~al.}(2006){Pety}, {Gueth}, {Guilloteau}, \&
  {Dutrey}}]{pety_hh30}
{Pety}, J., {Gueth}, F., {Guilloteau}, S., \& {Dutrey}, A. 2006, \aap, 458, 841

\bibitem[{{Picogna} {et~al.}(2019){Picogna}, {Ercolano}, {Owen}, \&
  {Weber}}]{Picogna_dispersal_2019}
{Picogna}, G., {Ercolano}, B., {Owen}, J.~E., \& {Weber}, M.~L. 2019, \mnras,
  487, 691

\bibitem[{{Plunkett} {et~al.}(2015){Plunkett}, {Arce}, {Mardones}, {van
  Dokkum}, {Dunham}, {Fern{\'a}ndez-L{\'o}pez}, {Gallardo}, \&
  {Corder}}]{plunkett_episodic_2015}
{Plunkett}, A.~L., {Arce}, H.~G., {Mardones}, D., {et~al.} 2015, \nat, 527, 70

\bibitem[{{Podio} {et~al.}(2011){Podio}, {Eisl{\"o}ffel}, {Melnikov}, {Hodapp},
  \& {Bacciotti}}]{podio_tracing_2011}
{Podio}, L., {Eisl{\"o}ffel}, J., {Melnikov}, S., {Hodapp}, K.~W., \&
  {Bacciotti}, F. 2011, \aap, 527, A13

\bibitem[{{Pudritz} {et~al.}(2007){Pudritz}, {Ouyed}, {Fendt}, \&
  {Brandenburg}}]{Pudritz_DW_2007}
{Pudritz}, R.~E., {Ouyed}, R., {Fendt}, C., \& {Brandenburg}, A. 2007, in
  Protostars and Planets V, ed. B.~{Reipurth}, D.~{Jewitt}, \& K.~{Keil}, 277

\bibitem[{{Riols} {et~al.}(2020){Riols}, {Lesur}, \&
  {Menard}}]{riols_dust_2020}
{Riols}, A., {Lesur}, G., \& {Menard}, F. 2020, \aap, 639, A95

\bibitem[{{Rodr{\'\i}guez} {et~al.}(2012){Rodr{\'\i}guez}, {Dzib}, {Loinard},
  {Zapata}, {Raga}, {Cant{\'o}}, \& {Riera}}]{Rodriguez_Radio_2012}
{Rodr{\'\i}guez}, L.~F., {Dzib}, S.~A., {Loinard}, L., {et~al.} 2012, \rmxaa,
  48, 243

\bibitem[{{Shang} {et~al.}(2006){Shang}, {Allen}, {Li}, {Liu}, {Chou}, \&
  {Anderson}}]{shang_unified_2006}
{Shang}, H., {Allen}, A., {Li}, Z.-Y., {et~al.} 2006, \apj, 649, 845

\bibitem[{{Shang} {et~al.}(2020){Shang}, {Krasnopolsky}, {Liu}, \&
  {Wang}}]{Shang_unified_2020}
{Shang}, H., {Krasnopolsky}, R., {Liu}, C.-F., \& {Wang}, L.-Y. 2020, \apj,
  905, 116

\bibitem[{{Shang} {et~al.}(1998){Shang}, {Shu}, \&
  {Glassgold}}]{Shang_synthetic_1998}
{Shang}, H., {Shu}, F.~H., \& {Glassgold}, A.~E. 1998, \apjl, 493, L91

\bibitem[{{Shu} {et~al.}(1991){Shu}, {Ruden}, {Lada}, \&
  {Lizano}}]{shu_star_1991}
{Shu}, F.~H., {Ruden}, S.~P., {Lada}, C.~J., \& {Lizano}, S. 1991, \apjl, 370,
  L31

\bibitem[{{Tabone} {et~al.}(2017){Tabone}, {Cabrit}, {Bianchi}, {Ferreira},
  {Pineau des For{\^e}ts}, {Codella}, {Gusdorf}, {Gueth}, {Podio}, \&
  {Chapillon}}]{tabone_alma_2017}
{Tabone}, B., {Cabrit}, S., {Bianchi}, E., {et~al.} 2017, \aap, 607, L6

\bibitem[{{Tabone} {et~al.}(2020){Tabone}, {Cabrit}, {Pineau des For{\^e}ts},
  {Ferreira}, {Gusdorf}, {Podio}, {Bianchi}, {Chapillon}, {Codella}, \&
  {Gueth}}]{Tabone2020}
{Tabone}, B., {Cabrit}, S., {Pineau des For{\^e}ts}, G., {et~al.} 2020, \aap,
  640, A82

\bibitem[{{Tabone} {et~al.}(2018){Tabone}, {Raga}, {Cabrit}, \& {Pineau des
  For{\^e}ts}}]{tabone_interaction_2018}
{Tabone}, B., {Raga}, A., {Cabrit}, S., \& {Pineau des For{\^e}ts}, G. 2018,
  \aap, 614, A119

\bibitem[{{Terquem} {et~al.}(1998){Terquem}, {Papaloizou}, {Nelson}, \&
  {Lin}}]{Terquem_tidal_1998}
{Terquem}, C., {Papaloizou}, J.~C.~B., {Nelson}, R.~P., \& {Lin}, D.~N.~C.
  1998, \apj, 502, 788

\bibitem[{{Ulrich}(1976)}]{ulrich_infall_1976}
{Ulrich}, R.~K. 1976, \apj, 210, 377

\bibitem[{{Wang} {et~al.}(2019){Wang}, {Bai}, \& {Goodman}}]{Wang_global_2019}
{Wang}, L., {Bai}, X.-N., \& {Goodman}, J. 2019, \apj, 874, 90

\bibitem[{{Wang} \& {Goodman}(2017)}]{Wang_hydrodynamic_2017}
{Wang}, L. \& {Goodman}, J. 2017, \apj, 847, 11

\bibitem[{{Wang} {et~al.}(2015){Wang}, {Shang}, {Krasnopolsky}, \&
  {Chiang}}]{Wang_temperature_2015}
{Wang}, L.-Y., {Shang}, H., {Krasnopolsky}, R., \& {Chiang}, T.-Y. 2015, \apj,
  815, 39

\bibitem[{{Wilgenbus} {et~al.}(2000){Wilgenbus}, {Cabrit}, {Pineau des
  For{\^e}ts}, \& {Flower}}]{Wilgenbus_ortho_2000}
{Wilgenbus}, D., {Cabrit}, S., {Pineau des For{\^e}ts}, G., \& {Flower}, D.~R.
  2000, \aap, 356, 1010

\bibitem[{{Zapata} {et~al.}(2015){Zapata}, {Lizano}, {Rodr{\'\i}guez}, {Ho},
  {Loinard}, {Fern{\'a}ndez-L{\'o}pez}, \& {Tafoya}}]{zapata_kinematics_2015}
{Zapata}, L.~A., {Lizano}, S., {Rodr{\'\i}guez}, L.~F., {et~al.} 2015, \apj,
  798, 131

\bibitem[{{Zhang} {et~al.}(2019){Zhang}, {Arce}, {Mardones}, {Cabrit},
  {Dunham}, {Garay}, {Noriega-Crespo}, {Offner}, {Raga}, \&
  {Corder}}]{zhang_episodic_2019}
{Zhang}, Y., {Arce}, H.~G., {Mardones}, D., {et~al.} 2019, \apj, 883, 1

\bibitem[{{Zhang} {et~al.}(2018){Zhang}, {Higuchi}, {Sakai}, {Oya},
  {L{\'o}pez-Sepulcre}, {Imai}, {Sakai}, {Watanabe}, {Ceccarelli}, {Lefloch},
  \& {Yamamoto}}]{Zhang_Rotation_2018}
{Zhang}, Y., {Higuchi}, A.~E., {Sakai}, N., {et~al.} 2018, \apj, 864, 76

\bibitem[{{Zhu}(2019)}]{Zhu_inclined_2019}
{Zhu}, Z. 2019, \mnras, 483, 4221

\end{thebibliography}

\begin{appendix}

\section{Determination of the redshifted outflow PA}
\label{sec:axis}

We show in Sect. \ref{sec:tomo} that the velocity difference between two radially symmetric positions, at a given altitude z in the flow, is directly linked to the rotation velocity of the outflow. However, this derivation is highly sensitive to the determination of the outflow radial center position. Because of the steep gradient in projected velocity versus radius observed in transverse PV diagrams, a small variation in radial position would cause an important variation in the velocity difference. Therefore, the determination of the position axis is a key concern in the characterization of the specific angular momentum of the outflow.

\begin{figure}
    \resizebox{\hsize}{!}{\includegraphics{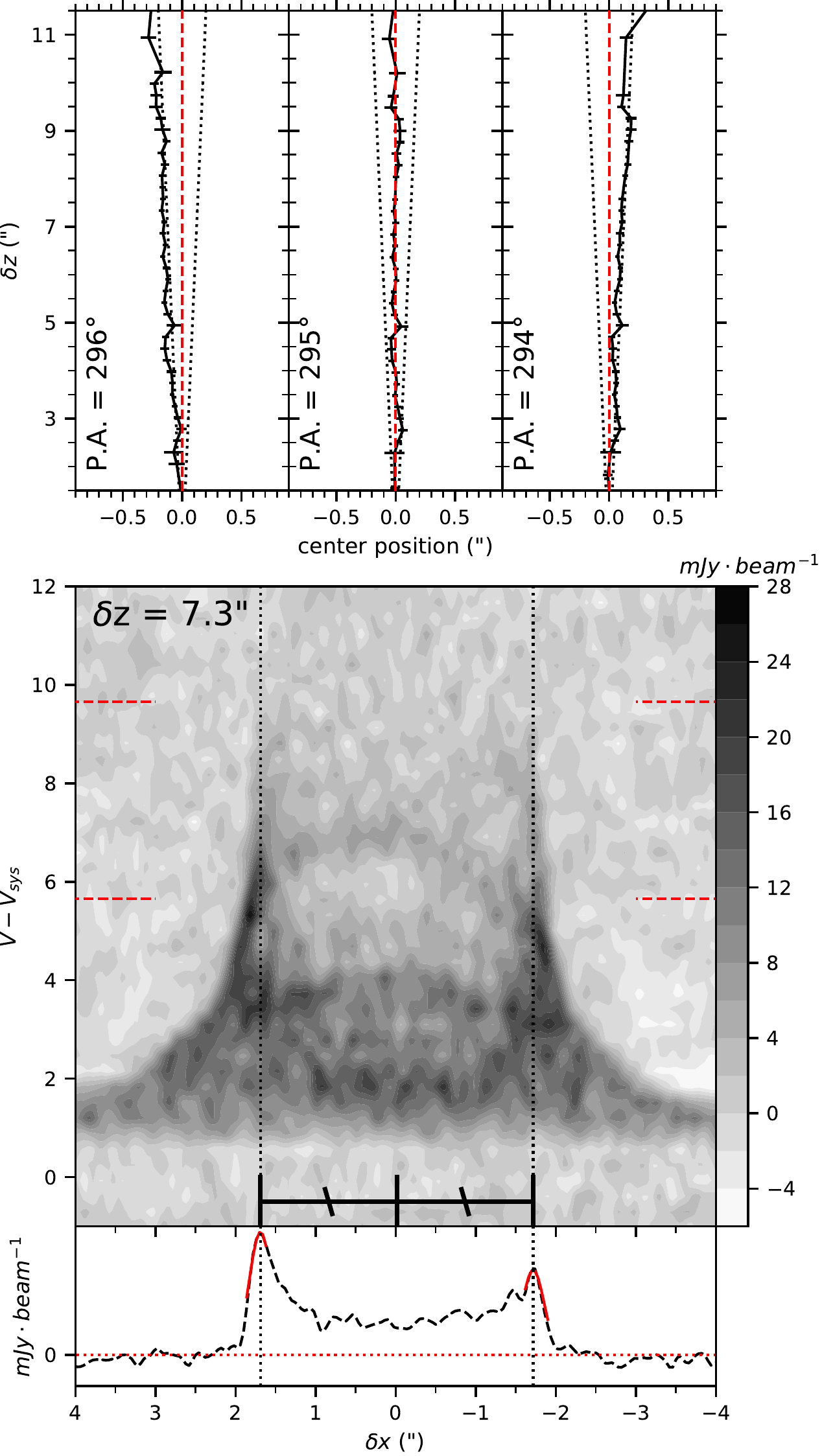}}
    \caption{Determination of the outflow axis PA. {\sl Top panels}: Determination of the high-velocity component radial center as a function of projected heights above the disk for different outflow axis PA orientations. The black dotted lines show an angle of $\pm 1^\circ$ with respect to the central axis. {\sl Middle panel}: $^{12}$CO PV diagram perpendicular to the outflow axis.  $(V-V_{\rm sys})$ unit is \kms. {\sl Bottom panel}: The black dashed profile shows the emissivity integrated from the PV diagram between $(V-V_{\rm sys}) = 6.64$~\kms~to $(V-V_{\rm sys}) = 9.77$~\kms, this domain is indicated by the red dashed lines in the middle panel. The red curves show Gaussian fits used to derive the radial positions of the two edge peaks. The vertical black dotted lines in the bottom and middle panels indicate the positions of these peaks. The average of these two radial positions give the radial center at this height.}
    \label{fig:pa}
\end{figure}

To accurately derive the flow center position as a function of the projected altitude, we adopted the following method, illustrated in Fig.~\ref{fig:pa}. At each projected altitude above the disk $\delta z$, we integrated the emissivity from $(V-V_{\rm sys}) = 6.64$~\kms~to $(V-V_{\rm sys}) = 9.77$~\kms. This gave the radial emissivity profile of the high-velocity emission tracing the limiting inner cone. Indeed as shown in DV20, at large projected velocities, the emission traces an inner cone with almost constant radius.
We derived the radial positions of the two peaks tracing the edges of the high-velocity component (See Fig. \ref{fig:pa}) using Gaussian fitting. 
From the median value of these two positions, we derived the radial center position and its associated uncertainty.
We applied this method after rotating the data cube with three different values of outflow PA (sampled around the disk PA), and the transverse PV diagram was obtained with a slice perpendicular to this axis. The effect of the different PV cut is completely negligible, as the considered variation of outflow P.A is small (less than 2 degrees).

Figure \ref{fig:pa} shows the derived radial offset 
as a function of the projected height for three values of the flow position angle (PA). With a PA of $296^{\circ}$ and $294^{\circ}$, the radial offset increases consistently with a miss-estimation of the outflow PA of $\pm 1^\circ$. The global offset is minimized with a PA of $295 \pm 1^{\circ}$. This is consistent with the atomic redshifted jet PA of 296$^{\circ}$ derived by \citet{mundt_collimation_1991} and the PA of the projected disk in the plane of the sky, determined at $25.7 \pm 0.3^{\circ}$ by DV20 and $24 \pm 1^{\circ}$ by \citet{guilloteau_dual-frequency_2011}. 

The outflow rotation could potentially introduce a bias in this method. Indeed, at a given projected velocity, the two sides of a rotating ring  fall at different projected radial offsets from the axis, inducing an artificial shift of the position centroid of the flow axis. 
However, since a rotating ring produces a tilted ellipse in the transverse PV diagram, centered at $V_{\rm Z} \times \cos(i)$, the radial shift has an opposite sense at velocities above and below $V_{\rm Z} \times \cos(i)$. By integrating emission over a broad range of velocity, we thus averaged out this effect. Moreover, if the variation of radial offset was due to flow rotation, the measured radial offsets should decrease with distance from the source, as the rotation velocity decreases (due to the conservation of angular momentum along conical streamlines). This is not consistent with our results in Fig.~\ref{fig:pa}  where the radial offsets stay constant (at our nominal PA) or increase linearly (at non-nominal PAs) with height.
Fig. \ref{fig:pa} also shows no clear signature of wiggling in the high-velocity component of DG~Tau~B. This is  consistent with the absence of wiggling in the atomic jet observed by \citet{mundt_jets_1987}.  From the maximum error bars observed, we derived an upper limit for the wiggling angle of $\theta \le 0.5^{\circ}$. We consider in the following that the redshifted outflow PA is constant for all layers (velocities)  at $\rm PA=295^{\circ}$, corresponding to the PA derived here for the high-velocity emission.

\section{Characterization of the cusps}
\label{sec:Caract_discrete}

\begin{figure*}[h]
    \resizebox{\hsize}{!}{\includegraphics{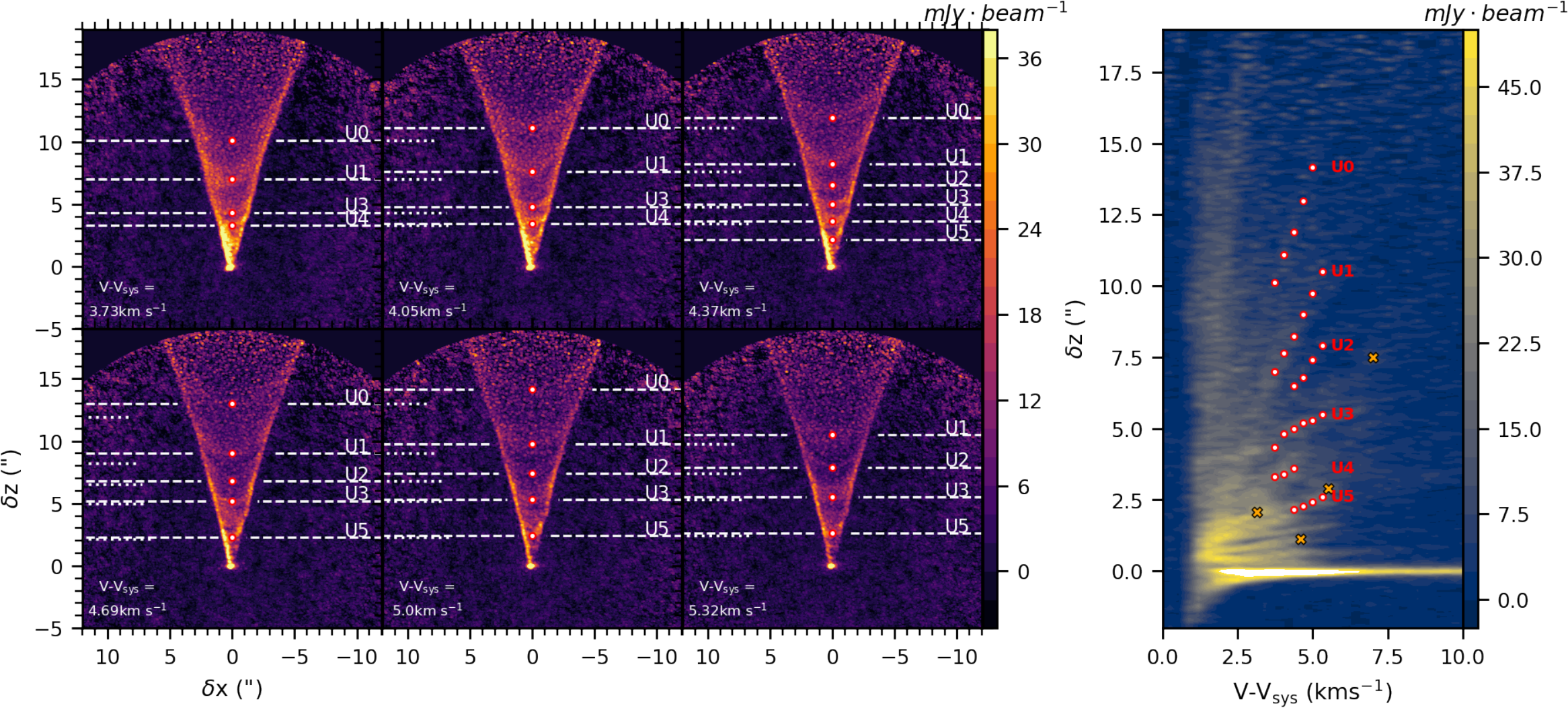}}
    \caption{Characterization of the cusps.
    {\sl Left panels}: Individual $^{12}$CO channel maps at different line-of-sight velocities illustrating the spatial evolution of the cusps. Six cusps, labeled U0-U5, are clearly identified (U4 is only visible on three channel maps). We identify the projected heights of the bottom of the cusps (red circles) and their evolution in the channel maps (white dashed lines).  {\sl Right panel}: Longitudinal PV diagram at $\delta x =0^{\prime\prime}$ averaged over a slice of $\Delta Z = 0.2^{\prime\prime}$. The velocity and height locations of the cusps derived in the channel maps are shown as red dots. The four orange crosses indicate the maximum on-axis projected velocity (back side) of the top of the fit four bright elliptical structures identified in transverse PV diagrams at the corresponding heights (see Fig. \ref{fig:ellipse_fit}).}
    \label{fig:u_tracking}
\end{figure*}

\section{Ellipse fits in the transverse PV diagrams}

\label{sec:ellipse_fit}

\begin{figure*}[h!]
    \resizebox{\hsize}{!}{\includegraphics{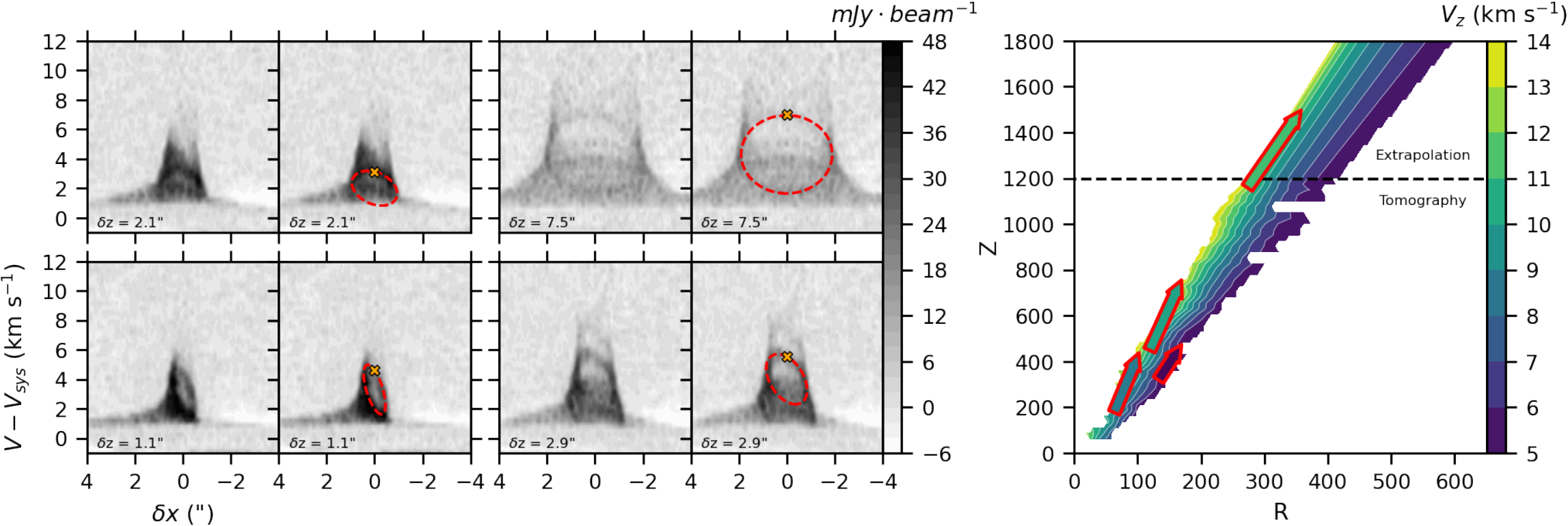}}
    \caption{Constraints on the radial velocity component $V_{\rm R}$.
    {\sl Left panels}: Transverse PV diagrams averaged over a slice of width $\Delta Z=0.2^{\prime\prime}$ at selected positions $\delta z$ along the flow. In red are shown the ellipse fits. We show side by side the PV diagram with and without the fit for more visibility. {\sl Right panel}: $V_{\rm Z}$ tomography of the outflow in the outflow referential, extrapolated beyond $Z > 1200$ au. The red arrows represent the velocity directions determined from the ellipse fits. The arrow colors trace the $V_{\rm Z}$ values determined from the ellipse fit. $R$ and $Z$ units are in au.}
    \label{fig:ellipse_fit}
\end{figure*}

In this section, we discuss an alternative method to derive the radial component of the velocity, not constrained by our tomographic method. 
At a few positions along the flow, transverse PV diagrams clearly show elliptical structures nested inside the main emission (see Fig. \ref{fig:ellipse_fit}). We assumed that each of these ellipses traces  one emitting layer of the outflow.
If the outflow velocity does not vary drastically with height,
one layer of the outflow of fixed radius  will be projected as an inclined elliptical structure in the PV diagram
\citep{louvet_hh30_2018}. The velocity width of the ellipse at $x=0$ is directly linked to the radial velocity component of the layer. We fit ellipses by the naked eye to the structures observed. From these fits, we recovered the radial velocity $V_{\rm R}$ as well as $V_{\rm Z}$ at a few specific positions along the outflow.

Figure \ref{fig:ellipse_fit} shows the $V_{\rm Z}$ profile of the outflow reconstructed with the tomographic technique and  extrapolated until Z $\approx$~2700 au. The red arrows trace the poloidal vectors derived from individual elliptical fits. The angle of the arrow is determined by the ratio between $V_{\rm Z}$ and  $V_{\rm R}$. The length and colors of the arrows correspond to $V_{\rm Z}$. The $V_{\rm Z}$ values derived from the elliptical fits appear consistent with the estimates  from the tomographic study. The poloidal velocity direction is also consistent with the conical lines of constant $V_{\rm Z}$ in the tomography. This suggests that the flow is indeed aligned with these conical lines.

\begin{figure}[h!]
    \resizebox{\hsize}{!}{\includegraphics{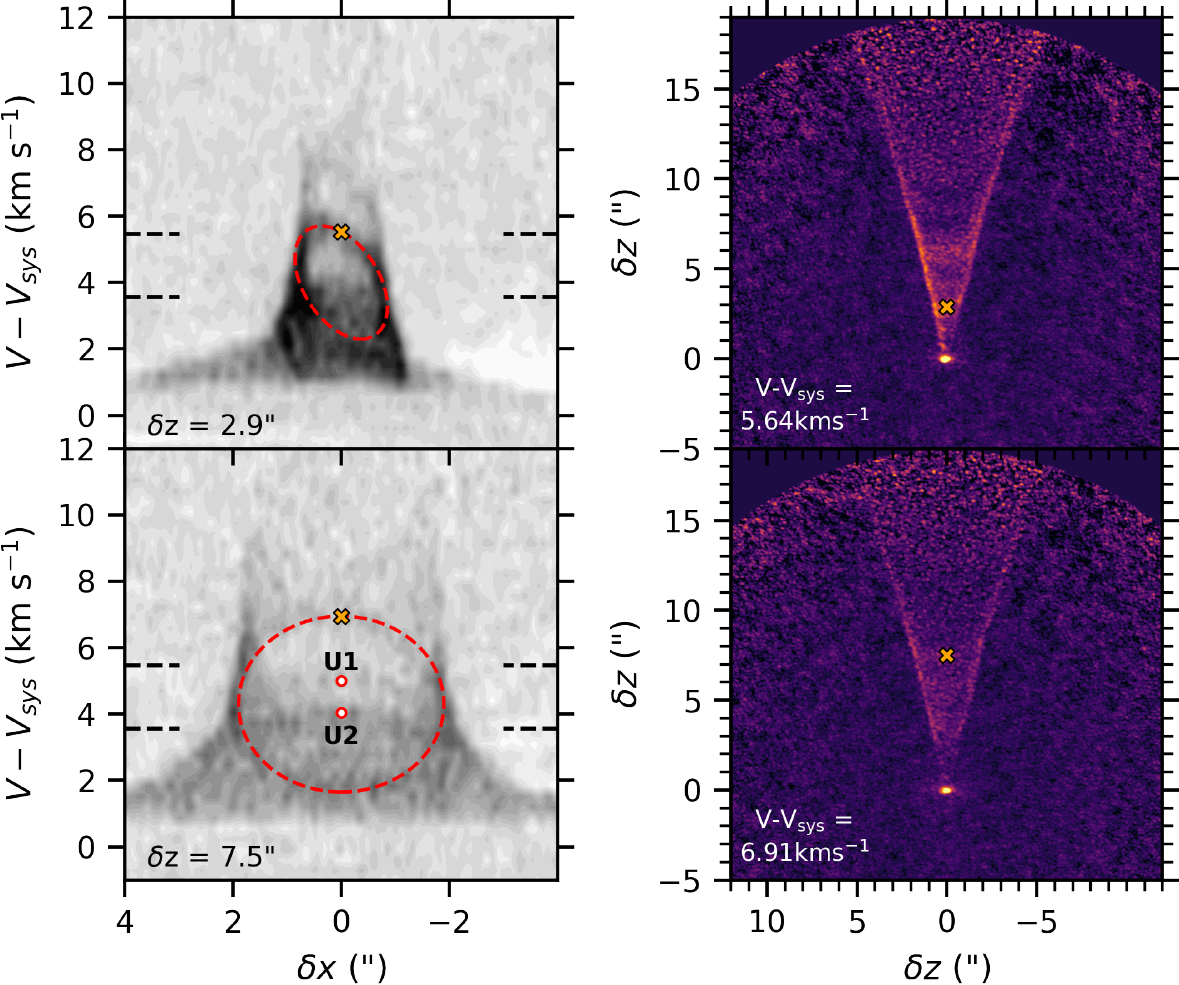}}
    \caption{Illustration of the correspondences between ellipses and cusps. Left panels: transverse PV diagrams at $\delta z = 2.9^{\prime \prime}$ and $7.5^{\prime \prime}$ (from Fig \ref{fig:ellipse_fit}), with red dots indicating the on-axis projected velocities of cusps U1 and U2 at $\delta z = 7.5^{\prime \prime}$. Each of these two cusps corresponds in position and velocity to the maximum velocity (back side) of an ellipse in the PV cut. Right panels: channel maps at the same line-of-sight velocity as the orange crosses in the left panels, showing that elliptical structures seem to correspond to fainter cusps in position-position space, but falling outside of the velocity domain where cusp characterization was possible.}
    \label{fig:discrete_XVCM}
\end{figure}

We studied the relation between the ellipses visible in the transverse PV diagrams and the cusps visible in the channel maps.
The on-axis maximal velocity of the fit ellipses was compared to the cusp location in the on-axis longitudinal PV diagram (see Fig. \ref{fig:u_tracking}). The (R,Z) positions of the fit ellipses were also compared to the cusp locations in the tomographic $V_Z$ map (see Fig. \ref{fig:tomo_discrete}). The ellipse located at $\delta z = 1.1^{\prime\prime}$ comes from a lower altitude region, not included in our cusp analysis because of crowding. The ellipses at $\delta z = 2.1^{\prime\prime}$ and $2.9^{\prime\prime}$ seem to extend the U5 cusp, at higher and lower line-of-sight velocities respectively. The ellipse at $\delta z = 7.5^{\prime\prime}$ is not so clearly associated with a cusp extension in  Fig. \ref{fig:u_tracking}. It could be the high velocity extension of the U3 cusp, or associated with a fainter cusp located between U2 and U3, not included in our analysis. Moreover, Fig. \ref{fig:discrete_XVCM} indicates that the tops of the ellipses located at $\delta z = 2.9^{\prime\prime}$ and  $\delta z = 7.5^{\prime\prime}$ are possibly consistent with a cusp structure which was not included in our study, due to their low S/N. Similarly, Fig. \ref{fig:discrete_XVCM} shows that the cusps U1 and U2 also coincide with internal structures  in the transverse PV diagram, but the stacking of ellipses in the PV diagram makes the identification more confusing than in the channel maps.

The cusps could be precisely located only at moderate projected velocities $< 5.3$~\kms, where they are sufficiently bright (see Fig. B.1), hence they probe outer streamlines; in contrast elliptical structures in transverse PV diagrams are best distinguished at high projected velocities $> 5.3$~\kms where they have less overlap with each other (see Fig. C.1), hence most of them probe inner faster flow streamlines. However, we see clear correspondences between faint cusps in channel maps and the on-axis high-velocity portion of some ellipses in transverse PV cuts, and vice-versa, which demonstrates that they trace different portions of the same underlying substructures, extending across the whole conical outflow. Therefore, assuming a radial flow to deproject the cusp apparent positions seems fully justified.

\section{Biases in the tomographic reconstruction}
\label{sec:bias_tomo}

We studied possible biases in the method presented in Sect. \ref{sec:tomo} to reconstruct poloidal maps of $V_{\rm Z}$ and $j=R \times V_{\phi}$. We first studied the bias introduced by projection effects considering one single layer of the outflow using the analytical solutions from Eqs. \ref{eqx},\ref{eqz},\ref{eqv}. To study the impact of beam convolution and multiple layers of the outflow,  we also applied our tomographic reconstruction method to  the synthetic data cube presented in Sect.~\ref{fig:synthetic} and estimate the difference between the  reconstructed $V_{\rm Z}$ and $j$ and the initial values of the model.

\subsection{Single shell}
\label{sec:ellipse_incl}

\begin{figure*}[h]
   \centering
    \resizebox{0.85\hsize}{!}{\includegraphics{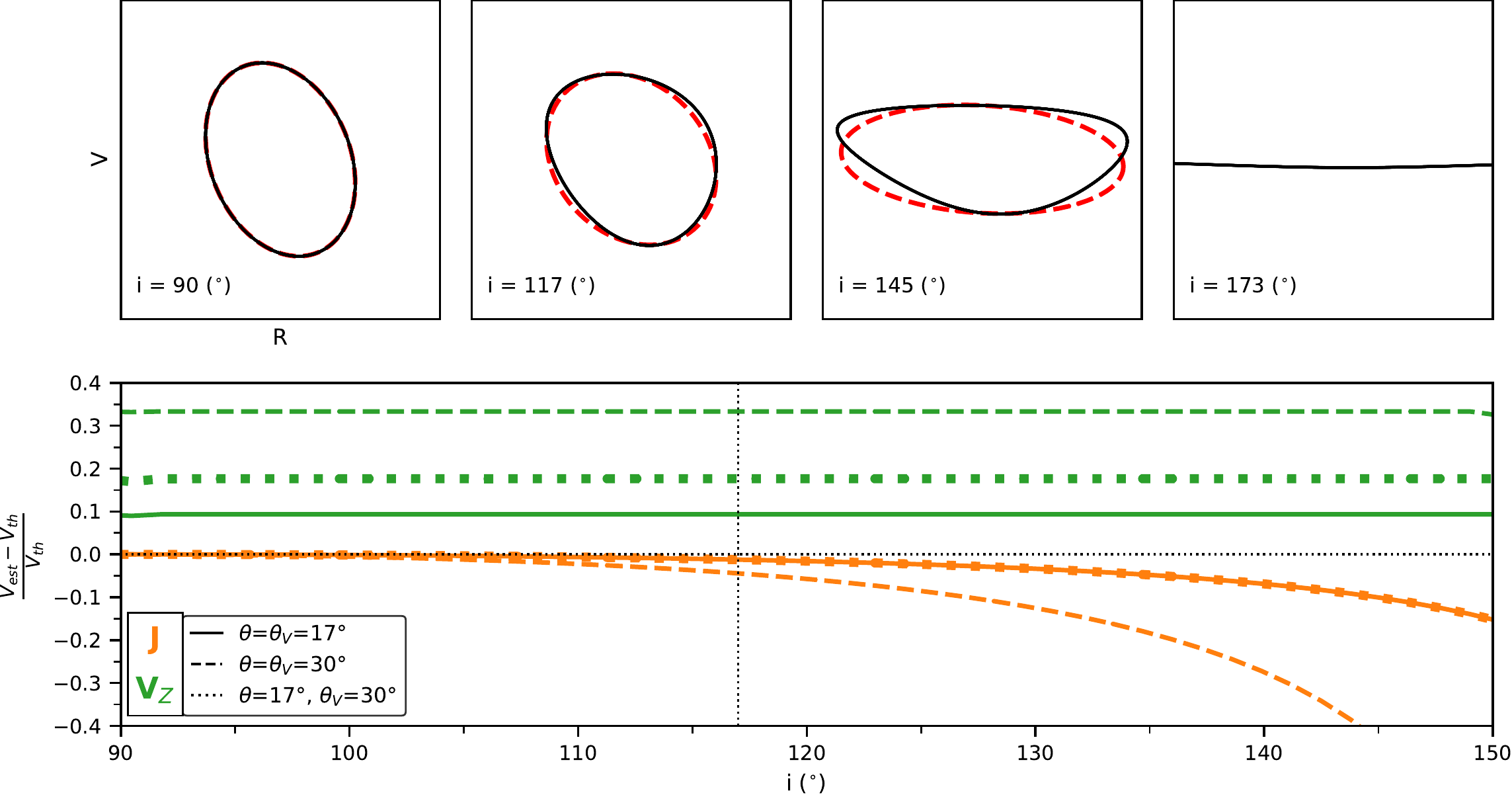}}
    \caption{Biases in the tomographic reconstruction for one single conical shell. {\sl Top panels}: Black curves show computed transverse PV diagrams for one  conical layer with constant $V_{\rm Z}$, $V_{\rm R}$, and $j$ seen at different inclinations  $i$ and with $\theta_V=\theta=17^{\circ}$. The red dashed curve traces an elliptical fit. {\sl Bottom Panel}: Computed relative bias in  $V_{\rm Z}$ (green curves) and $j$ (orange curves) for one conical layer as a function of the inclination and for different combinations of $\theta$ and $\theta_v$ : solid curves: $\theta=\theta_v=17^{\circ}$, dotted curves: $\theta=17^{\circ}$ \& $\theta_v=30^{\circ}$, and dashed curves: $\theta= \theta_v=30^{\circ}$.}
    \label{fig:ellipse_incl}
\end{figure*}

In our tomographic study, we assumed that one shell will be projected as an ellipse in transverse PV diagrams. We also assumed that the projected velocities at the extrema radii allowed us to recover $V_{\rm Z}$ and $V_{\phi}$. We determine in this section biases introduced by these two assumptions.

We assumed a conical shell of radius R(Z) in the outflow referential and with local opening angle $\theta$, such as $R(Z)=R_0 + Z \tan{\theta}$. 
For the DG Tau B inner conical outflow, the opening angle $\theta$ of the layers vary between 17$^{\circ}$ and 12$^{\circ}$, increasing with decreasing velocities. The estimated opening angle of the lower velocity emission contributing to the pedestal is $\theta \simeq 30^{\circ}$.
We defined the parameter $\theta_v$ corresponding to the angle of the velocity vector with the Z axis in the poloidal plane ($V_{\rm R}=V_{\rm Z}\tan{\theta_V}$). For a flow parallel to the conical surface, $\theta=\theta_V$. We also assumed a constant specific angular momentum ($V_{\phi}=j/R(\phi)$) as well as constant $V_{\rm Z}$ and $V_{\rm R}$ over the transverse slit width. Solving Eq. \ref{eqz} with constant $\delta z$ gave:
\begin{equation}
    \label{eqsolR}
    R(\phi)=\frac{\delta z \tan{\theta} + R_0\sin{i}}{\sin{i}-\sin{\phi}\cos{i}\tan{\theta}}
.\end{equation}

In the case $i=\theta$ and $i<\theta$, the cut of the conical outflow will be respectively a parabola and an hyperbola. Consequently, $R(\phi)$ will tend toward infinite. This is not the case for the DG~Tau~B conical outflow.
We then implemented this solution into Eqs. \ref{eqv},\ref{eqx} for multiple values of the inclination. The resulting transverse PV diagrams are
represented in Fig. \ref{fig:ellipse_incl} for $\theta_V=\theta=17^{\circ}$. The PV diagrams were generated using Eqs. \ref{eqv}, \ref{eqz} with the solution from Eq. \ref{eqsolR}. The difference with an ellipse increases with the inclination but is expected to be small for the DG Tau B case  ($i=117^{\circ}$).

We assume in Sect.~\ref{sec:tomo} that the radial edges of the ellipse correspond to $\phi=0-\pi$. However, this is an approximation. The radial extrema of the ellipse correspond to solutions of the equation: $\frac{\partial \delta x}{\partial \phi}=0$. For a conical layer, the solution is obtained for: 
\begin{equation}
    \sin{\phi}=\frac{\tan{\theta}}{\tan{i}}
.\end{equation}

This difference is small in our situation. However, this leads to bias in the estimation 
of $R$, $Z$, $V_{\rm Z}$ and $V_{\phi}$. 
We determined the estimated value of $V_{\rm est,Z}$ and $V_{\rm est,\phi}$  using Eqs. \ref{eqvz},\ref{eqvphi}. We then computed the relative differences with the real values $R_{\rm real}$, $Z_{\rm real}$, $V_{\rm real,Z}$, and $V_{\rm real,\phi}$:
\begin{equation}
    \frac{R_{\rm est}-R_{\rm real}}{R_{\rm real}}=\sqrt{1-\frac{\tan^2{\theta}}{\tan^2{i}}}-1
\end{equation}
\begin{equation}
    \frac{Z_{\rm est}-Z_{\rm real}}{Z_{\rm real}}=-\frac{\tan^2\theta}{\tan^2 i}
\end{equation}
\begin{equation}
    \frac{V_{\rm est,Z}-V_{\rm real,Z}}{V_{\rm real,Z}}=\tan{\theta}\times\tan{\theta_V}
\end{equation}
\begin{equation}
    \frac{V_{\rm est,\phi}-V_{\rm real,\phi}}{V_{\rm real,\phi}}=\sqrt{1-\frac{\tan^2{\theta}}{\tan^2{i}}}-1
.\end{equation}

Hence, $R$, $Z$, and $j$ are systematically underestimated, while $V_{\rm Z}$ is overestimated. 
Figure \ref{fig:ellipse_incl} shows the relative biases due to inclination and projection effects in the estimation of $V_{\rm Z}$ and $V_{\phi}$  for different values of $\theta$ and $\theta_v$. In the conical outflow, where $\theta_V=\theta \le 17^{\circ}$, the inclination bias is expected to be $\leq 10 \%$ in $V_{\rm Z}$ and $Z$, $\leq$ 1 \% in $j$ and $R$.

\subsection{Multiple layers}
\label{sec:ellipse_stack}

\begin{figure*}[h]
   \centering
    \resizebox{0.81\hsize}{!}{\includegraphics{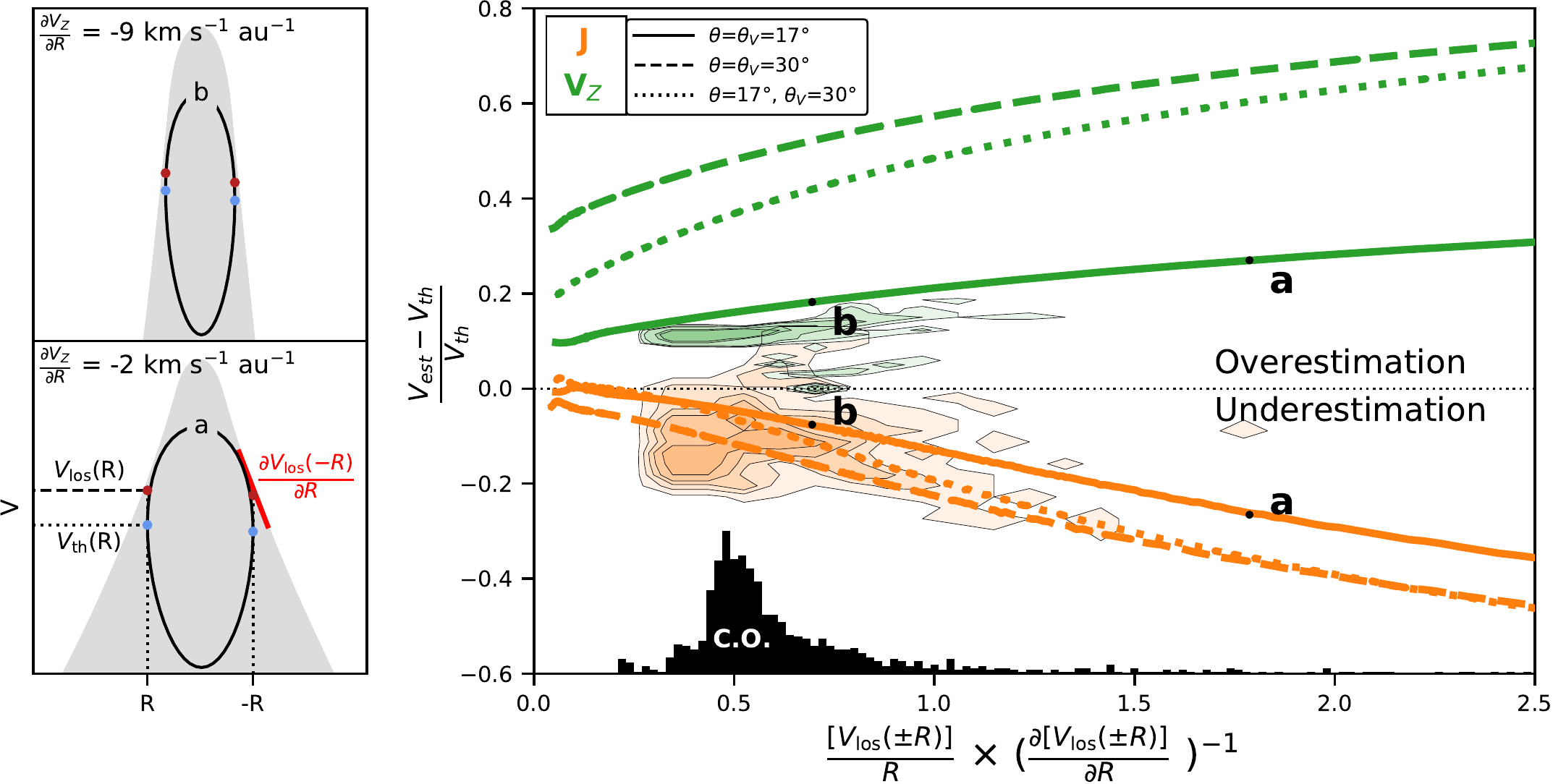}}
    \caption{ Biases in the tomographic reconstruction for a stacking of conical shells. {\sl Left panels}: The gray area shows the limits of the transverse PV diagram predicted for a stacking of conical shells with a shearing factor $f_{\rm sh}$ (see text). In black is shown the ellipse corresponding to a single layer of maximal radius $R$. The blue and red dots correspond  respectively to the velocity of the ellipse at $\pm R$ and the maximal velocity of the PV diagram at $\pm R$. The red dots are the ones used in our method, while the blue dots correspond to the true measurements. The differences in velocities will lead to a bias in our $j$ and $V_{\rm Z}$ estimation. The red line shows the slope of the PV diagram $\frac{\partial V_{\rm los}(\pm R)}{\partial R}$, see text for more details.  {\sl Right panel}: Plot of the relative bias in $V_{\rm Z}$ (green curves) and $j$ (orange curves) due to the shearing aspect of the PV diagram as a function of the $x$ parameter (see text). The solid, dashed, and dotted lines correspond to different configurations of $\theta$ and $\theta_v$. The black histogram, labeled C.O., shows the distribution of the $x$ parameter derived at different heights Z and radii R in the conical outflow. 
   The relative biases corresponding to the a and b ellipses shown in the left panels are represented. The orange (resp. green) colored contours trace the relative biases derived from applying the tomographic method to the synthetic data cube. Contours outline 30 to 90 \% of the distributions in step of 15 \%.}
    \label{fig:ellipse_stack}
\end{figure*}

In the previous section, we estimated the bias in the estimation of $V_{\rm Z}$, $V_{\rm \phi}$ due to projection effects for one single conical layer. However, the DG Tau B transverse PV diagrams shows a clear shear-like velocity structure suggesting a stacking of layers. We modeled this effect directly in the transverse PV diagrams by stacking the elliptical projections for conical layers of increasing radii at origin $R_0$ with the same opening angle $\theta$.
The velocity shear in $V_{\rm Z}$ of the conical layers was defined by a shearing parameter $f_{\rm sh}= \frac{\partial V_{\rm Z}}{\partial R}$. The specific angular momentum of each layer, $j$, was assumed constant with $Z$ and vary between 20 and 90 au~\kms, increasing with increasing radius, such as $j \times V_{\rm P}$ was kept constant to mimic the DG~Tau~B observations.

Due to the stacking, the maximal velocity at $\pm R$ does not perfectly describe the velocity of the ellipse of radius R. This effect is larger when $f_{\rm sh}$ is small. Using a similar procedure as before, we derived the relative difference between the estimation of $V_{Z}$ and $j$  with the tomographic method and the input theoretical values. This was achieved for a range of $f_{\rm sh}$ values between 0.2 to 20 \kms~au$^{-1}$ and with the three  ($\theta$, $\theta_V$) configurations studied in the previous subsection. 

In order to study efficiently this bias, we defined the a-dimensional parameter $x= \frac{[V_{\rm los}(\pm R)]}{R}$ $\times$ ($\frac{\partial [V_{\rm los}(\pm R)]}{\partial R}$ )$^{-1}$,
where we defined $[V_{\rm los}(\pm R)] = 1/2(V_{\rm los}(R)+V_{\rm los}(-R))$.  
This a-dimensional parameter can be derived from the observations. We show in Fig.~\ref{fig:ellipse_stack} the predicted relative biases in $V_{\rm Z}$ and $j$ as a function of this parameter $x$. Biases increase with increasing $x$ values illustrating the effect of the velocity shear. We also show the distribution of observed x values computed at each (R,Z) position in the conical outflow. The x values are concentrated around $\simeq 0.5$, suggesting moderate biases in both $V_{\rm Z}$ and $j$ for the conical outflow where $\theta_V=\theta \le 17^{\circ}$ .

However, this modeling did not include the effect of beam smearing nor the impact of our polynomial fitting method to describe the shape of the PV diagram. In order to study these effects, we directly applied our tomographic method to the synthetic data cube presented in Sect. \ref{fig:synthetic} and determined the relative differences between the computed and the input  $V_{\rm Z}$ and $j$ values at each (R,Z) position along the conical flow. We show in Fig. \ref{fig:ellipse_stack}  the derived relative biases in $V_{\rm Z}$ and $j$ as a function of the $x$ parameter. Their distributions are
broader and flatter than predicted, especially in $j$, likely due to the combination of velocity shear and beam smearing effects.

From this study, we estimate that in the conical part of the outflow (at $(V-V_{\rm sys}) \ge 2$ \kms), the tomographic method suffers from a relative bias  $\leq 15$\% in the estimation of $V_{\rm Z}$ and $\leq 20$\% in the estimation of $j$. The X values extend up to $\approx 1.3$ in the pedestal region. With an opening angle $\theta$, $\theta_v > 30^{\circ}$, a tomographic study of the low velocity component would suffer from a relative bias larger than 60\% and 30\% in the estimation of $V_{\rm Z}$ and $j$ respectively.
 
\section{Wind-driven shell analytical solutions}
\label{sec:Aspect_ratio}

In this section, using Eqs. \ref{eqx}, \ref{eqz},\ref{eqv} and \ref{eq:wds2}, we derive an analytical solution for predicted channel maps in the case of the generalized WDS model introduced in Sect. \ref{sec:WDS}. The WDS model is defined by three parameters: $C$, $\tau$ and $\eta$, see Eqs. \ref{eq:wds2}. The projection on the plane of the sky  ($\delta x$, $\delta z$) for the emissivity map at $(V-V_{\rm sys})=V_{\rm CM}$ could be recovered by solving Eq.~\ref{eqv} with $V_{\rm los} = V_{\rm CM}$ and using Eqs.~\ref{eq:wds2} for $V_{\rm Z}$ and $V_{\rm R}$:
\begin{eqnarray}   
    \label{AppRho}
    \zeta &= &\frac{4V_{\rm CM}\tau C \cos{i}}{\eta^{2} \sin^2{i}} \\
    \label{AppR}
    R&=&-\eta\frac{\tan{i}}{2C}(\sin{\phi} \pm \sqrt{\sin^2{\phi}-\zeta} ) \\ 
    \label{Appx1}
    \delta x &= & R\cos{\phi} \\
    \label{Appz1}
    \delta z & =&CR^2\sin{i}- R\sin{\phi}\cos{i}. 
\end{eqnarray}

Equation \ref{AppR} has no solution in the case $\zeta > 1$ ($V_{\rm CM} \cos{i} >\frac{\eta^{2} \sin^2{i}}{4 \tau C}$). This corresponds to the case where no emission is predicted at $V_{\rm los}=V_{\rm CM}$. Similarly, in the situation $0 \ge \zeta \ge 1$, only a fraction of $\phi$ will be projected such as $\sin^2{\phi} >\zeta$. Developing Eq. \ref{Appz1}, we obtained the following:
\begin{eqnarray}    
    \label{Appz2}
    \delta z & =&\eta\frac{\sin^2{\phi}\sin{i}}{2C}(\eta\tan^2{i} +1) - V_{CM}\tau \tan{i}  \\
    && \pm \eta\frac{\sin{\phi}\sin{i}}{2C}(\eta \tan^2{i} +1)\sqrt{\sin^2{\phi}-\zeta}
.\end{eqnarray}

We reformulated $\delta z$ as $z_0 + \Delta Z \sin{\beta}$ with:
\begin{eqnarray}    
    \label{Appz0}
    z_0 & =&\eta\frac{\sin{i}}{2C}(\eta\tan^2{i} +1) - V_{CM}\tau \tan{i} \\
    \label{Appdz}
    \Delta Z &= &  \eta\frac{\sin{i}}{2C}(\eta\tan^2{i} +1)\sqrt{1-\zeta} \\
    \label{Apptheta}
    \sin{\beta}&=&\frac{-1+\sin^2{\phi}\pm\sin{\phi} \sqrt{\sin^2{\phi}-\zeta}}{\sqrt{1-\zeta}}
.\end{eqnarray}

Similarly, manipulating Eq.~\ref{Appx1}, $\delta x$ could be reformulated as  $\delta x = \Delta X \cos{\beta}$, with the following:
\begin{equation}
    \label{Appdeltar}
    \Delta X = \mp \eta \frac{\tan{i}}{2C}\sqrt{1-\zeta}
.\end{equation}    

Hence ($\delta x$, $\delta z$) trace an ellipse of center (0, $z_0$) and aspect ratio:
\begin{equation}
    \label{AppAspectRatio}
    \Bigl|\frac{\Delta Z}{\Delta X}\Bigr|  =(\eta \tan^{2}{i}+1)|\cos{i}|
.\end{equation}    

Therefore, in the classical WDS models with radial velocity vectors ($\eta=1$), the aspect ratio of the ellipse on the channel maps only depends on the inclination. We represent on the channel maps in Figs. \ref{fig:outer}, \ref{fig:loops}, and \ref{fig:Vmax_HL} the two limiting ellipses computed at $V-\delta V/2$ and $V+ \delta V/2$ to take into account the  width of the channel map. 

\section{Global model}
\label{sec:Global_model}

\begin{figure}
    \resizebox{\hsize}{!}{\includegraphics{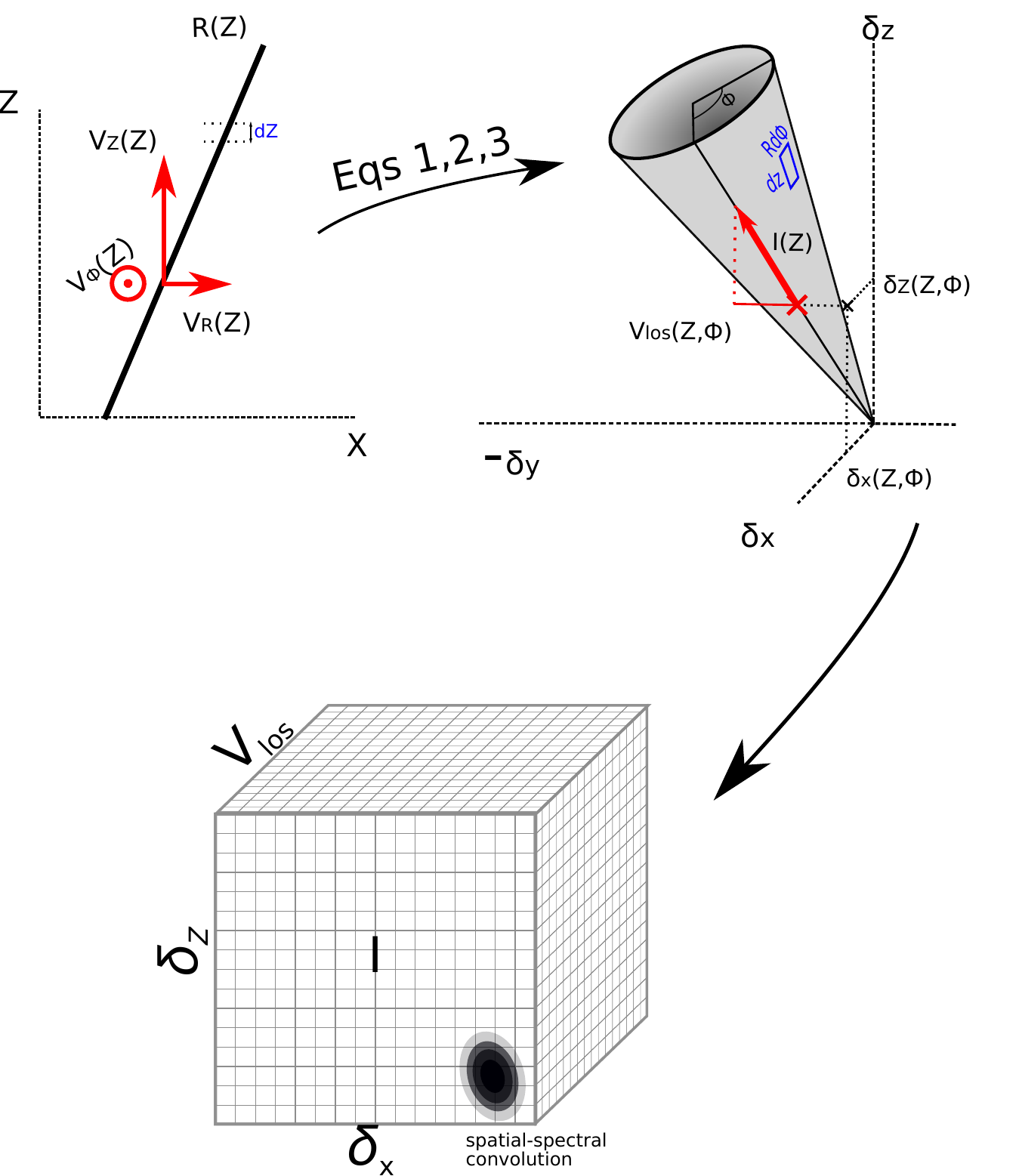}}
    \caption{Schematic representation of the method used to create synthetic data cube.}
    \label{fig:Platypus2}
\end{figure}

In order to comprehend the impact of projection or convolution effects on the observations, we developed a locally axisymmetric code that allows us to simulate optically thin observations of simple outflow models. We took advantage of the axisymmetric hypothesis that permits to reduce the complexity of the model by defining at each height the radius $R(Z)$ and the cylindrical velocities $V_{\rm z}$, $V_{\rm r}$ and $V_{\phi}$ (See the schematic view of Fig.~\ref{fig:referential}). The dependency between the height and the radius or the velocities vary with the model used. We then created at each height an emitting ring with azimuth parametrized
with $\phi$. As the morphology is axisymmetric, neither the radius nor velocities depend on $\phi$. Under the assumption of optically thin emission, the emissivity at position ($Z$,$\phi$) is proportional to the elementary volume $dV=R(Z)d \phi dRdZ$. We added an additional variation of the emissivity with the height and radius as a power-law with  parameters $\alpha$ and $\beta$ respectively. Proper modeling of the emissivity would require the temperature and chemistry to be solved, which is well beyond the scope of this model. The positions of the outflow emission on the data cube ($\delta x$,$\delta y$, $V_{\rm proj}$) were then defined by Eqs. \ref{eqx},\ref{eqz},\ref{eqv}.

We then created a data cube with the same spectral and spatial resolution than our observations, and placed on each point ($\delta x$,$\delta y$, $V_{\rm proj}$) the emissivity I. Under the assumption of optically thin emission, we summed each emissivity corresponding to the same positions on the data cube.
We set a step size of 1$^{\circ}$ for $\phi$ and a fraction of the spatial pixel for Z. We then convolved the data cube by a 2D Gaussian matching the spatial beam characteristics in order to fully simulate the ALMA observations. The code, written in Python 3, is publicly online\footnote{https://github.com/Alois-deValon/Axoproj}.

\section{Channel maps of disk-wind models}
\label{sec:CM_DW}

\begin{figure*}
    \centering
    \resizebox{0.95\hsize}{!}{\includegraphics{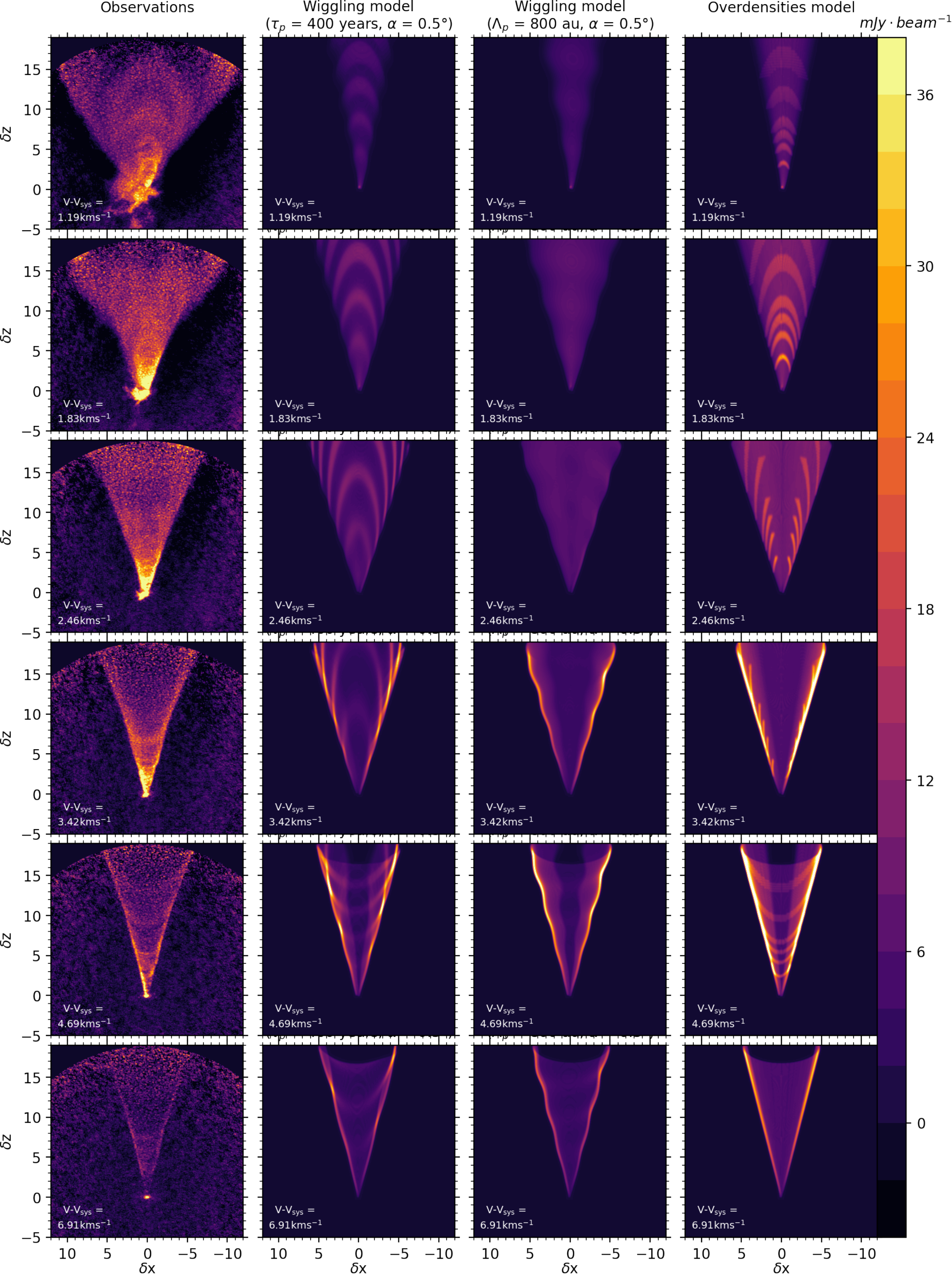}}
    \caption{$^{12}$CO observed (first column) and synthetic  disk-wind channel maps at selected line-of-sight velocities. The second and third columns correspond to the two disk-wind models with axis precession presented in Sect. \ref{sec:wigg}. The last column corresponds to a  disk-wind model with axisymmetric density enhancements, as presented in Sect. \ref{sec:Density_DW}. $\delta x$ and $\delta z$ units are  arcseconds.}
    \label{fig:CM_DW}
\end{figure*}

\end{appendix}

\end{document}